\documentclass[apj,twocolumn]{emulatearxiv}
\usepackage{amsmath}

\begin{document}

\def\lsim{\mathrel{\rlap{\lower 4pt \hbox{\hskip 1pt $\sim$}}\raise 1pt
\hbox {$<$}}} 
\def\gsim{\mathrel{\rlap{\lower 4pt \hbox{\hskip 1pt $\sim$}}\raise 1pt
\hbox {$>$}}}

\title{Electron Capture Supernovae of Super-AGB Stars: Sensitivity on Input Physics}

\author{Shing-Chi Leung\thanks{Email address: shingchi.leung@ipmu.jp}}

\affiliation{Kavli Institute for the Physics and 
Mathematics of the Universe (WPI), UTIAS, The 
University of Tokyo, Kashiwa, Chiba 277-8583, Japan}

\affiliation{TAPIR, Walter Burke Institute for Theoretical Physics, 
Mailcode 350-17, Caltech, Pasadena, CA 91125, USA}

\author{Ken'ichi Nomoto\thanks{Email address: nomoto@astron.s.u-tokyo.ac.jp}}

\affiliation{Kavli Institute for the Physics and 
Mathematics of the Universe (WPI), UTIAS, The 
University of Tokyo, Kashiwa, Chiba 277-8583, Japan}

\author{
Tomoharu Suzuki\thanks{Email address: tsuzuki@isc.chubu.ac.jp}}

\affiliation{College of Engineering, Chubu University, 
1200 Matsumoto-cho, Kasugai, Aichi 487-8501, Japan}

\date{\today}

\begin{abstract}

Stars of $\sim$ 8 -- 10 $M_{\odot}$ on their main-sequence form strongly
electron-degenerate oxygen-neon-magnesium (ONeMg) cores and become super-AGB stars.
If such an ONeMg core grows to 1.38 $M_\odot$, 
electron captures on $^{20}$Ne$(e,\nu_e)^{20}$F$(e,\nu_e)^{20}$O
take place and ignite O-Ne deflagration around the center.
In this work, we perform two-dimensional hydrodynamics simulations of
the propagation of the O-Ne flame to see whether such a flame 
triggers a thermonuclear explosion or 
induces a collapse of the ONeMg core due to subsequent electron capture behind the
flame.
We present a series of models to explore how the outcome depends on
model parameters for a central density ranging between $10^{9.80}$
to $10^{10.20}$ g cm$^{-3}$, flame structures of both centered and
off-centered ignition kernels, special and general relativistic
effects, turbulent flame speed formulae and the treatments of laminar
burning phase.
We find that the ONeMg core obtained from stellar evolutionary
models has a high tendency to collapse into a neutron star.
We obtain bifurcation between the electron-capture induced collapse and
thermonuclear explosion. We discuss the implications of the ECSNe in 
chemical evolution 
and the possible observational signals of this class of supernovae.

\end{abstract}



\section{Introduction}
\label{sec:intro}

\subsection{Formation and Evolution of Degenerate ONeMg Cores}

Stars with a mass between 8 and 10 $M_{\odot}$ have an 
interesting transition from massive white dwarf (WD) 
formation \citep[e.g.,][]{Sugimoto1980, Nomoto1988, Nomoto2013}
to core collapse supernova (CCSN) \cite[e.g.,][]{Arnett1996}.
WDs with masses below $M_{\rm up, C} = 7 \pm 2 ~M_{\odot}$
can form a carbon-oxygen (CO) WD \cite[e.g.,][]{Nomoto1982, Karakas2017}.
Above that, C-burning in the core produces an oxygen-neon-magnesium (ONeMg) core.
The helium shell expands and is dredged up by surface convection \citep{Nomoto1987}. 
The final ONeMg core mass depends on the competition between the mass
deposition from H-burning in the envelope and the mass 
loss by thermal pulses \citep[e.g.,][]{sie07,pumo09,langer12}). 
In the transition mass, Ne can burn spontaneously in an ONeMg core 
above 1.37 $M_{\odot}$ \citep{Nomoto1984}, while a hybrid CO-ONeMg WD 
can form near this transition mass \citep{Doherty2015, Woosley2015}.

Once the ONeMg core reaches a central density of $10^9$ g cm$^{-3}$,
the odd number isotope pairs ($^{25}$Mg, $^{25}$Na), ($^{23}$Na,
$^{23}$Ne) and ($^{25}$Na, $^{25}$Ne) undergo URCA processes 
(electron captures and $\beta$ decays, see e.g. \cite{Schwab2017}
for the CO WD case) with their rates computed in for example \cite{Toki2013, Suzuki2016}.
At $10^{9.6}$ g cm$^{-3}$, electron capture on $^{24}$Mg
may further create a steep electron fraction $Y_{{\rm e}}$ gradient,
which may trigger semi-convection. The lowered $Y_{{\rm e}}$ makes
the core further contract \citep{Miyaji1980,nom82crab,Nomoto1987}. 
Meanwhile, electron captures heat the core by its gamma-ray deposition.
Depending on the treatment of convection, one can use
Schwarzschild criterion \citep{Miyaji1980,Nomoto1987,taka13,Jones2014} or
Ledoux criterion \citep{miya85,hashi93,Jones2013,Schwab2015}.
They give a range of O/Ne ignition densities from $10^{9.95}$ g cm$^{-3}$ (Ledoux 
criterion) to $10^{10.2}$ g cm$^{-3}$ (Schwarzschild criterion).
However, the exact runaway density is unclear because even
with Ledoux criterion, convection is unstable which may delay the 
nuclear runaway by transporting the nuclear energy from O- and 
Ne-burning away. In this sense, $10^{9.95}$ g cm$^{-3}$ is the 
lower limit of the runaway density.

Electron capture supernovae (ECSNe) are one of the channels for low-mass neutron 
star (NS) formation, similar to the accretion-induced collapse \citep{Canal1976}.
However, the full picture of how such low mass NS forms
remains a matter of debate due to the limited observational constraints 
\cite[see e.g.,][]{Mochkovitch1989,Yoon2007,Dessart2006}.

\subsection{Physics of ONe-Deflagration}

Near the end of the super-AGB star evolution, the ONeMg core of can  
attain a central density $\sim 10^{10}$ g cm$^{-3}$ where weak 
interactions are important \citep{Nomoto1984}.
Above $\sim 10^9$ K, the typical burning timescale of O in the core becomes
shorter than the hydrodynamics timescale $t_{{\rm hyd}}$. 
The nuclear reactions are no longer regulated by heat loss or expansion.
A nuclear runaway takes places and its location depends
on the convection timescale $t_{{\rm conv}}$.
When the timescales form a hierarchy $t_{{\rm nuc}} < t_{{\rm hyd}} < t_{{\rm conv}}$,
thermonuclear runaways can take place near the
center in the form of a nuclear deflagration wave \citep{Timmes1992}.
The rapid electron captures in the burnt ash lower the 
electron fraction $Y_{{\rm e}}$.
The detailed evolution is dependent upon the initial model
and related input physics, including the runaway
density, position, and geometry of the O-Ne deflagration, 
the turbulent flame physics and the transition from laminar
flame to turbulent flame regime.
Therefore the final fate of the ECSN is less obvious because   
electron captures can slow down the propagation of the nuclear flame
or can even trigger the collapse. 
To model the turbulent flame properly,
multi-dimensional simulations are necessary.

Nuclear deflagration has been extensively
studied and modeled in the Type Ia supernova literature
\citep{Reinecke1999b, Reinecke2002a, Reinecke2002b, Roepke2005a, Roepke2005b,
Roepke2007, Ma2013, Fink2014, Long2014}. 
By electron conduction, the deflagration wave propagates
with a sub-sonic velocity and the speed increases with density 
\citep{Timmes1992}. Deflagration is susceptible to fluid advection
and hydrodynamical instabilities including Rayleigh-Taylor
instabilities, Kelvin-Helmholtz instabilities and 
Landau-Derrrieius instabilities \citep{Timmes1992,Livne1993,Roepke2004a,Roepke2004b,
Bell2004a, Bell2004b}. 
In general, the flame has a complex geometry,
and an explicit front-capturing scheme is often essential to
accurately describe the evolution of the deflagration wave \citep{Osher1988}.
Due to the sub-sonic nature of the flame, the burnt matter may
have sufficient time to expand and relax isobarically \citep{Khokhlov1997}, which creates
a density contrast in the fuel. Matter with a high density ($>5 \times 10^9$ K)
may release sufficient energy to make the matter enter the nuclear statistical 
equilibrium (NSE). The photo-disintegration of iron-peak elements in the 
ash and its further electron capture may also alter the structure of 
the laminar deflagration wave.

\subsection{Motivation}

The uncertainties of the input physics in stellar evolution 
near the ignition of the ONeMg core result in uncertainties
about the initial models. 
The uncertainties originate from the needs of an extensive nuclear
network for the weak interaction process, the treatment of URCA
process and its associated convection, and the possibility of 
(semi-)convection near the core before the onset of nuclear runaway. 
As a result, the ignition density of ECSN, the position, and size 
of the nuclear runaway are not yet well constrained.
Early work shows that the results are sensitive to the ignition density \citep{Gershtevin1977, Chechetkin1980}.
Furthermore, the results depend on the nature of the turbulent flame \citep{Nomoto1991},
where multi-dimensional simulations are naturally required.
The first three-dimensional model of the deflagration
phase \citep{Jones2016} demonstrates
the importance of the input physics. Their
models show that the Coulomb corrections 
in the equation of state can result in different explosion strengths.
The choices of the convection criteria, which affect the ignition density,
can also alter the final explosion strength. 
In \cite{Jones2018}, the nucleosynthesis based on their previous work
is computed with a large nuclear network including 5234 isotopes.
Their models can reproduce features of a recently observed
Mn-enhanced low mass WD LP 40-365 \citep{Raddi2018}.
These results inspire us to examine carefully the role of the initial model 
and various input 
physics of the ECSN to
determine the final fate of the ECSN. 
We use the two-dimensional hydrodynamics code for the computation.
Two-dimensional
models allow us to explore the parameter space systematically
in reasonable computational time.

In Section \ref{sec:methods} we briefly
outline our hydrodynamics code and the updates employed to 
model the pre-collapse phase. In Section \ref{sec:results} we present
our parameter study, which includes an array of models
which follow the evolution of ONeMg cores
with different configurations.
This aims at studying the post-runaway evolution 
of the ONeMg core at different (1) central densities,
(2) initial flame structures, (3) initial flame positions,
(4) gravity models, 
(5) flame physics,
(6) pre-runaway configurations and
(7) initial composition. 
In Section \ref{sec:discussion} we discuss
how our results can be understood collectively
for future models given by stellar evolution. 
We also compare our results with
the representative models in the literature.
Then, we discuss the possible observational 
constraints on ECSN. At last, we present our conclusions.
In the appendix we provide the 
resolution study of our ECSN models. We also
present briefly the possible observational
consequences when the ECSN collapses to form a NS by carrying out one-dimensional 
simulations with neutrino transport (the advanced leakage scheme).

\section{Methods}
\label{sec:methods}

We use the two-dimensional hydrodynamics code developed for 
supernovae and nucleosynthesis. We refer readers to 
\cite{Leung2015a, Leung2015b, Leung2016, Nomoto2017a, Leung2017d} for a detailed 
description of the code and its previous applications. 
We also refer the readers to \cite{Nomoto2017b}
for the evolutionary path of an ECSN before the onset of nuclear runaway.  
In general, the input physics of an ECSN is similar to a Type Ia supernova
since nuclear reactions and electron captures are the 
principle input physics. In Table \ref{table:input_physics}
we tabulate the governing physics and their typical values
for these two types of simulations to characterize the principle 
similarities and differences.

\begin{table*}
\begin{center}
\caption{Comparison between the input physics of ECSN and Type Ia supernova.}
\begin{tabular}{|c|c|c|}
\hline
input physics & ECSN & Type Ia supernova \\ \hline
central density & $\sim 10^{10}$ g cm$^{-3}$ & $10^{7} - 10^{10}$ g cm$^{-3}$ \\ \hline
mass & 1.38 & 0.9 - 1.38 \\ \hline
$Y_{{\rm e}}$ range & 0.37 - 0.50 & 0.44 - 0.50 \\ \hline
composition & ONe-rich matter & CO-rich matter \\ \hline
peak temperature & $\sim 10^{10}$ K & $\sim 10^{10}$ K \\ \hline
energy production & ONe- and Si-burning and NSE & CO-burning, Si-burning and NSE \\ \hline
electron capture & NSE matter & NSE matter \\ \hline
\end{tabular}
\label{table:input_physics}
\end{center}
\end{table*}

\subsection{Hydrodynamics}

The code solves the Euler equations in the cylindrical 
coordinates. The simulation box uses a uniform
$400 \times 400$ grid mesh with a size $\sim 4$ km in both $r-$ and $z-$directions. 
Courant factor is chosen to be 0.25.
Only a quadrant of the sphere is modeled where the inner (outer) boundaries
are chosen to be reflective (outgoing).
We use the fifth-order weighted-essentially 
non-oscillatory (WENO) scheme for the spatial discretization \citep{Barth1999}
and the five-step third-order
non-strong stability-preserving Runge-Kutta (NSSP-RK) scheme \citep{Wang2007} 
for the time-discretization. We use the Helmholtz equation of 
state \citep{Timmes1999b}. This equation of state includes 
the contributions of an ideal electron gas at
arbitrarily degenerate and relativistic levels, 
ions in the form of a classical ideal gas, 
photons with the Planck distribution and the 
electron-positron pairs. The level-set method is 
used for tracking the flame geometry inside the ECSN.

We use the same turbulent
flame prescription used in our SN Ia work. 
The effective flame propagation speed is a function of 
to the laminar flame speed $v_{{\rm lam}}$ and 
the local velocity fluctuations due to turbulence $v'$
(see also \cite{Pocheau1994, Niemeyer1995b, Schmidt2006b, Leung2015a}
for the general formulation of a turbulent nuclear flame).
In this work, we choose
the flame models proposed in \cite{Schmidt2006b}.
The laminar speed is a function of density and $^{16}$O mass
fraction given in \cite{Timmes1992}.
The one-equation model \citep{Niemeyer1995b} is used 
for modeling the growth and the decay of sub-grid scale turbulence.
We define the specific kinetic energy density in the sub-grid
scale $q_{{\rm turb}} = |\vec{v}'|^2/2$. This energy density
is a scalar which follows fluid advection
and exchanges energy with the internal energy of the fluid.
Depending on the context, the source terms of sub-grid turbulence $\dot{q}_{{\rm turb}}$ can contain
different terms. In a star, 
$\dot{q}_{{\rm turb}} = \dot{q}_{{\rm prod}} + \dot{q}_{{\rm diss}} + \dot{q}_{{\rm comp}} + 
\dot{q}_{{\rm RT}} + \dot{q}_{{\rm diff}}$. The terms on the 
right hand side stand for the source terms by shear
stress, turbulence dissipation, turbulence production
by compression, Rayleigh-Taylor instabilities 
and turbulent diffusion. 

\subsection{Microphysics}

\begin{figure*}
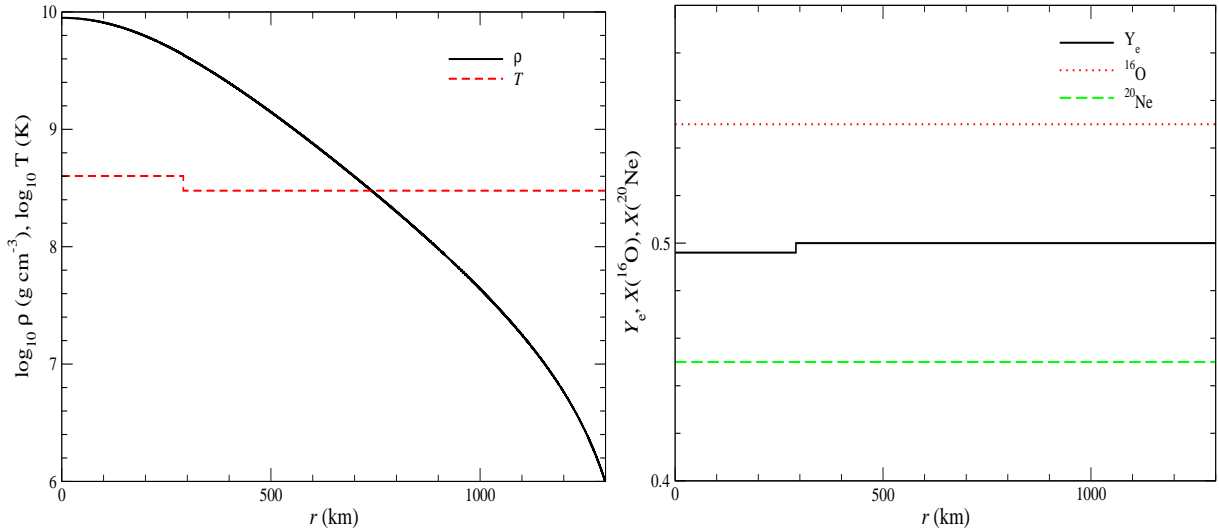

\centering
\includegraphics*[width=8cm,height=7cm]{fig1a.eps}
\includegraphics*[width=8cm,height=7cm]{fig1b.eps}
\caption{(left panel) Initial density and temperature profiles of Model c3-09950-N.
(right panel) Same as the left
panel, but for the $Y_{\rm e}$ and abundances of major isotopes.}
\label{fig:initial_model_plot}
\end{figure*}

In this article, we follow the  
burning scheme prescription proposed in 
\cite{Townsley2007}. This improves the description
of the chemical composition in the ash, which can be very different
from the currently used 7-isotope network (including 
$^{4}$He, $^{12}$C, $^{16}$O, $^{20}$Ne, $^{24}$Mg, $^{28}$Si and $^{56}$Ni).
We introduce the quantities 
$\bar{Y}$, $\bar{q}_B$ and $\phi_i$ ($i = 1, 2, 3)$. They 
represent the inverse of mean atomic mass ($1 / \bar{A}$), 
binding energy and the burning progress variables, and  
they follow the fluid advection.
Operator splitting is used to solve separately 
fluid advection and nuclear reactions within and behind
the deflagration wave. In the 
hydrodynamics phase, we solve the left-hand side of the below equations
without the source terms, including
\begin{eqnarray}
\frac{\partial \bar{Y}}{\partial t} + \vec{v} \cdot \nabla \bar{Y} = \dot{\bar{Y}}, \\
\frac{\partial \bar{q}_B}{\partial t} + \vec{v} \cdot \nabla \bar{q}_B = \dot{\bar{q}}_B.
\end{eqnarray}
After each step, the mean atomic mass $\bar{A}$ and mean atomic 
number $\bar{Z}$ are reconstructed by $1 / \bar{Y}$ and 
$Y_{\rm e} / \bar{Y}$. $\bar{A}$ and $\bar{Z}$ are passed to the 
equation of state subroutine for finding other
thermodynamics quantities including the pressure
and its derivatives with respect to the local density and 
temperature. 

After the hydrodynamics substep, we solve the nuclear burning phase.
$\phi_1$, $\phi_2$ and $\phi_3$ represent the 
burning of $^{20}$Ne, burning until nuclear quasi-statistical equilibrium (NQSE)
and that from NQSE to NSE. The level-set method is used 
for controlling the energy release by $\phi_1$.
To prevent burnt matter from
repeatedly releasing energy due to numerical diffusion, 
$\phi_1$, $\phi_2$ and $\phi_3$ are restricted to be
monotonically increasing and $\phi_2$ ($\phi_3$) is 
allowed to evolve only when the burning represented by $\phi_1$ ($\phi_2$) 
is completely finished. Their evolution also satisfies
the following equation
\begin{equation}
\frac{\partial \phi_i}{\partial t} + \vec{v} \cdot \nabla \phi_i = \dot{\phi_i},
\end{equation}
where $i = 1, 2, 3$. 
We also apply the operator splitting between the advection
term and the source term. 
The source term is solved analytically. 
We remark that when the fluid elements are
not in NSE, no electron capture takes place. 
This is a good approximation because the electron capture
rates below $5 \times 10^9$ K are in general much slower
than the hydrodynamical timescale.

\subsection{NSE and weak interactions}

To couple the hydrodynamics with an extended nuclear reaction
network for matter with a low $Y_{{\rm e}}$ matter typical in an ECSN,
we prepare the NSE composition by the 
495-isotope network with isotopes from $^{1}$H to $^{91}$Tc,
\citep{Timmes1999a} as a function of density $\rho$, temperature $T$ 
and $Y_{{\rm e}}$.
The network also includes the Coulomb correction
factor \citep{Kitamura2000}.
Matter with a temperature above $5 \times 10^9$ K
is assumed to be in the NSE. 
We require the new composition $X_{{\rm new}}$, the new
temperature $T_{{\rm new}}$ and the 
new specific internal energy $\epsilon_{{\rm new}}$ satisfying
\begin{eqnarray}
\frac{\epsilon_{{\rm new}} - \epsilon }{\Delta t} =
N_A (m_n - m_p - m_e) \frac{\Delta Y_{\rm e}}{\Delta t} + \dot{q}_{\nu} + \nonumber \\
\frac{q_{B}(X_{{\rm NSE,~new}}) - q_{B}(X_{{\rm NSE}})}{\Delta t}.
\end{eqnarray}
We remind that the composition in NSE is a function
of density, temperature and $Y_{\rm e}$ that
$X_{{\rm NSE,~new}} = X_{{\rm NSE}}(\rho_{{\rm new}},T_{{\rm new}},Y_{{\rm e,new}})$.
The source terms on the right-hand side are the 
change of the binding energy when the composition
changes, the energy loss due to neutron-proton
mass difference and
the energy loss by neutrino emissions during 
electron captures. 

To obtain the electron capture rates at 
low $Y_{\rm e}$, we follow \cite{Seitenzahl2010,Jones2016}
and extend the electron capture rate table by including 
neutron-rich isotopes. Individual electron capture rates
given in \cite{Langanke2001} and \cite{Nabi1999} are used. 
We solve 
\begin{equation}
\frac{dY_{\rm e}}{dt} =  \sum_i X_i \frac{m_B}{m_i}(\lambda^{{\rm ec}}_i + \lambda^{{\rm pc}}_i + \lambda^{{\rm bd}}_i + \lambda^{{\rm pd}}_i),
\end{equation}
where $m_B$ and $m_i$ are the baryon mass
and the mass of the isotope $i$. $\lambda^{{\rm ec}}_i$,
$\lambda^{{\rm pc}}_i$,
$\lambda^{{\rm bd}}_i$ and $\lambda^{{\rm pd}}_i$ are the 
rates of electron capture, positron capture, 
beta-decay and positron-decay by the isotope $i$
respectively in units of s$^{-1}$.

\section{Models and Results}
\label{sec:results}

\subsection{Initial Model}

In this section, we describe how we prepare 
the initial models for the hydrodynamics run. 
Each ONeMg core is modeled by the 
two-layer structure presented in
\cite{Schwab2015}. We obtain the
necessary data (temperature, $Y_{\rm e}$ and 
composition) by extracting the numerical 
values from Figure 5 in their work. The inner part
imitates the zone where electron captures
take place. It has a lower $Y_{{\rm e}}$ 
and higher temperature in the inner part
and vice verse for the outer part. The inner part has
$(Y_{{\rm e}},T) = (0.496, 4 \times 10^8$ K)
and the outer part has $(Y_{{\rm e}},T) = (0.5, 3 \times 10^8$ K).
We assume that the chemical composition
variation is small enough that it remains 
$X(^{16}$O)$=0.55$ and $X(^{20}$Ne$)=0.45$
throughout the star. To maintain a
high level of hydrostatic equilibrium, we do not 
map the initial model directly from the stellar evolutionary model,
instead, we build the initial model by solving the equations for 
hydrostatic equilibrium
using the given temperature and $Y_{{\rm e}}$ profiles
in the mass coordinate.
In Figure \ref{fig:initial_model_plot} we plot the
initial density, temperature, $Y_{\rm e}$ and 
abundance profiles for Model c3-09950-N.

\subsection{Numerical Models}

\subsubsection{Uncertainties in Stellar Evolutionary Models}

\begin{figure}
\centering
\includegraphics*[width=8cm,height=7cm]{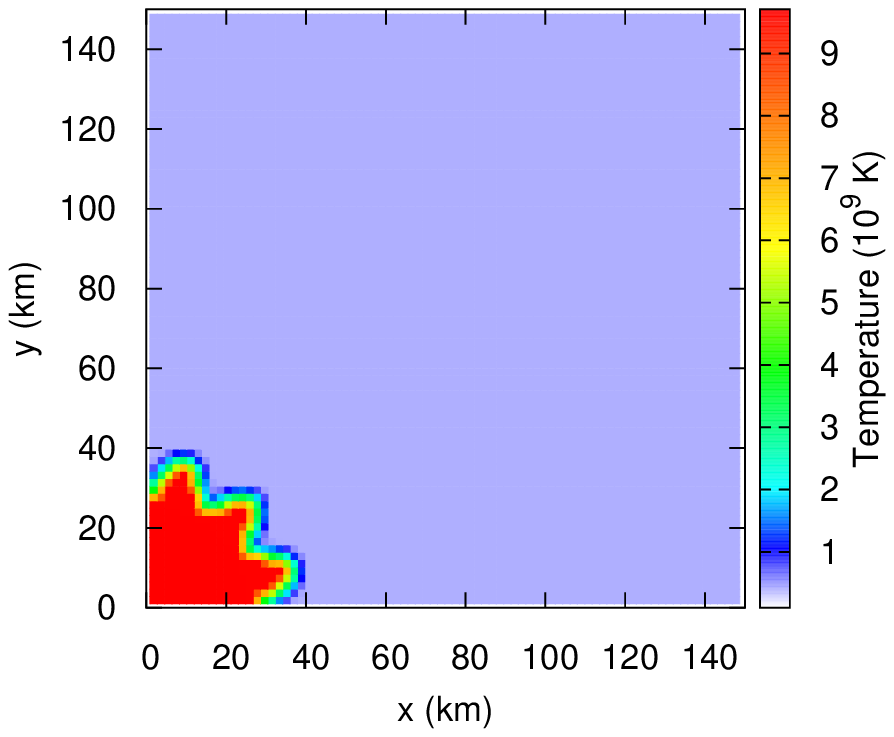}
\includegraphics*[width=8cm,height=7cm]{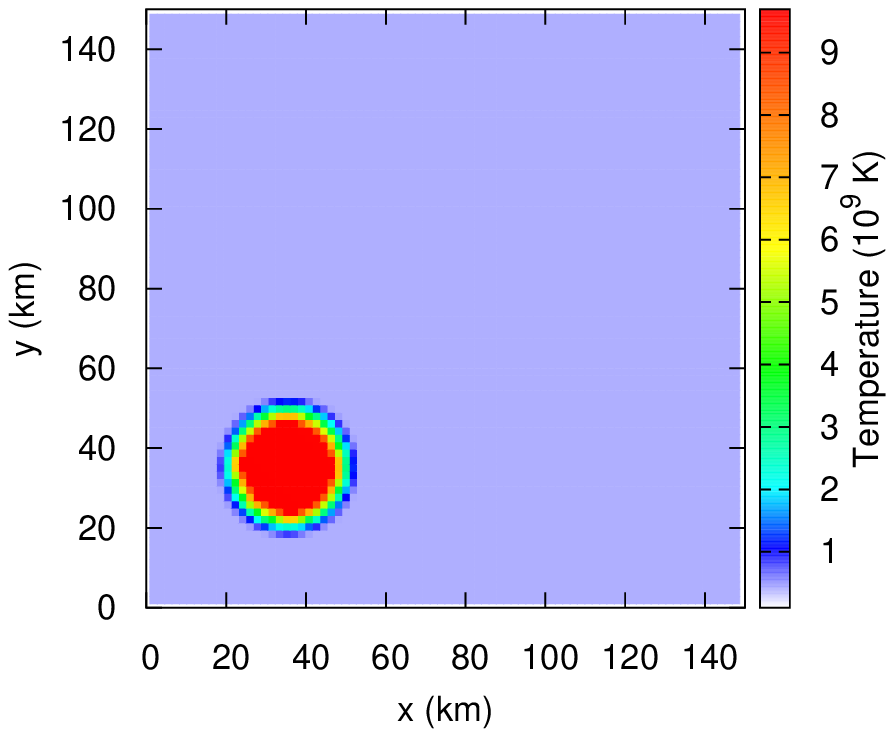}
\includegraphics*[width=8cm,height=7cm]{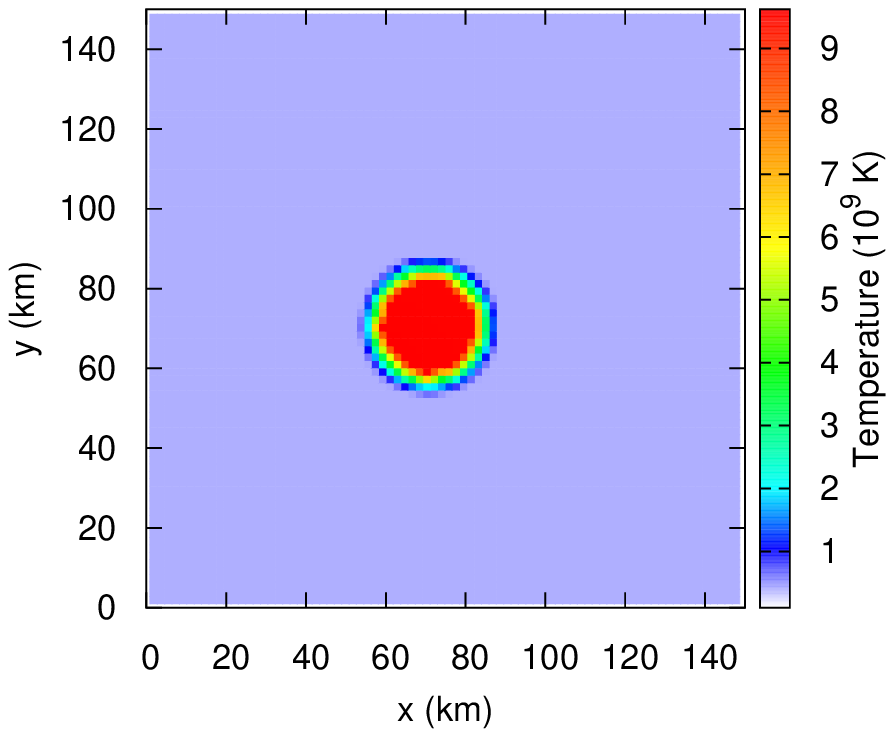}
\caption{(top panel) Temperature colour plot for the initial flame 
profile $c3$ using Model c3-09950-N. (middle panel) Same as the top panel, but 
for the initial flame profile $b1a$. (bottom panel) Same as the top panel, but
for the initial flame profile $b1b$.}
\label{fig:initial_flame_plot}
\end{figure}

The uncertainties in the evolution of an ONeMg core
lead to the ambiguity of the ECSN evolution.
The first one is the convection triggered by electron captures
and O-burning before nuclear runaway takes place.
Depending on the efficiency of core convection after O-burning
has started, the ONe deflagration
density in the ONeMg increases from $\sim 10^{9.95}$ 
(Ledoux criterion) to $\sim 10^{10.2}$ g cm$^{-3}$
(Schwarzschild criterion).
More efficient mixing leads to a higher central density \citep{taka13}.
Therefore, $10^{9.95}$ g cm$^{-3}$
set by the Ledoux-criterion is the lower limit to the deflagration density.
The exact density depends on the competition between the heat
generation by the hydrostatic O-burning and the heat transport
by the core convection.

The second uncertainty
is the initial flame structure. The development of the initial
flame is sensitive to the internal turbulent and convective
motion of the star.  
In stellar evolution, which assumes spherical symmetry, the 
non-radial motion of matter is not included. Local
turbulence can create velocity and temperature fluctuations, 
which are important to trigger the nuclear runaway. 
Efficient convection may smooth out 
the temperature fluctuations in the core and promote centered burning.
The initial
flame in the ONeMg core, similar to SNe Ia, cannot be constrained
without the pre-supernova convective structure.

The third uncertainty is the relativistic correction of gravity.
The impact of such correction is unclear.
In an ONeMg core, the density in the core is sufficiently high that the 
electrons are ultra-relativistic. The contribution of
the pressure and internal energy as a gravity source can be
non-negligible. We want to understand how such corrections
affect the dynamics, and whether the collapse criteria change
with them. 

The fourth uncertainty is how turbulence couples with 
flame propagation. The turbulent flame formalism assumes that 
the effective flame propagation speed is a function of velocity 
fluctuations from eddy motion. However, no experimental data is 
available for flame at such high Reynolds number 
Re $\sim 10^{14}$. There are limited experiments using
the terrestrial flame. In the literature of SNe Ia,
theoretical arguments based on 
self-similarity \cite[see e.g.][]{Pocheau1994,Hicks2015} are often used.
The asymptotic velocity of turbulent flame 
remains unclear.

\subsubsection{Model Description}

The model parameters spanned in this work attempt to overlap
the uncertainties in the stellar evolution modeling. 
In Table \ref{table:models}, we tabulate the initial setting of our
hydrodynamics models. The initial models are built by 
referring to the pre-deflagration model computed in 
\cite{Schwab2015}. In their models, the pre-deflagration
ONeMg core consists of three parts, 1. the outer envelope where no
burning occurs, 2. the outer core where hydrostatic burning of $^{24}$Mg
takes place, and 3. the inner core where electron capture 
and faster nuclear reactions occur
\footnote{We remark that some of the features in the 
stellar evolutionary models are omitted for numerical flexibility. 
The details of some minor elements 
such as $^{24}$Mg are ignored. We do that because the flame 
burning algorithm does not fully distinguish the $^{24}$Mg which 
appears in the original fuel and in the ash after $^{20}$Ne 
is first burnt. To completely avoid doubly releasing energy
from the burning of $^{24}$Mg, we decided to set the remaining 
$^{24}$Mg abundance into $^{16}$O, as both isotopes are burnt 
later than $^{20}$Ne. Also, as we will show in 
coming sections, the distribution of $Y_{\rm e}$ plays a more
important role to the evolution of the ONe WD. 
In general such an approach
might over-estimate the energy production of the flame. In 
Section \ref{sec:composition} we further study how the initial
composition affects the collapse-explode bifurcation. 
Future works with more extensive on-site nuclear reaction network
will be essential to distinguish this degeneracy.}. We use the temperature
and $Y_{{\rm e}}$ profiles to construct our initial models
at different initial central densities. However, we do not resolve the 
innermost core around $10^{-4}$ $M_{\odot}$ which
is equivalent to less than a few grid points in our simulations.

The initial flame configuration is
be where vigorous hydrostatic O-burning takes place. 
We remind 
that the precise geometry of the initial deflagration
requires full multi-dimensional simulations
right after the first nuclear runaway has started. 
We therefore implemented
different flame structures to mimic different
possible scenarios. In particular, we include the $c3$, 
$b1a$, $b1b$ and $b5$ flame structures (See Figure \ref{fig:initial_flame_plot}
of \cite{Reinecke1999b} for graphical illustrations).
The $c3$ flame is the same 
"three-finger" structure as in \cite{Niemeyer1995b}. 
The "finger shape" can enhance the development of
Rayleigh-Taylor instabilities. Also, this shape
prevents the development of enhanced flow along the boundary, which 
might not be physical. 
A $c3$ flame includes an outer radius 
of $\sim 40$ km and an inner radius of 20 km. The 
flame structure is similar to what we have used to trigger
the deflagration phase in \cite{Leung2015a, Leung2015b}
but with a smaller size. 
The $b1a$ flame assumes
a bubble of radius 15 km located at 50 km away from
the center. In Figure \ref{fig:initial_flame_plot}
we plot the temperature colour plot to 
show the initial flame structure
$c3$, $b1a$ and $b1b$ respectively.
We also include variations
of the $c3$ flame by changing
its size to achieve different initial burnt masses
$M_{{\rm burn,ini}}$. This attempts to overcome the uncertainties in the 
unresolved region during the final hydrostatic oxygen burning 
before the onset of thermonuclear runaway.

We also do not keep the details of the innermost part $(\sim 10 - 20$ km) 
in the initial model because there is competition between the
very late-phase electron captures during off-center O-burning and 
its related convective mixing. The exact $Y_{\rm e}$ profile in that
region is unclear. The question is further complicated by the 
initial flame. Despite that, they correspond to a few grids in
the simulation box. In this work, we assume a flat $Y_{\rm e}$ profile
in the core and patch the flame directly on the initial model. The
effect of the initial $Y_{\rm e}$ profile can be refereed from 
Section \ref{sec:time_lapse}.

\begin{table*}

\caption{The initial configurations and the final results
of the simulations. log$_{10} ~\rho_c$ is the logarithmic 
of the initial central density in units of g cm$^{-1}$. $Y_{e,~{\rm in}}$
and $Y_{e,~{\rm out}}$ are the initial electron fraction
of the core and envelope. $Y_{e{\rm, min}}$ is the minimum 
electron fraction reached in the simulation. $t_{{\rm coll}}$
is the time lapse from the beginning of simulation to 
the moment where the central density exceeds $10^{11}$ g cm$^{-1}$.
No $t_{{\rm coll}}$ is given for models which expand.
$M$ and $M_{{\rm burn}}$ are the initial mass
and  the amount of matter burnt by deflagration
in units of $M_{\odot}$. $R$ is the initial radius of
the star in units of $10^3$ km. $E_{{\rm tot}}$ and $E_{{\rm nuc}}$ are the 
final energy and the energy released by nuclear reactions
in units of $10^{50}$ erg. $E_{{\rm tot}}$ is not
recorded for models which collapse. 
"Gravity" means
the choice of gravity source term assuming 
Newtonian ("N") and with relativistic corrections ("R").
"Results" stand for the final fate
of the ONeMg core, where "C" ("E") means that 
the core collapses (expands)
when the simulation is stopped.}

\begin{center}
\label{table:models}
\begin{tabular}{|c|c|c|c|c|c|c|c|c|c|c|c|c|c|}
\hline
Model & ${\rm log_{10}} \rho_c$ & flame & $Y_{e,~{\rm in}}$ & $Y_{e,~{\rm out}}$ & $M$ & $R$ & 
$Y_{e,{\rm min}}$ & $t_{{\rm coll}}$ & $M_{{\rm burn}}$ & $E_{{\rm tot}}$ & $E_{{\rm nuc}}$ & Gravity & Results  \\ \hline
 c3-09800-N & 9.80   & $c3$  & 0.496 & 0.5 & 1.38 & 1.54 & 0.397 & N/A  & 1.12 & -0.16 &  8.19 & N & E  \\
 c3-09850-N & 9.85   & $c3$  & 0.496 & 0.5 & 1.38 & 1.49 & 0.387 & N/A  & 1.21 & 0.23  &  9.67 & N & E  \\
 c3-09900-N & 9.90   & $c3$  & 0.496 & 0.5 & 1.39 & 1.45 & 0.357 & 0.96 & 1.00 & N/A   &  7.92 & N & C  \\
 c3-09900-R & 9.90   & $c3$  & 0.496 & 0.5 & 1.39 & 1.45 & 0.357 & 0.96 & 1.00 & N/A   &  8.68 & R & C  \\
 c3-09925-N & 9.925  & $c3$  & 0.496 & 0.5 & 1.39 & 1.42 & 0.354 & 0.76 & 0.52 & N/A   &  6.83 & N & C  \\
 c3-09950-N & 9.95   & $c3$  & 0.496 & 0.5 & 1.39 & 1.40 & 0.353 & 0.69 & 0.40 & N/A   &  6.83 & N & C  \\
 c3-09975-N & 9.975  & $c3$  & 0.496 & 0.5 & 1.39 & 1.38 & 0.353 & 0.63 & 0.34 & N/A   &  6.70 & N & C  \\
 c3-10000-N & 10.0   & $c3$  & 0.496 & 0.5 & 1.39 & 1.36 & 0.353 & 0.59 & 0.30 & N/A   &  6.56 & N & C  \\ 
 c3-10000-R & 10.0   & $c3$  & 0.496 & 0.5 & 1.39 & 1.36 & 0.353 & 0.59 & 0.30 & N/A   &  6.56 & R & C  \\ 
 c3-10200-N & 10.2   & $c3$  & 0.496 & 0.5 & 1.39 & 1.19 & 0.351 & 0.37 & 0.18 & N/A   &  4.78 & N & C  \\
 c3-10200-R & 10.2   & $c3$  & 0.496 & 0.5 & 1.39 & 1.19 & 0.351 & 0.37 & 0.18 & N/A   &  4.78 & R & C  \\ \hline
b1a-09875-N & 9.875  & $b1a$ & 0.496 & 0.5 & 1.38 & 1.47 & 0.395 & N/A  & 1.20 & 0.25  & 10.18 & N & E  \\
b1a-09900-N & 9.90   & $b1a$ & 0.496 & 0.5 & 1.39 & 1.45 & 0.382 & N/A  & 1.32 & 0.26  & 12.39 & N & E  \\
b1a-09900-R & 9.90   & $b1a$ & 0.496 & 0.5 & 1.39 & 1.45 & 0.358 & N/A  & 1.28 & 0.39  & 11.94 & R & E  \\
b1a-09925-N & 9.925  & $b1a$ & 0.496 & 0.5 & 1.39 & 1.42 & 0.364 & 0.73 & 0.68 & N/A   &  6.21 & N & C  \\
b1a-09950-N & 9.95   & $b1a$ & 0.496 & 0.5 & 1.39 & 1.40 & 0.363 & 0.62 & 0.48 & N/A   &  5.47 & N & C  \\
b1a-10000-N & 10.0   & $b1a$ & 0.496 & 0.5 & 1.39 & 1.36 & 0.360 & 0.51 & 0.34 & N/A   &  4.37 & N & C  \\ \hline
b1b-09900-N & 9.90   & $b1b$ & 0.496 & 0.5 & 1.39 & 1.45 & 0.395 & N/A  & 1.17 & 0.13  &  9.91 & N & E  \\
b1b-09950-N & 9.95   & $b1b$ & 0.496 & 0.5 & 1.39 & 1.40 & 0.388 & N/A  & 1.37 & 0.27  & 13.47 & N & E  \\
b1b-09975-N & 9.975  & $b1b$ & 0.496 & 0.5 & 1.39 & 1.38 & 0.364 & 0.58 & 0.74 & N/A   &  6.98 & N & C  \\
b1b-10000-N & 10.0   & $b1b$ & 0.496 & 0.5 & 1.39 & 1.36 & 0.357 & 0.49 & 0.54 & N/A   &  5.94 & N & C  \\ \hline
mc3-09850-N & 9.85   & $mc3$ & 0.496 & 0.5 & 1.38 & 1.49 & 0.395 & N/A  & 1.10 & 0.23  &  9.17 & N & E  \\ 
mc3-09900-N & 9.90   & $mc3$ & 0.496 & 0.5 & 1.39 & 1.45 & 0.375 & N/A  & 1.36 & -0.37 & 10.07 & N & E  \\
mc3-09925-N & 9.925  & $mc3$ & 0.496 & 0.5 & 1.39 & 1.42 & 0.355 & 0.64 & 0.53 & N/A   &  5.86 & N & C  \\ 
mc3-09950-N & 9.95   & $mc3$ & 0.496 & 0.5 & 1.39 & 1.40 & 0.355 & 0.56 & 0.40 & N/A   &  5.06 & N & C  \\ \hline
bc3-09925-N & 9.925  & $bc3$ & 0.496 & 0.5 & 1.39 & 1.42 & 0.395 & N/A  & 1.14 & 0.13  & 10.12 & N & E  \\ 
bc3-09950-N & 9.95   & $bc3$ & 0.496 & 0.5 & 1.39 & 1.40 & 0.386 & N/A  & 1.26 & 0.48  & 12.12 & N & E  \\ 
bc3-09975-N & 9.975  & $bc3$ & 0.496 & 0.5 & 1.39 & 1.38 & 0.354 & 0.54 & 0.73 & N/A   &  6.94 & N & C  \\ \hline
b1b-09950-N-Lam & 9.95  & $b1b$ & 0.496 & 0.5 & 1.39 & 1.40 & 0.375 & 1.15  & 0.02 & N/A  & 0.07 & N & C  \\
b1b-09975-N-Lam & 9.975 & $b1b$ & 0.496 & 0.5 & 1.39 & 1.38 & 0.377 & 1.34 & 0.04 & N/A   &  0.88 & N & C  \\
b1b-10000-N-Lam & 10.0  & $b1b$ & 0.496 & 0.5 & 1.39 & 1.36 & 0.374 & 0.97 & 0.02 & N/A   &  0.16 & N & C  \\ \hline
bc3-09950-N-vf025 & 9.95 & $bc3$ & 0.496 & 0.5 & 1.39 & 1.40 & 0.368 & 0.55 & 0.07 & N/A & 2.42 & N & C \\
bc3-09950-N-vf050 & 9.95 & $bc3$ & 0.496 & 0.5 & 1.39 & 1.40 & 0.368 & 0.56 & 0.71 & N/A & 5.74 & N & C \\
bc3-09950-N       & 9.95 & $bc3$ & 0.496 & 0.5 & 1.39 & 1.40 & 0.386 & N/A  & 1.26 & 0.48  & 12.12 & N & E  \\ \hline
bc3-09950-N-B025 & 9.95 & $bc3$ & 0.496 & 0.5 & 1.39 & 1.40 & 0.365 & 0.61 & 0.51 & N/A & 4.83 & N & C \\
bc3-09950-N-B050 & 9.95 & $bc3$ & 0.496 & 0.5 & 1.39 & 1.40 & 0.367 & 0.55 & 0.53 & N/A & 5.14 & N & C \\ 
bc3-09950-N-B075 & 9.95 & $bc3$ & 0.496 & 0.5 & 1.39 & 1.40 & 0.367 & 0.54 & 0.70 & N/A & 6.28 & N & C \\
bc3-09950-N & 9.95 & $bc3$ & 0.496 & 0.5 & 1.39 & 1.40 & 0.386 & N/A  & 1.26 & 0.48  & 12.12 & N & E  \\ \hline


\end{tabular}
\end{center}
\end{table*}

\subsection{Effects of Central Density}

\subsubsection{Model with a Centered Ignition Kernel}

In this part we discuss the global behavior of the 
ONeMg cores with different initial central densities and a centered flame
at the beginning of simulations.

In all our simulations, we follow the evolution of each model until 
the central density $\rho_c$ reaches $10^{11}$ g cm$^{-3}$ (collapse case) or 
when the total time reaches 1.5 s (expansion
case).
For models with a heading of $c3$ (Models c3-09800-N, c3-09850-N,
c3-09900-N, c3-09925-N, c3-09950-N, c3-09975-N, c3-10000-N), 
we compute the deflagration phase of the ONeMg core at different 
initial central densities but with the same flame structure 
of $c3$ using the Newtonian gravity. 
(We postpone the comparison of flame development in Section \ref{sec:lam}).
In this series of models, 
when the initial central density increases, the total mass increases 
from 1.38 to 1.39 $M_{\odot}$. Only a mild rise in mass
is observed because of the highly relativistic and degenerate electron gas. On
the other hand, the radius decreases from $1.54 \times 10^3$ 
to $1.36 \times 10^3$ km, showing that the ONeMg core is becoming more
compact as the central density increases. The minimum $Y_{\rm e}$ also
drops when $\rho_c$ increases, because the typical electron capture rate
increases when $\rho_c$ increases for the same $Y_{\rm e}$. The collapse time,
which is related to how fast the $Y_{\rm e}$ drops, also decreases. Similarly, we 
observe a decrease in the burnt mass. 

For models which expand, i.e., 
Models c3-09800-N and c3-09850-N, $\sim 1$ $M_{\odot}$ is burnt.
In the collapsing models, the faster they collapse, the 
smaller amount of fuel is burnt. The 
final energy ($\sim 10^{49}$ erg) is much lower than typical Type Ia supernovae ($\sim 10^{50}$ erg).

\begin{figure}
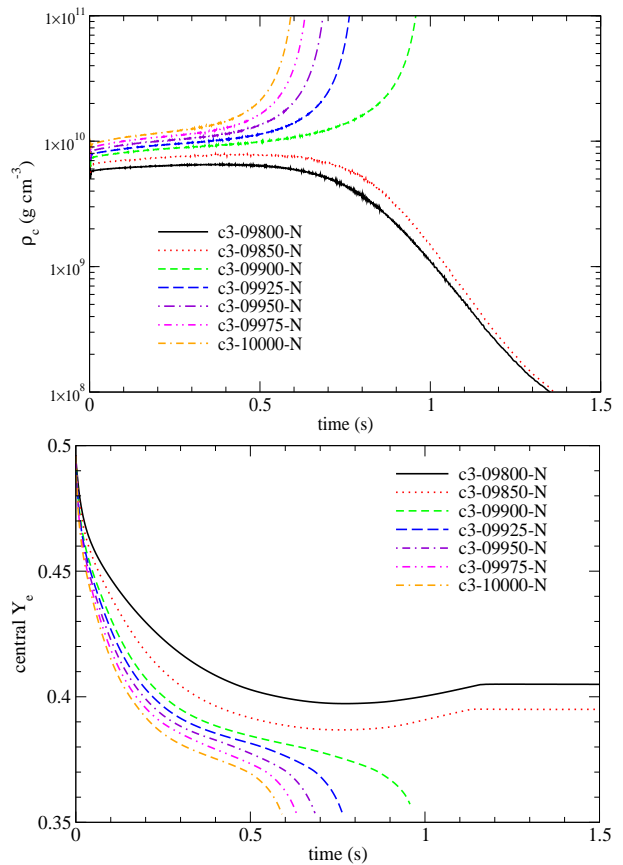

\centering
\includegraphics*[width=8cm,height=5.7cm]{fig3a.eps}
\includegraphics*[width=8cm,height=5.7cm]{fig3b.eps}
\caption{(top panel) Time evolution of the central densities
of Models c3-09800-N ($\rho_c = 10^{9.8}$ g cm$^{-3}$), 
c3-09850-N ($\rho_c = 10^{9.85}$ g cm$^{-3}$), 
c3-09900-N ($\rho_c = 10^{9.9}$ g cm$^{-3}$), 
c3-09925-N ($\rho_c = 10^{9.925}$ g cm$^{-3}$), 
c3-09950-N ($\rho_c = 10^{9.95}$ g cm$^{-3}$), 
c3-09975-N ($\rho_c = 10^{9.75}$ g cm$^{-3}$) 
and c3-10000-N ($\rho_c = 10^{10}$ g cm$^{-3}$). 
All models assume Newtonian gravity and the initial flame geometry $c3$.
Refer to 
Table \ref{table:models} for the
details of the configurations. 
(bottom panel) The evolution of central $Y_{\rm e}$ for the same 
models shown in the upper panel, which compares
the effects of the initial central density (also the initial mass)
on the final evolution.}
\label{fig:rhoc_plot}
\end{figure}

In the upper panel of Figure \ref{fig:rhoc_plot} we plot 
the central densities for Models c3-09800-N, c3-09850-N,
c3-09900-N, c3-09925-N, c3-09950-N, c3-09975-N, c3-10000-N.
In all models, the central densities increase in the
first 0.5 - 0.7 s where the electron captures 
dominate the dynamics. Models with a central density greater
than $10^{9.9}$ g cm$^{-3}$ collapse directly
within 0.5 -- 1.0 s, where the contraction rate increases
with the central density. Models with a lower initial central density
expand after $\sim$ 0.6 s, showing that the energy released
by the deflagration wave is sufficient to balance the loss of
pressure after electron captures.

In the lower panel of Figure \ref{fig:rhoc_plot} we plot the central
electron fraction ($Y_{\rm e}$) as a function of time for 
the same models as the upper panel.
Unlike the central densities, the central 
electron fraction drops drastically in the first 
0.5 s and then the decrease slows down.
The equilibrium $Y_{\rm e}$ decreases while 
the initial central density increases. For the models
which directly collapse, the drop of central $Y_{{\rm e}}$ slows 
down at $Y_{\rm e} \approx 0.38$ around 0.3 to 0.5 s.
Then, it further decreases to 0.36,
as the central densities of these models further 
increase to $10^{11}$ g cm$^{-3}$. For models which expand, 
the central electron fraction
drops similar to the collapsing models, 
but they reach a higher intermediate $Y_{\rm e}$ compared
to those models. In particular, Models c3-09800-N
and c3-09850-N show an equilibrium $Y_{\rm e}$ of 0.39 and
0.40 respectively at $t \approx 0.7 - 0.8$ s after the deflagration
has started. Following the expansion of the star,
the central $Y_{\rm e}$ gradually increases and reaches
the equilibrium value of $\sim$ 0.40 at $t \approx 1.1$ s.

\subsubsection{The $b1a$ Series}

For models with a heading $b1a$ (Models 
b1a-09800-N, b1a-09875-N, b1a-09900-N, b1a-09925-N, 
b1a-09950-N, and b1a-10000-N), they are the ONeMg core models
similar to above, but with an initial flame $b1a$, which
means a flame bubble (a ring in the three-dimensional
projection) of a radius 15 km at 50 km away from the 
ONeMg core center. The initial models are 
the ONeMg cores in different initial central
densities in hydrostatic equilibrium. The initial masses
and radii are the same as those in the $c3$ series. 
Models b1a-09800-N, b1a-09875-N, b1a-09900-N are 
exploding while the others are collapsing. 
In general, the trends of the $Y_{\rm e}$ at the end of simulations
are similar that a higher initial central density implies
a lower $Y_{\rm e}$ at the end of simulations. 
However, for models with the initial same central 
density, $Y_{\rm e}$ is higher for the $b1a$ flame than the 
$c3$ flame. Also, less mass is burnt and the direct collapse
occurs faster, for the same central density, with an 
exception of Model b1a-09875-N. 
As there is a shorter time
for the deflagration wave to sweep the fuel before the 
core collapse, less energy is released by nuclear
reactions when the initial $\rho_c$ increases. 
The general pattern for the $b1a$ series is similar to the $c3$ series. 

\begin{figure}
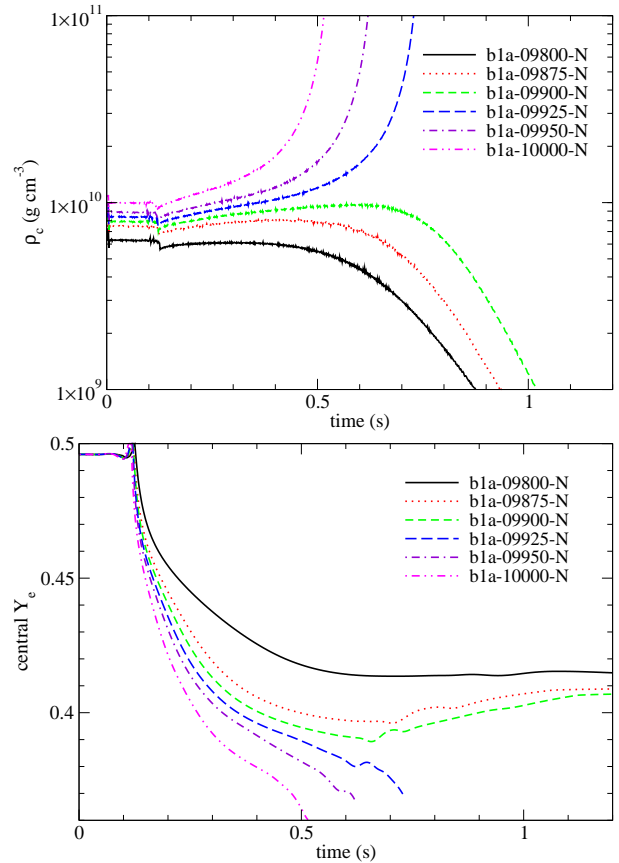

\centering
\includegraphics*[width=8cm,height=5.7cm]{fig4a.eps}
\includegraphics*[width=8cm,height=5.7cm]{fig4b.eps}
\caption{(top panel) Time evolution of the central densities
of Models b1a-09875-N ($\rho_c = 10^{9.875}$ g cm$^{-3}$), 
b1a-09900-N ($\rho_c = 10^{9.9}$ g cm$^{-3}$), 
b1a-09925-N ($\rho_c = 10^{9.925}$ g cm$^{-3}$), 
b1a-09950-N ($\rho_c = 10^{9.95}$ g cm$^{-3}$)
and b1a-10000-N ($\rho_c = 10^{10.0}$ g cm$^{-3}$).  
The models share the same setting of Newtonian gravity
and the initial flame geometry $b1a$. 
(bottom panel) The central $Y_{{\rm e}}$ evolution for the 
the same set of models shown in the upper panel.}
\label{fig:rhoc_b1a_plot}
\end{figure}

In the upper panel of Figure 
\ref{fig:rhoc_b1a_plot} we plot the central 
density against time similar to Figure
\ref{fig:rhoc_plot}. Due to the off-
center burning, there is no change in the central density
before 0.1 s. Once the flame reaches the center, 
the central density drops rapidly due to the expansion
of matter. After the initial expansion, the central density of all models 
increases. Models with an initial central density
greater than $10^{9.925}$ g cm$^{-3}$ reach the threshold density
between 0.4 -- 0.7 s. Again, the collapse time decreases when the 
central density increases. On the contrary, Models 
b1a-09800-N, b1a-09875-N and b1a-09900-N expand 
at about 0.5 -- 0.7 s. In particular, the central density
of Model b1a-09900-N can reach as high as $10^{10}$ g cm$^{-3}$,
before the expansion takes place. 
Such a high central density can be observed for 
models near the bifurcation density, where the flame requires
more time to grow until it can balance the electron capture effects.

\begin{figure}
\centering
\includegraphics*[width=8cm,height=5.7cm]{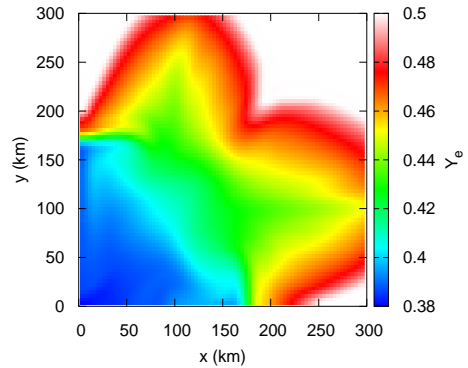}
\caption{The $Y_{{\rm e}}$ distribution of Model
b1a-09925-N at $t = 0.625$ s. The aspherical distribution
distribution of $Y_{{\rm e}}$ can produce bumps in the 
evolution of $Y_{{\rm e}}$.}
\label{fig:ye_plot}
\end{figure}

In the lower panel of Figure \ref{fig:rhoc_b1a_plot}
we plot similar to the upper panel the
central $Y_{\rm e}$ evolution of Model b1a-10000-N.
Similar to the central density, there is no 
change in central $Y_{\rm e}$ before 0.1 second, when the 
flame has not reached the core. After that, it quickly drops
with a rate proportional to the central density and slows
down after it reaches $\sim$ 0.38 -- 0.41. For models
which directly collapse, the central $Y_{\rm e}$ quickly falls rapidly
again and reaches 0.35 -- 0.36 at the end of
the simulations. In Models b1a-09925-N and b1a-09950-N, 
there are mild bumps in the central $Y_{\rm e}$ at $t \approx 0.6$ s.
This is because the off-center burning has led to an uneven
distribution of $Y_{\rm e}$. Unlike the models with $c3$ flame,
the central ignition allows that the matter with a higher
density to be burnt for a longer time, thus having more
time for electron capture and a lower $Y_{\rm e}$. This 
creates a distribution of increasing $Y_{\rm e}$ along the radial 
outward direction. For the $b1a$ cases, the region which undergoes
the longest duration of electron capture is away from the center.
Furthermore, in the core, before the 
homologous expansion fully develops, mixing from neighboring cells may 
also affect the $Y_{{\rm e}}$ distribution. 
The temporary inward
flow to the center can also increase the central $Y_{{\rm e}}$.
In Figure \ref{fig:ye_plot} we plot the $Y_{{\rm e}}$ distribution
of the Model b1a-09925-N at $t = 0.625$ s.
Near the center $Y_{{\rm e}}$ is not 
completely spherically symmetric. Such asymmetry may
give rise to small scale bumps in the $Y_{{\rm e}}$ evolution.
However, for Model b1a-10000-N, the direct collapse occurs
without reaching any intermediate $Y_{\rm e}$. Therefore, the 
electron capture around all the 
regions is similar. $Y_{\rm e}$ only drops monotonically with time.

\subsubsection{The $b1b$ Series}

In this series, we further study the density dependence
of an ONeMg core with an off-center flame placed at 100 km from
the stellar center. The models include
Models b1b-09900-N, b1b-09950-N, b1b-09975-N
and b1b-10000-N. The flame structure in this series of models
is similar to $b1a$, but the flame "ring" is located at
100 km apart from the core. Similar to the $b1a$ series, the
initial profiles are the same as the $c3$ series
that the models have the same masses and radii. 
In this series, Models b1b-09900-N and b1b-09950-N
are expanding while the others are directly collapsing. 
Similar to the two series above, when the central density
is higher, the final $Y_{\rm e}$ at the end of
the simulation is lower and the model has a faster collapse. 
Less nuclear 
energy is released owing to a smaller mass of fuel burnt by
the deflagration wave. 

\begin{figure}
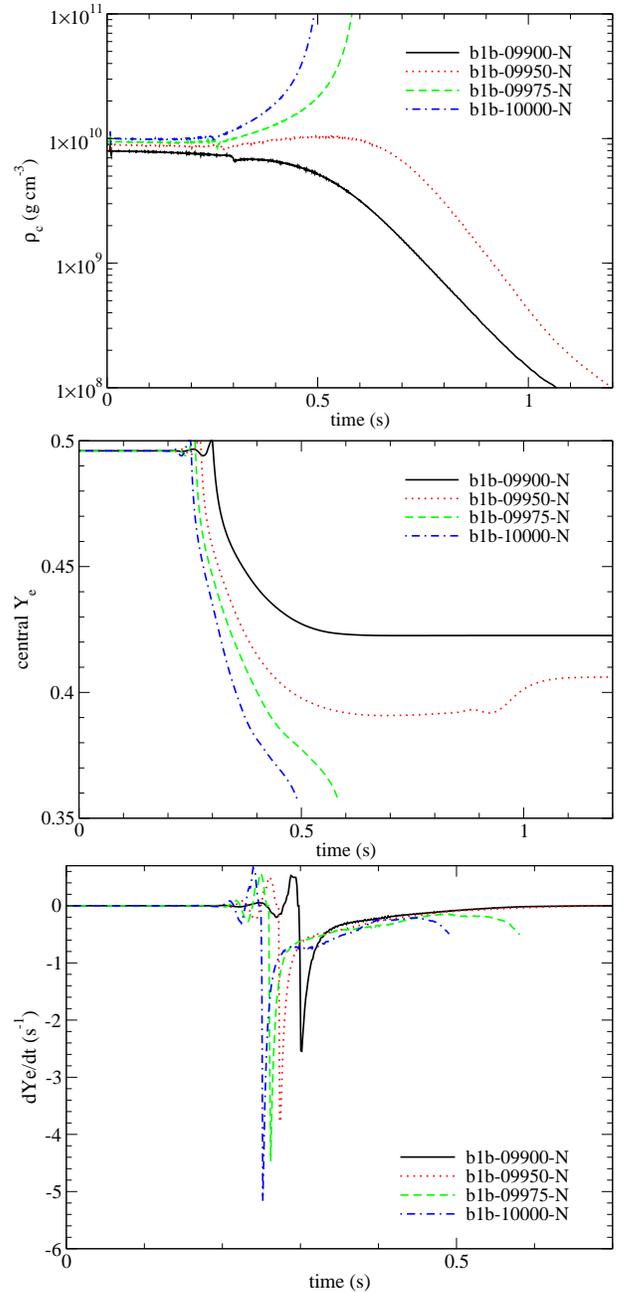

\centering
\includegraphics*[width=8cm,height=5.7cm]{fig6a.eps}
\includegraphics*[width=8cm,height=5.7cm]{fig6b.eps}
\includegraphics*[width=8cm,height=5.7cm]{fig6c.eps}
\caption{(top panel) Time evolution of the central densities
of Models b1b-09900-N ($\rho_c = 10^{9.9}$ g cm$^{-3}$), 
b1b-09950-N ($\rho_c = 10^{9.95}$ g cm$^{-3}$), 
b1b-09975-N ($\rho_c = 10^{9.975}$ g cm$^{-3}$)
and b1b-10000-N ($\rho_c = 10^{10.0}$ g cm$^{-3}$).
All models assume Newtonian gravity and the initial flame geometry $b1b$.
(middle panel) Same as the upper panel, 
but for the evolution
of the central $Y_{\rm e}$ for the same set of models.
(bottom panel) Time derivative of $Y_{\rm e}$ for 
the same set of models as in the upper panel. }
\label{fig:rhoc_b1b_plot}
\end{figure}

In the upper panel of Figure \ref{fig:rhoc_b1b_plot} we plot
the central density against time for the four models similar
to Figures \ref{fig:rhoc_plot} and \ref{fig:rhoc_b1a_plot}.
With a flame bubble located farther from the center, the flame
takes $\sim$ 0.3 s to reach the center, which creates a 
small drop in the central density. At $\sim$ 0.5 s, Models 
b1b-09975-N and b1b-10000-N begin its collapse. The 
central density of Model b1b-09950 also increases above 
$10^{10}$ g cm$^{-3}$ at $\sim$ 0.5 s, but drops again when 
the star expands at 0.7 s.  Model b1b-09900 shows almost no 
contraction when the electron captures take place at the core. 
This is because the typical density is low and
the initial flame is sufficiently
far. The burnt matter can expand
before the flame reaches the center.

In the middle panel of Figure \ref{fig:rhoc_b1b_plot} we plot the time evolution 
of the central $Y_{\rm e}$ for the same series of models 
as in the upper panel.
There is no change in $Y_{\rm e}$ in the first 0.3 s. 
This is because the flame has not arrived the core.
Thus, the cold matter cannot carry out efficient electron captures
compared to the burnt ash.
After the deflagration wave has arrived at the center, 
$Y_{\rm e}$ drops immediately. When the initial $\rho_c$
is higher, its rate of $Y_{\rm e}$ decrease is higher. 
In Models b1b-09750-N and b1b-10000-N,
the electron captures mildly slow down when $Y_{\rm e} \approx 0.37$,
and then the drop resumes again until the end of simulations, down
to a value of $\approx 0.36$. In contrast, $Y_{\rm e}$ shows 
temporary values at 0.39 and 0.42 for Models
b1b-09900-N and b1b-09950-N. The latter one remains 
the same value after the expansion has started, while the former
one slightly increases to 0.41, as the matter in the core
begins to mix with the surrounding material, which has 
a higher $Y_{\rm e}$.  

We also notice that at early time there is a mild 
drop in the central density before the flame arrives the 
center. It is not because the model is not in good equilibrium
during construction, but because the initial off-center flame
and its subsequent electron captures disturb the pressure
gradient. The core slowly expands to adjust to the 
presence of the flame.

In the bottom panel of Figure \ref{fig:rhoc_b1b_plot} we plot the 
time derivative of the central $Y_{\rm e}$ to illustrate the
density dependence of the electron capture rate. Depending on
the central density, the arrival time of the deflagration wave
differs by at most 0.05 s. Such small time difference
can provide the time for the deflagration wave to grow larger
and hence burn more matter, which suppresses the contraction 
after the flame reaches the center.
Once the center is burnt, 
the sharp drops of $d Y_{\rm e,c}/dt$ show that the weak
interactions rapidly occur in the high-density core. Furthermore,
the rate of decrease increases when the initial $\rho_c$ increases.
This shows that the rate of decreases is an increasing function
of the progenitor mass, i.e., the initial runaway density.

\subsection{Effects of General Relativity}

Here we study how the relativistic corrections in the gravity can 
affect the bifurcation of the ONeMg core evolution.
In the simulations, we study the counterpart models
of c3-10000-N and c3-10200-N, i.e.,
Models c3-10000-R, c3-10200-R. These models are the 
most compact ONeMg cores constructed in this work, we therefore
expect that the relativistic effects in these cores are the most
pronounced. In general, 
embedding the physics of relativistic gravity requires 
a complete restructuring of the code due to the necessary inclusion of the metric tensor. 
We look for corrections of the Newtonian gravity as
the first step.
We follow the prescription in \cite{Kim2012}. Based on the 
Poisson equation for the gravitational potential 
$\nabla^2 \Phi = 4 \pi G \rho$, where $\Phi$ and $\rho$
are the gravitational potential and matter mass density.
We replace $\rho$ by $\rho_{{\rm active}}$, where 
\begin{equation}
\rho_{{\rm active}} = \rho h \frac{1 + v^2}{1 - v^2} + 2 P,
\end{equation}
and $P$, $v^2$ are the fluid pressure and the magnitude square
of the velocity. $h = 1 + \epsilon + P / \rho$ is the 
specific enthalpy of the matter. In this sense, the 
extra mass-energy contributions by the internal energy and the 
kinematics of the matter are included. 

\begin{figure}
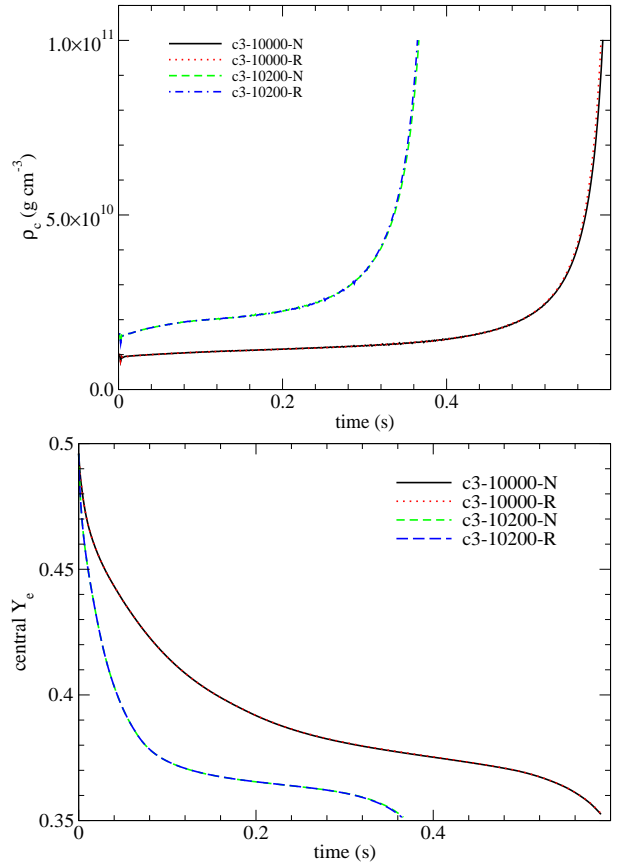

\centering
\includegraphics*[width=8cm,height=5.7cm]{fig7a.eps}
\includegraphics*[width=8cm,height=5.7cm]{fig7b.eps}
\caption{(top panel) Central densities against time 
for Models 
c3-10000-N ($\rho_c = 10^{10}$ g cm$^{-3}$, Newtonian gravity),
c3-10000-R ($\rho_c = 10^{10}$ g cm$^{-3}$, with relativistic corrections),
c3-10200-N ($\rho_c = 10^{10.2}$ g cm$^{-3}$, Newtonian gravity) and 
c3-10200-R ($\rho_c = 10^{10.2}$ g cm$^{-3}$, with relativistic corrections).
(bottom panel) Same as the upper panel, but for the 
central $Y_{\rm e}$ for the same set of models.}
\label{fig:rhoc_GR_plot}
\end{figure}

To demonstrate the effects of the relativistic corrections in the 
gravitational potential, in the upper panel of Figure \ref{fig:rhoc_GR_plot}
we plot the central density for the models with a centered
flame with an initial geometry $c3$
and initial central densities $10^{10.0}$ and $10^{10.2}$
g cm$^{-3}$ respectively. 
In the lower panel of Figure \ref{fig:rhoc_GR_plot}
we plot the same as the upper panel but for the central
$Y_{\rm e}$. In both cases, a direct collapse 
is observed. The evolution of the central density
is not sensitive to the relativistic corrections
in gravity. Models c3-10000-N and c3-10000-R 
overlap with each other in the figure throughout the simulation,
and so as Models  c3-10200-N and c3-10200-R.
Similar results can be found for the central $Y_{\rm e}$.

By combining these models, we show that
when GR correction terms in gravity are included,
no observable change in the evolution
even for the most compact models with the
initial $\rho_c = 10^{10.2}$ g cm$^{-3}$. This 
suggests that Newtonian gravity
is sufficient in following the runaway phase of an ONeMg core
accurately before its onset of collapse.

\subsection{Effects of Initial Flame Size}

The exact extent of the nuclear runaway is not well constrained because it depends on the competition between the convective efficiency and the hydrostatic O-burning. Numerically, it is difficult to implement due to the sharp $Y_{{\rm e}}$ contrast and complications from the URCA process. In general, efficient convection leads to a faster transport of the heat produced during electron captures. This smooths the temperature profile, allows for a larger initial flame, and raises the ignition density accordingly.

Without knowing the exact details of the initial flame
evolution, we try to span the parameter space by considering
different flame sizes for the central ignition model. They include $c3$, $mc3$ and $bc3$. The latter two flame
structures are the same as the $c3$
flame, but with a size 2 times or 4 times larger.
The width of the reaction front is kept fixed
as indicated by the level-set scheme.
The case $bc3$ is so extended that it might be incompatible with typical stellar evolution. We use it as a qualitative 
comparison in this work.

\begin{figure}
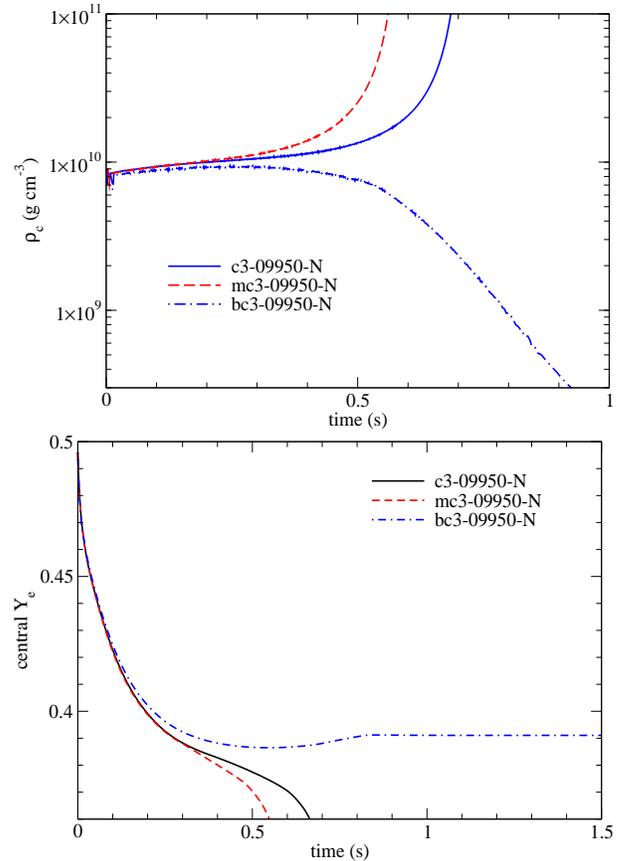

\centering
\includegraphics*[width=8cm,height=5.7cm]{fig8a.eps}
\includegraphics*[width=8cm,height=5.7cm]{fig8b.eps}
\caption{(top panel) Time evolution of the central densities
in Models c3-09950-N ($c3$ initial flame), 
mc3-09950-N ($mc3$ initial flame) 
and bc3-09950-N ($bc3$ initial flame).
All models share the same initial central density ($\rho_c = 10^{9.95}$ g cm$^{-3}$)
and are without relativistic corrections. 
(bottom panel) Same as the upper panel 
but for the central $Y_{\rm e}$ for the same set of models.}
\label{fig:rhoc_flamesize_plot}
\end{figure}

In the upper panel of Figure \ref{fig:rhoc_flamesize_plot}, we plot the evolution of central densities against time for models of different initial flame masses. 
The initial mass being burnt $M_{{\rm burn,ini}}$
range from $10^{-4}$ to $10^{-2}$ $M_{\odot}$. For 
Model c3-09950-N with $M_{{\rm burn,ini}} \sim 10^{-4}$ $M_{\odot}$, the 
central density increases for the first 0.1 s.
The models deviate at $\sim$ 0.3 s.
Beyond $t = 0.7$ s, the ONeMg core collapses. 
On the other hand, when $M_{{\rm burn,ini}} \sim 10^{-3} ~M_{\odot}$, 
a similar evolution occurs but the collapse starts earlier, 
at 0.5 s after the simulation. When 
$M_{{\rm burn,~ini}} \sim 10^{-2}$ $M_{\odot}$, a similar contraction
occurs at the beginning, but after $t = 0.5$ s, the star's 
central density decreases due to expansion of the ONeMg core. It produces a low-energy supernova explosion, similar to a ``Type 1.5x" supernova (when a realistic progenitor model including an H-envelope is considered). 
 
In the lower panel of Figure \ref{fig:rhoc_flamesize_plot}, we plot the same as the upper panel but for the central $Y_{\rm e}$. Similar to previous models, all three models show a rapid drop of the central $Y_{\rm e}$ once the core is burnt to NSE. It drops to about 0.39 within 0.3 s, until the capture rate decreases. The equilibrium $Y_{\rm e}$ of c3-09950-N is slightly higher than that in Models mc3-09950-N and bc3-09950-N. At $t = 0.5$--0.6 s, $Y_{\rm e}$ drops rapidly again for  Models c3-09950-N and mc3-09950-N. However, in Model c3-09950-N, due to expansion, mixing occurs in the core with the matter in the outer zones, which has on average a higher $Y_{\rm e}$. The $Y_{\rm e}$ slowly increases to 0.39 and remains unchanged after $t = 0.8$ s.
 
These figures show that the initial flame size also plays a role in determining the collapse-expansion
bifurcation. In particular, a small flame $\sim 10^{-4} - 10^{-3}$ $M_{\odot}$
favours the collapse scenario, while a large
flame favours the expansion scenario. 
In Section \ref{sec:discussion}, we further discuss
the non-monotonic variations of the collapse time 
among models with the $c3$, $mc3$ and $bc3$ flame structures.

\subsection{Effects of Flame Physics}
\label{sec:flamephysics}

In order to model a turbulent flame, a formula 
describing the relation between the turbulent velocity $v'$
and the effective flame propagation speed $v_{{\rm turb}}$
is necessary.
However, only a statistical description is 
available due to the stochastic nature 
of turbulent motion. Also, a terrestrial experiment
cannot create such extreme
environment. 
How the turbulent motion can 
enhance the propagation of flame and also
the effective flame speed remains
unclear. In the 
literature of Type Ia supernova where turbulent
flame models are used, the models assume self-similar
flames. With the renormalization scheme \citep{Pocheau1994},
the general formula writes
\begin{equation}
v_{{\rm turb}} = v_{{\rm lam}} \left[ 1 + C_n \left( \frac{v'}{v_{{\rm lam}}} \right)^n \right]^{1/n},
\end{equation}
where $v_{{\rm lam}}$ is the laminar flame propagation speed
while $C_n$ and $n$ are the constants derived 
from experiments. The velocity spectra
of the turbulence structure determine $n$. 
This formula has two asymptotic properties that are expected
experimentally. 
1. The effective propagation speed 
reduces to the laminar flame speed, when $v' \rightarrow 0$.
This corresponds to the case that, when there is no
perturbation to the surface structure of the flame,
the flame propagates as a laminar wave. 
2. The effective propagation speed has an asymptotic value $\approx \sqrt[n]{C_n} v'$ (Given $v' > v_{\rm lam}$).
This means that when the fluid motion is highly turbulent, the 
flame no longer depends on the laminar flame speed, but solely on
the velocity fluctuations inside the fluid. 

However, one shortcoming of this model is that
in order to derive this formula, isotropic turbulence is 
assumed by the renormalization procedure.
Gravity makes the radial direction distinctive from 
the angular directions. Furthermore, the Rayleigh-Taylor
instabilities enhance the flame propagation along 
the radial direction.

Numerically, one has different $C_n$ and $n$ based on 
the context. In \cite{Peter1999, Schmidt2006b}, $C_n = 4/3$
and $n = 2$ correspond to the Gaussian 
distribution in the velocity fluctuations. 
In \cite{Hicks2015}, it is shown numerically that 
for a premixed flame with a one-way reaction such as H$_2$-air mixture, 
the relation has a best fit of $C_n = 0.614$ when $n = 2$,
while $v_{{\rm turb}} = v_{{\rm lam}} (1 + 0.4321 \tilde{v}'^{1.997})$
is the best fit with $\tilde{v}'$ being the 
scaled $v'$. 
The variations of this formula demonstrate that 
the scaling factor $C_n$ and the scaling power $n$
are not yet well constrained. 

To understand the effects of this quantity on the ONeMg
core evolution, we vary the original value of $C_n$ (denoted as $C_{n0}$) 
by considering $C_n = 0.25$ $C_{n0} $ and 0.50 $C_{n0}$.
They correspond to the turbulent flame where 
turbulent production is less effective in disturbing
the flame structure.

\begin{figure}
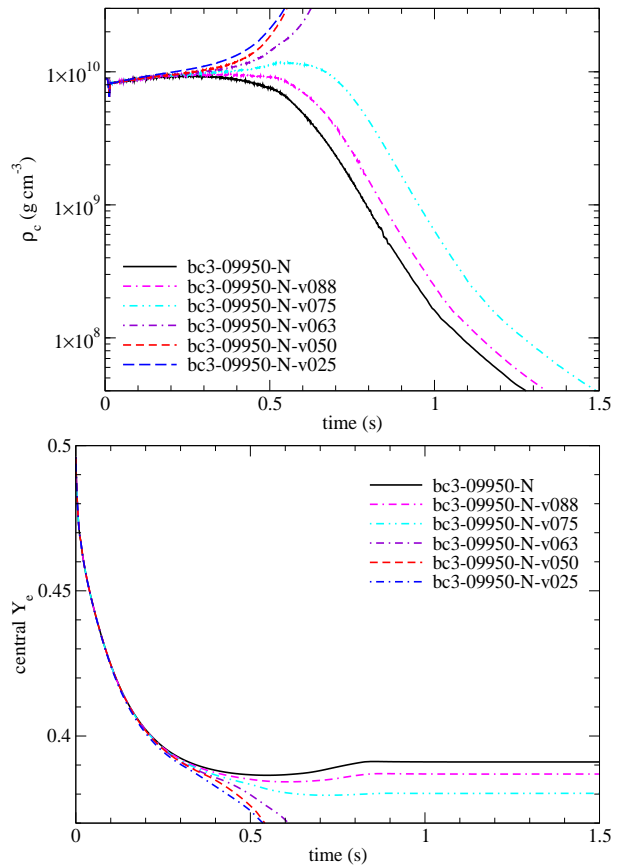

\centering
\includegraphics*[width=8cm,height=5.7cm]{fig9a.eps}
\includegraphics*[width=8cm,height=5.7cm]{fig9b.eps}
\caption{(top panel) Time Evolution of central densities against time 
for Models bc3-09950-N ($C_n = C_{n0}$), 
bc3-09950-N-vf050 ($C_n = 0.5 C_{n0}$) and 
bc3-09950-N-vf025 ($C_n = 0.25 C_{n0}$).
Extra models are plotted including Models 
bc3-09950-N-vf063 ($C_n = 0.63 C_{n0}$), 
bc3-09950-N-vf075 ($C_n = 0.75 C_{n0}$)
and bc3-09950-N-vf088 ($C_n = 0.88 C_{n0}$).
All models share the same initial flame geometry $bc3$,
central density of $10^{9.95}$ g cm$^{-3}$ and assume
Newtonian gravity.
(bottom panel) Same as the upper panel
but for the central $Y_{\rm e}$ for the same set of models.
For the expanding models (Models bc3-09950-N, bc3-09950-N-vf088 and 
bc3-09950-N-vf075), the flame speed affects final $Y_{{\rm e}}$.}
\label{fig:rhoc_vflame_plot}
\end{figure}

In the upper panel of Figure \ref{fig:rhoc_vflame_plot} we plot the central
density against time for Models bc3-09950-N, bc3-09950-vf050 and
bc3-09950-vf025. We choose the basis model with the $bc3$ flame 
structure because it has a bifurcation density near $10^{9.95}$ g cm$^{-3}$.
The effects of the flame physics are more pronounced 
near the bifurcation density. We note that 
in our models, the transition density for the c3 flame is
$\sim 10^{9.90}$ g cm$^{-3}$, using a slower flame does
not change the fate of the model at $10^{9.95}$ g cm$^{-3}$
from its collapse into a neutron star. Thus, we consider
the bc3 flame, where the transition
occurs at a central density of $10^{9.95}$ g cm$^{-3}$.
We use this flame structure because . In fact similar
effects can also be demonstrated by the ONeMg models with
the $c3$ flame and a central density near $10^{9.90}$ g cm$^{-3}$.
In this series of models, Model bc3-09950-N
explodes while Models bc3-09950-vf050 and bc3-09950-vf025 collapse.
We also plot the results from additional models (not included in Table \ref{table:models})
for demonstrating the sensitivity of our models on the flame speed
by including Models bc3-09950-N-vf063, bc3-09950-N-vf075
and bc3-09950-N-vf088, which are 63~\%, 75~\% and 88~\% 
of the default asymptotic flame speed.

The central density of
all models mildly increase for the first 0.3 s.
For the collapsing models, the increase of $\rho_c$ 
resumes at $t \approx 0.4$ s. On the other hand, 
for the expanding models, such as
Model bc3-09950-N, $\rho_c$ slowly drops till $t = 0.6$ s. 
Accompanying with the expansion, its central density
rapidly drops after $t = 0.6$ s. At $t = 1$ s, 
the central density drops to about 1 \% of its initial value.
When the flame speed is high, the conversion from
contraction to expansion becomes fast and so is the 
expansion of the core.

In the lower panel of Figure \ref{fig:rhoc_vflame_plot} we plot the central $Y_{\rm e}$
similar to Figure \ref{fig:rhoc_vflame_plot}.
In the exploding Model bc3-09950-N, the central $Y_{\rm e}$
again quickly drops from 0.5 to 0.38 within 0.3 s.
Unlike the previous test on the effects of the initial flame size, 
the large initial flame we used is less changed
by the surrounding. Beyond $t = 0.3$ s, the drop of $Y_{{\rm e}}$
accelerates again and the central
$Y_{\rm e}$ drops below 0.37 at $t = 0.5$ s. 
On the other
hand, for the models which expand, during their
expansion, the central $Y_{\rm e}$ drops until it reaches 
its asymptotic value 0.38 -- 0.39 after $t = 0.8$ s.
We remark that the asymptotic $Y_{{\rm e}}$ 
increases when the flame speed is faster. Notice that 
the final $Y_{{\rm e}}$ determines the characteristic 
abundance of the ash, especially when it is ejected. 
The low-$Y_{\rm e}$ ejecta contains a significant overproduction 
of neutron-rich isotopes, e.g. $^{50}$Ti, $^{54}$Cr, 
$^{60}$Fe and $^{64}$Ni with respect to $^{56}$Fe.
Such overproduction can be strongly constrained by 
the galactic chemical evolution. We will discuss 
further the ejecta properties in Section \ref{sec:failed}.

Combining these three plots, 
the effective formula of the turbulent flame prescription
also plays a role in the ONeMg collapse condition
similar to the initial flame size and the properties of the
flame kernel. In particular, models tend to collapse (expand)
when the flame is slow (fast). This is because the slower
flame provides more time for the electron captures, thus allowing
the star to contract faster before the flame can burn the matter 
in the outer regions.
On the other hand, the faster flame allows a faster growth of 
its surface area, which can balance the effects of 
decreasing $Y_{\rm e}$. At last, we remark that 
such flip of results from an expanding model to a collapsing
model can be seen only for those near the transition.
In Section \ref{sec:discussion} we further explore
the effects of flame physics on other models with different 
flame geometry.

\subsubsection{Extension: Effects of Laminar Flame Propagation}
\label{sec:lam}

We remark that the treatment of a nuclear flame in the 
literature does not always assume sub-grid scale
turbulent motion \citep[see e.g.][]{Plewa2007}.
The flame is only distorted by the smallest resolvable length
scale by the simulation and it is assumed that 
the fluid motion below the resolvable scale is 
laminar (except for the perturbations by Rayleigh-Taylor instabilities). 
This forms another limit 
in the flame propagation.

To demonstrate this limit, we pick an ONeMg core configuration with an initial
flame size which expands in the default setting. Our models assuming
flame speed much slower than the speed of sound can have the flame propagation 
more enhanced along the symmetry axis.
So, an off-center flame is preferred. We choose the Models b1b-09950-N-Lam, 
b1b-09975-N-Lam and b1b-10000-N-Lam. 
(An ending "-Lam" corresponds
to the flame which only propagates without sub-grid acceleration.)
For the effects of the slower flame in general we discuss in Section \ref{sec:global}.

Models b1b-09950-N-Lam, b1b-09975-N-Lam
and b1b-10000-N-Lam collapse into a neutron star. 
In contrast, we compare Model b1b-09950-N-Lam with Model b1b-09950-N.
They have the same configurations but the latter is modeled with a turbulent
flame prescription. Model b1b-09950-N expands
like a Type 1.5 supernova. On the other hand, when the
laminar flame prescription is used, the star directly
collapses. This shows that whether or not the 
flame geometry interacts with the
sub-grid scale eddy motion, or only interacts
with the buoyancy smearing, changes the collapse-explode
bifurcation of the benchmark model $\rho_c = 10^{9.95}$ g cm$^{-3}$. 

\begin{figure}
\centering
\includegraphics*[width=8cm,height=7cm]{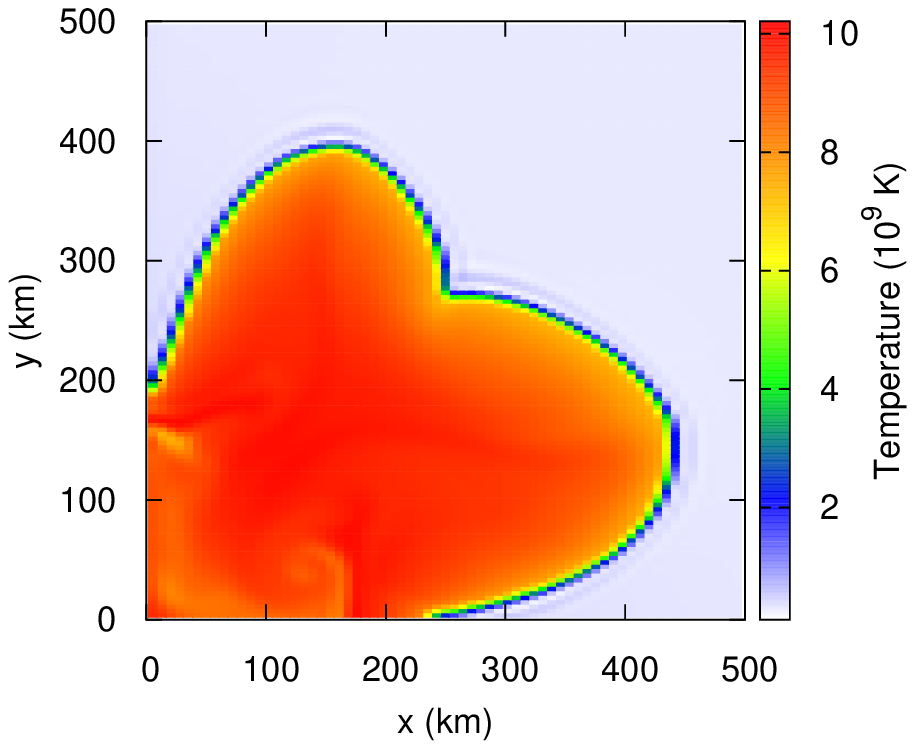}
\includegraphics*[width=8cm,height=7cm]{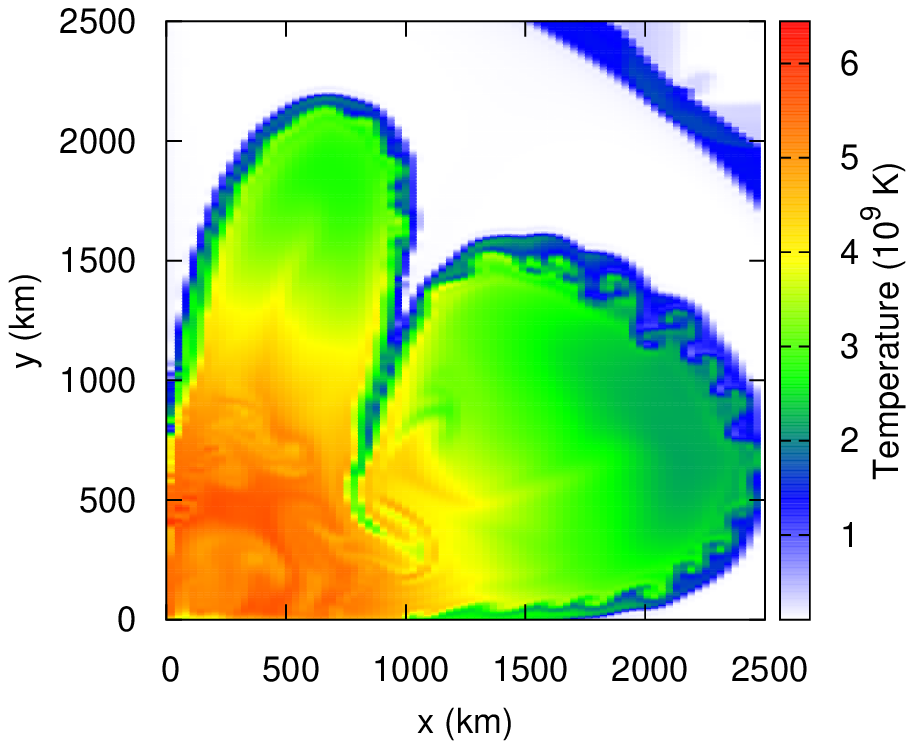}
\includegraphics*[width=8cm,height=7cm]{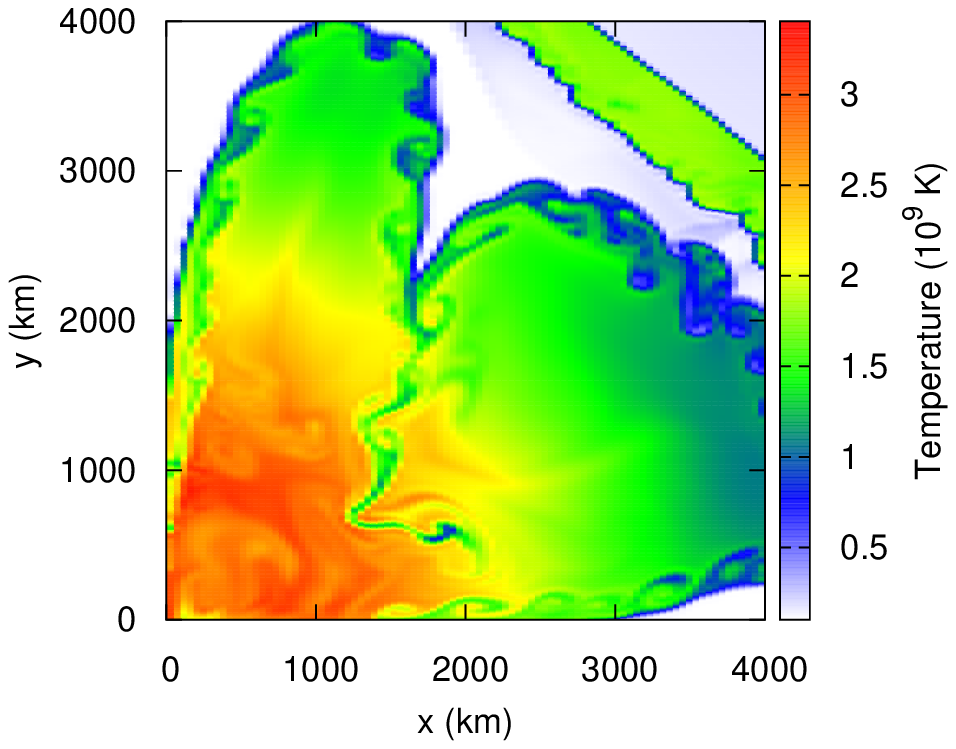}
\caption{Temperature colour plots of Model b1b-09950-N
at 0.5, 1.0 and 1.25 s of the simulations. 
The hot regions also represent those being burnt by the 
ONe deflagration.}
\label{fig:b1b_09950_plot}
\end{figure}

To characterize the differences of the flame propagation by the turbulent
flame and the laminar flame, we plot in Figure \ref{fig:b1b_09950_plot}
the temperature colour plots of Model b1b-09950-N from 
0 to 1.25 s at selected time points. The hot elements also trace
the flame structure. The turbulent flame allows
the structure to grow rapidly. Within the first 0.5 s, there 
is a two-bump structure developed and the size has grown to 
$\sim$ 450 km. At $t = 0.75$ s onward, the large-scale structure
freezes and the two-"finger" shape emerges. 
At $t = 1.0$ s, the flame expands rapidly to 2000 km, 
where the surface shows more features when the hydrodynamics
instabilities become pronounced. 

\begin{figure}
\centering
\includegraphics*[width=8cm,height=7cm]{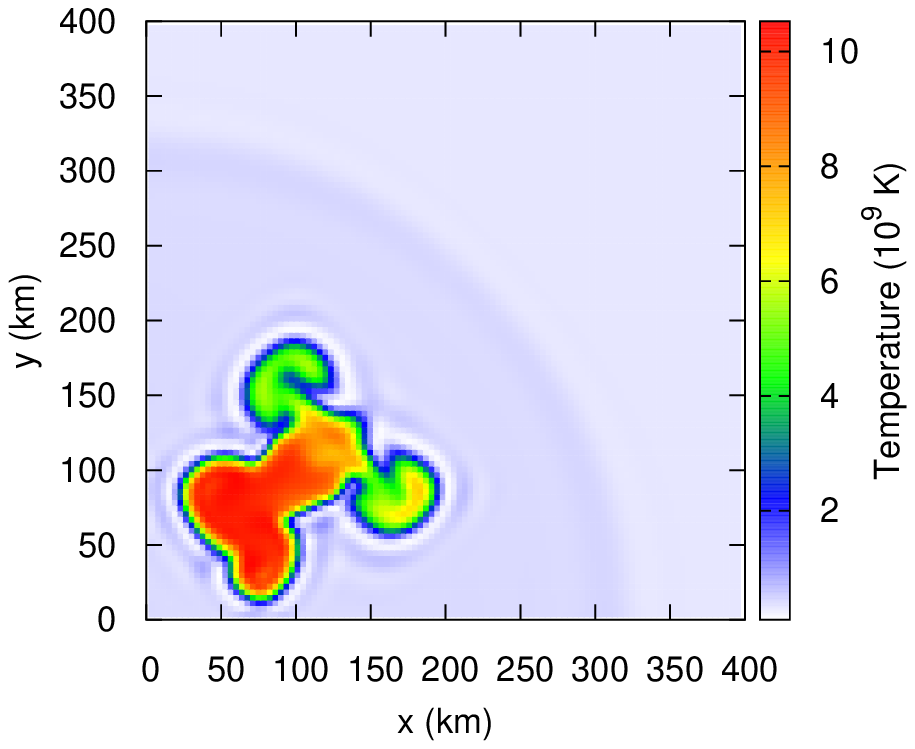}
\includegraphics*[width=8cm,height=7cm]{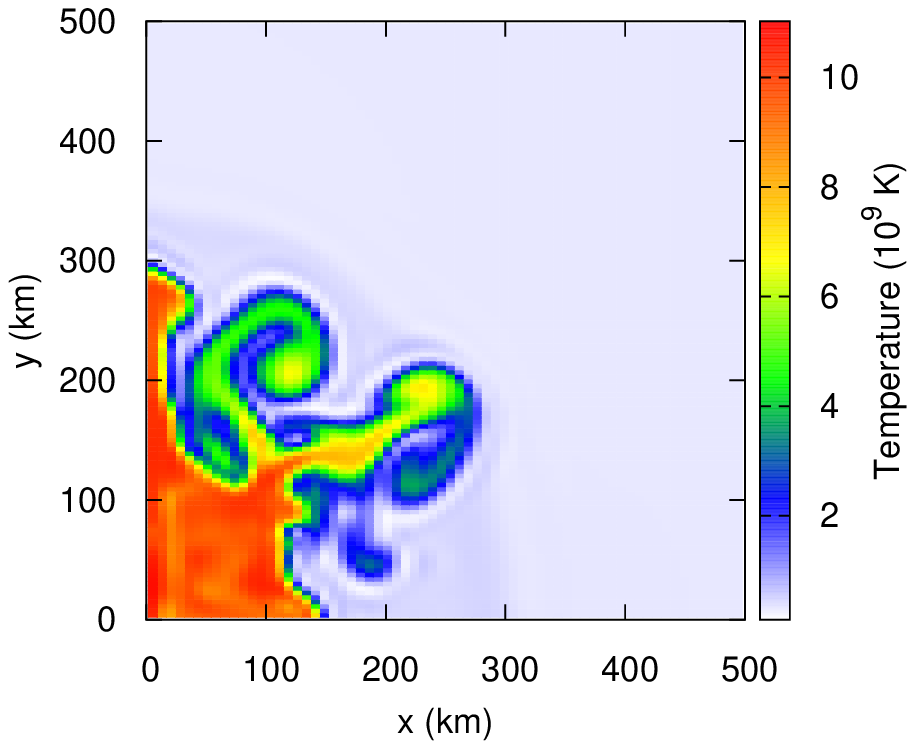}
\includegraphics*[width=8cm,height=7cm]{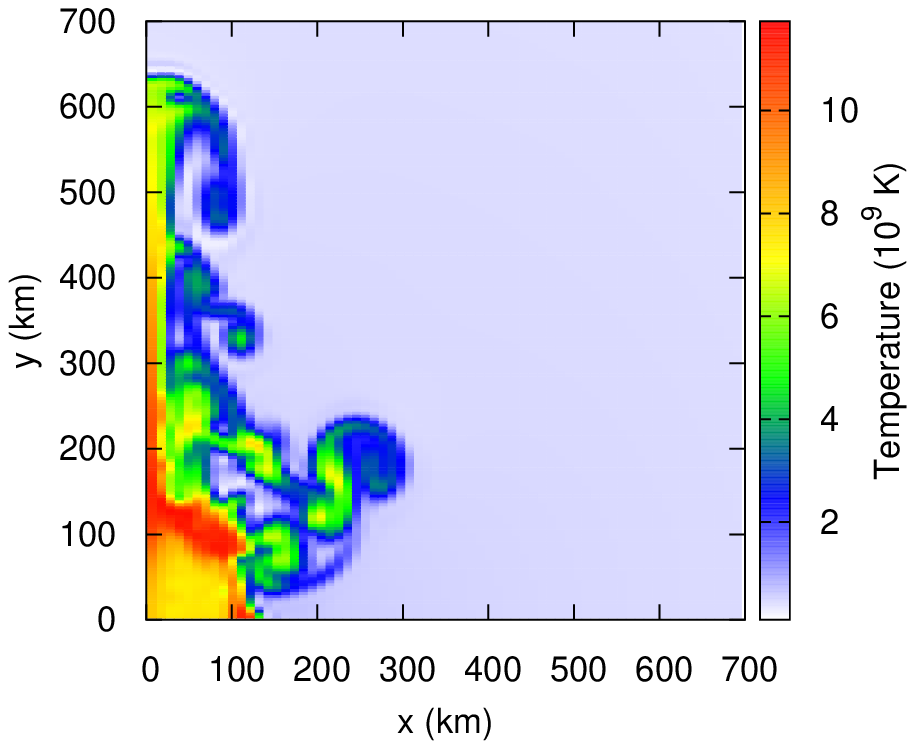}
\caption{Temperature colour plots of Model b1b-09950-N-Lam
at 0.4, 0.8 and 1.2 s of the simulations. 
The hot regions also represents the regions being burnt by the 
ONe deflagration.}
\label{fig:b1b_09950_lam_plot}
\end{figure}

In Figure \ref{fig:b1b_09950_lam_plot} we plot the same as 
Figure \ref{fig:b1b_09950_plot} but for Model b1b-09950-N-Lam
from 0.2 - 1.2 s at selected time points. A qualitative 
comparison of the flame structure already demonstrates 
drastic differences between the propagation of the laminar flame
and the turbulent flame. At early time before 0.4 s, the 
fluid motion has largely reshaped the original spherical 
flame structure. Many small-scale "mushroom shapes" swarm out 
as a manifestation of the Rayleigh-Taylor instabilities. 
At $t = 0.6$ s, the flame has finally reached the core,
where a hot core of size 150 km can be seen. After that,
the core does not grow significantly. However, there
is a hot flow along the rotation axis. This is the 
mentioned enhancement due to Rayleigh-Taylor instabilities
along the symmetry boundary. However, this enhancement does not
affect the results as we checked that the burnt mass does not 
increase significantly. Within another 0.2 s, the core 
directly collapses. 

\begin{figure}
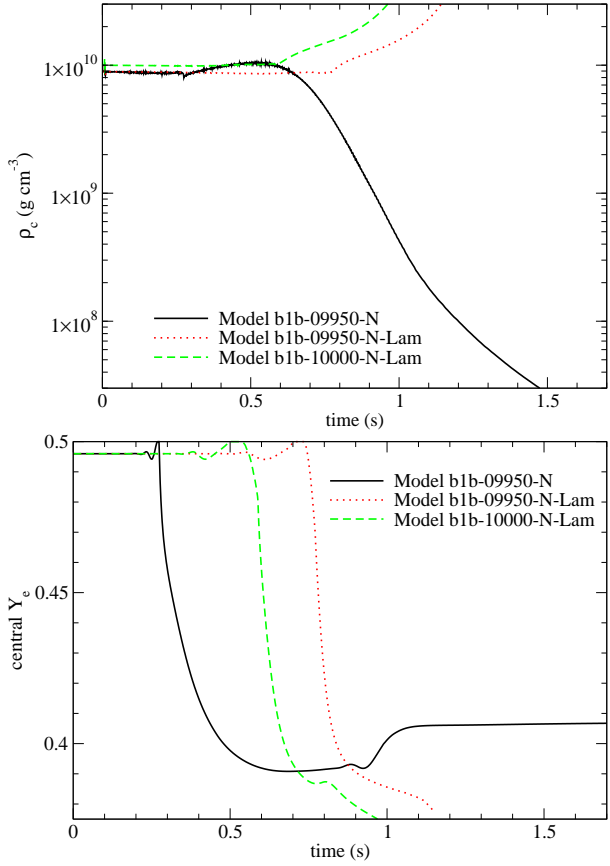

\centering
\includegraphics*[width=8cm,height=5.7cm]{fig12a.eps}
\includegraphics*[width=8cm,height=5.7cm]{fig12b.eps}
\caption{(top panel) Central densities against time for Models
b1b-09950-N ($\rho_c \ 10^{9.95}$ g cm$^{-3}$, turbulent flame), 
b1b-09950-N-Lam ($\rho_c \ 10^{9.95}$ g cm$^{-3}$, laminar flame)
and b1b-10000-N-Lam ($\rho_c \ 10^{10}$ g cm$^{-3}$, laminar flame).
(bottom panel) Same as the upper panel, but for 
the central $Y_{\rm e}$ for the same set of models.}
\label{fig:rhoc_lam_plot}
\end{figure}

We further examine their evolution by the quantities in the core.
In the upper panel of Figure \ref{fig:rhoc_lam_plot} we plot the central density
against time for Models b1b-09950-N-Lam, b1b-09975-N-Lam, b1b-10000-N-Lam.
For comparison we also include the data from Model b1b-09950-N.
The central densities of Models b1b-09950-N and b1b-09950-N-Lam
are the same before $t = 0.4$ s, when the flame has not reached
the core. Once it reaches the core, namely at $t = 0.4$ s 
for Model b1b-09950-N and at $t = 0.8$ s for Model
b1b-09950-N-Lam, they deviate from each other. Both 
models show an increase in central density due to 
the softening effect by electron capture. However, 
for Model b1b-09950-N, the central density starts to 
drop beyond $t = 0.6$ s, showing that the turbulent flame
has released sufficient energy to support against the 
inward flows. On the other hand, in Model b1b-09950-N-Lam,
the increase in the central density leads to the collapse
where there is no sign for the core to reach a 
temporary equilibrium. 
A similar evolution can be seen in Model b1b-10000-N-Lam. 
After $t = 0.8$ s where the flame reaches the core, 
the increase of the central density triggers the collapse. 

In the lower panel of Figure \ref{fig:rhoc_lam_plot} we plot similar to 
the upper panel but for the central $Y_{\rm e}$. 
After the flame has reached the core, which can be 
noted by the sudden drop of $Y_{\rm e}$, the electron captures
of the expanding Model b1b-09950-N
slows down at $t \approx 0.5$ s snd the central $Y_{\rm e}$ stays at 
$\approx$ 0.39. It later returns to a high value when
the core material begins to mix with 
the higher $Y_{\rm e}$ material in the 
outer zone. On the other hand, the $Y_{\rm e}$ does not
reach any equilibrium value once the core is burnt. 
The local electron capture rate slows down
at $t = 0.9$ s. 
Model b1b-10000-N-Lam also has a similar pattern. But
the fall of $Y_{\rm e}$ slows down at 0.8 s, showing that
the inner part does not collapse directly; while
the outer matter, which continues to flow inwards, 
as implied by the growth of the central density, 
triggers further electron captures which make the ONeMg core collapse. 

\begin{figure}
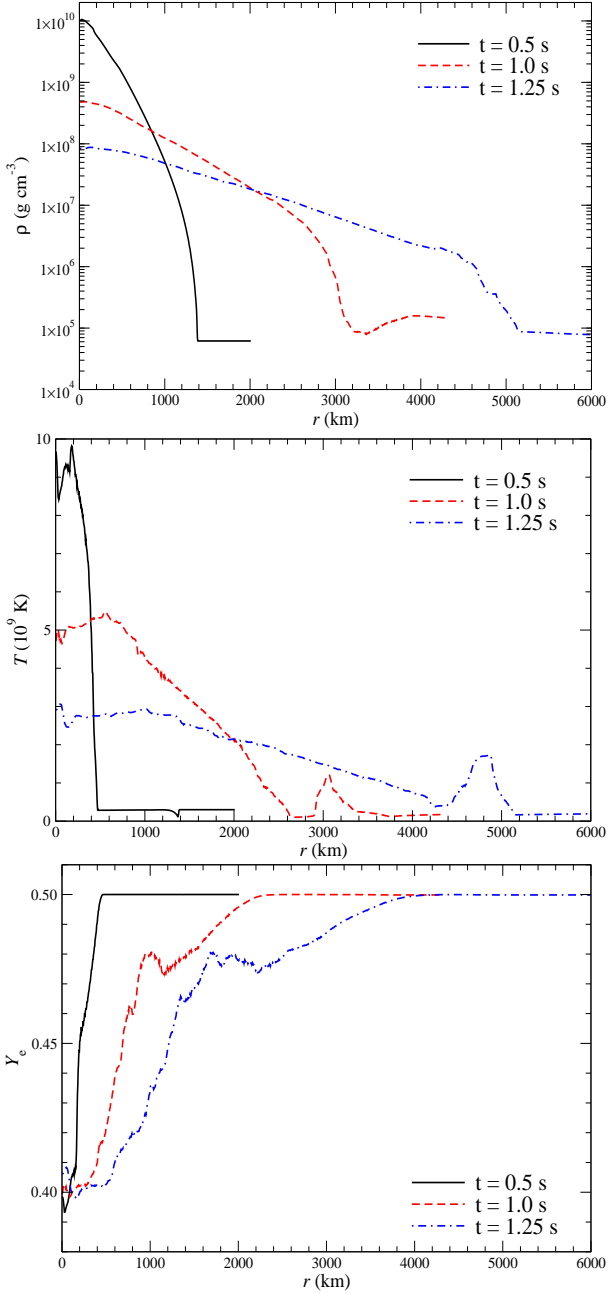

\centering
\includegraphics*[width=8cm,height=5.7cm]{fig13a.eps}
\includegraphics*[width=8cm,height=5.7cm]{fig13b.eps}
\includegraphics*[width=8cm,height=5.7cm]{fig13c.eps}
\caption{(top panel) The angular averaged radial density profiles of Models b1b-09950-N
at $t = 0.5$, 1.0 and 1.25 s. (middle panel) Same as 
the left panel but for the temperature profiles. 
(bottom panel) Similar to the left panel but for the $Y_{{\rm e}}$ profiles.}
\label{fig:profiles_turb_plot}
\end{figure}

\begin{figure}
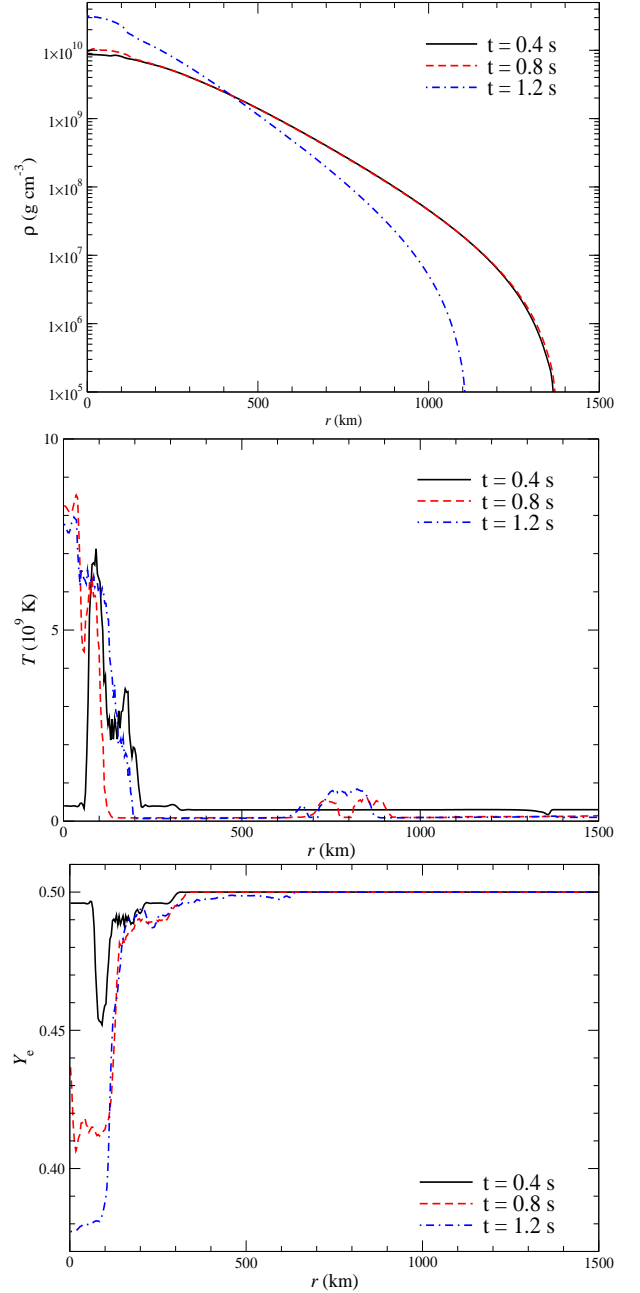

\centering
\includegraphics*[width=8cm,height=5.7cm]{fig14a.eps}
\includegraphics*[width=8cm,height=5.7cm]{fig14b.eps}
\includegraphics*[width=8cm,height=5.7cm]{fig14c.eps}
\caption{(top panel) Angular averaged radial density profiles of Models b1b-09950-N-Lam
at $t = 0.5$, 1.0 and 1.25 s. (middle panel) Same as the 
the upper panel but for the temperature profiles. 
(bottom panel) Similar to the left panel but for the $Y_{{\rm e}}$ profiles.}
\label{fig:profiles_lam_plot}
\end{figure}

Then, we compare the evolution of the two models by 
plotting the radial profiles. The radial profiles
are obtained by doing an angular average of 
the related quantities. This allows us to compare 
directly how the ONeMg core responds under different
types of flame, and furthermore how the ONeMg core looks
dynamically when it expands or collapses.

In Figure \ref{fig:profiles_turb_plot} we plot the 
density, temperature and $Y_{{\rm e}}$ radial profiles for the 
Model b1b-09950-N in the top, middle and hottom panels 
respectively. We plot in Figure \ref{fig:profiles_lam_plot}
similar to Figure \ref{fig:profiles_turb_plot} but for 
the Model b1b-09950-N-Lam.

Model b1b-09950-N is an expanding model. The central density 
of the star quickly drops by two orders of magnitude in 
$\approx$ 1 s. However, the monotonic variation of the density profile
in the inner core does not change throughout the 
simulation. This shows that the deflagration we modeled 
is quiet enough to suppress acoustic wave generation.
On the other hand, there is almost no change in the 
profile in Model b1b-09950-N-Lam, which is a collapsing model.
The star shows to contract homologously until the end of simulation.

In Model b1b-09950-N, the temperature profiles 
show more features compared to the 
density profiles. The off-center burning allows the temperature
peaks at 100 and 500 km at $t = 0.5$ and 1.0 s.
When the star begins its expansion, the off-center
temperature peak is smoothed out. Besides that,
the initial injection of flame creates a small pulse
which heats the near-surface matter and creates
a small temperature bump at 3000 (5000) km 
at 1.0 (1.25) s. For Model b1b-09950-N-Lam, the flame 
is still off-center at 0.4 s. A small temperature 
bump is observed at $\sim 700$ km due to the perturbation of 
initial flame. Until the end of the simulation, the 
high temperature region ($T > 3 \times 10^9$ K)
is confined within the innermost 200 km.

In Model b1b-09950-N, the initial electron captures
are confined to the innermost 200 km. Accompanying with the expansion,
the shape of the $Y_{{\rm e}}$ profile is frozen beyond 1.0 s, 
where expansion elongates the profile. 
For Model b1b-09950-N-Lam, the slow "laminar" allows more
transport of $Y_{{\rm e}}$ before rapid electron captures 
take place. The electron captures at $t = 0.8$ and 1.2 s
are localized in the innermost 100 km and carry on until the 
end of simulation.

\subsection{Effects of Pre-Runaway Time Lapse}
\label{sec:time_lapse}

In our simulations, the initial flame we impose 
is limited by the size of the resolution grid ($\sim 4$ km). 
However, it is unclear whether the flame is triggered
at this size, or at a size smaller than the grid resolution.
In fact, in \cite{Timmes1992}, the size of flame in mass
can be as small as ($10^{3}$ - $10^{17}$ g),
depending on the local temperature, such that the 
runaway can occur spontaneously. This means that the initial runaway can have a size much smaller than the 
typical resolution ($\sim$ km) when the first 
nuclear runaway starts. 
Therefore, there can be a time-lapse between the 
"first" nuclear runaway and the flame structure we used.
The time-lapse allows the $Y_{{\rm e}}$ inside the
runaway to be different from its initial value. 
To account for this lapse, we prepared models with a much smaller $c3$ flame (for a few grids to make 
the flame shape well resolved by the level-set
method). The flame is then allowed to expand self-similarly until it becomes the size and the shape of the $bc3$ flame. Meanwhile, all nuclear reactions,
such as photo-disintegration of $^{56}$Ni into $^{4}$He,
and electron capture, can proceed.
After the flame reaches the size of the $bc3$ flame, the fluid advection of the flame is resumed.
This attempts to mimic the laminar phase where the 
flame grows self-similarly without being
perturbed by the fluid motion.

In this series of models, we change the initial size of the flame from 25\% to 75\% of the original flame used in the $c3$ Model series. We choose the largest flame model because we want to contrast the effects of the time-lapse in the initial laminar phase. Again we use the bc3 as the template
because it has a sufficiently large size such that 
we can construct a similar flame structure of smaller size for comparison. Also, the effects of this treatment can be more clearly observed near the bifurcation density, which is near $10^{9.95}$ g cm$^{-3}$ for the $bc3$ flame. When a smaller flame is used, models with lower initial $\rho_c$ are necessary to see the changes. We stick to $10^{9.95}$ g cm$^{-3}$ because it is the typical runaway density predicted from the stellar evolutionary models using the Ledoux criterion.

\begin{figure}
\centering
\includegraphics*[width=8cm,height=7cm]{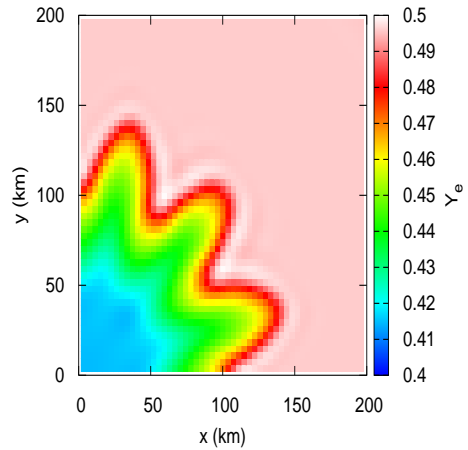}
\caption{The $Y_{\rm e}$ colour plot of Model c3-09950-N-B050 after including the laminar propagation phase.}
\label{fig:ye_lamphase_plot}
\end{figure}

In Figure \ref{fig:ye_lamphase_plot}, we
plot the $Y_{\rm e}$ of the Model c3-09950-N-B050
at the moment we allow the deflagration to 
follow the fluid motion when the flame reaches the 
required size. It takes $\sim$125 ms 
for the flame to reach from half of its size (about 
60 km) to the current size. The $c3$-flame is chosen as described above. Near the flame
surface, since the weak interaction is slow, most matter keeps its initial $Y_{\rm e}$. Around $r = 80$ km, the $Y_{\rm e}$ quickly drops from 0.50 to $\sim 0.44$. Within the innermost 40 km, the $Y_{\rm e}$ can drop as low as 0.40 -- 0.42. 

\begin{figure}
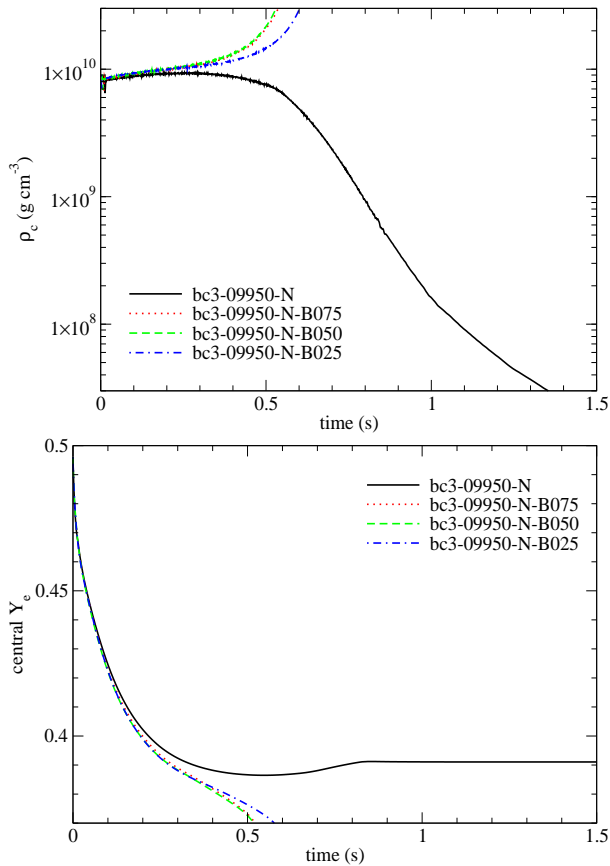

\centering
\includegraphics*[width=8cm,height=5.7cm]{fig16a.eps}
\includegraphics*[width=8cm,height=5.7cm]{fig16b.eps}
\caption{(upper panel) 
Evolution of central density against time 
for Model bc3-09950-N (default flame size), 
bc3-09950-N-B075 (75\% flame size), 
bc3-09950-N-B050 (50\% flame size) and 
bc3-09950-N-B025 (25\% flame size). All models share the same initial flame geometry $bc3$ and 
initial central density $10^{9.95}$ g cm$^{-3}$, and they assume Newtonian gravity.
(lower panel) Same as the upper panel, 
but for the central $Y_{\rm e}$.}
\label{fig:rhoc_lamphase_plot}
\end{figure}

In the upper panel of Figure \ref{fig:rhoc_lamphase_plot}, we plot the 
evolution of central density against time for Models bc3-09950-N, bc3-09950-N-B075, bc3-09950-N-B050 and bc3-09950-N-B025. Model bc3-09950-N explodes while the other three models collapse. In the first 0.2 s, all four models share a similar $\rho_c$ evolution. However, beyond that time, the $\rho_c$ in the latter three models are 
slightly higher, which leads to their later collapse at 0.5 -- 0.6 s.

In the lower panel of Figure \ref{fig:rhoc_lamphase_plot}, we plot 
the central $Y_{\rm e}$ evolution for the same set of models. The three collapsing models show a qualitatively similar pattern as those in previous sections. However, they all share a lower $Y_{\rm e}$ compared to the exploding model bc3-09950-N.
This is related to the difference in the relaxation of the initial flame by isobaric expansion. 

The models considering the effects of 
pre-runaway time-lapse show that the ONe core evolved from the stellar evolutionary
model is likely to collapse into a neutron star and create an ECSN, but the exact details still strongly depend on the pre-runaway scenario, where the electron captures in the sub-grid scale are important for the initial $Y_{\rm e}$ profile and also its subsequent dynamics. We also remark that despite the flame structure of flame c3-09950-N and bc3-09950-N-B025 being the same, they are not identical because bc3-09950-N-B025 has more time for electron captures during the enforced laminar flame phase. Also, the frozen flame shape during the laminar phase in Model bc3-09950-N-B025 causes a different turbulent energy distribution when the 
flame can propagate freely compared with Model c3-09950-N.

\subsection{Effects of Initial $^{24}$Mg}
\label{sec:composition}

In Section \ref{sec:methods} we discussed that we do not include $^{24}$Mg
in the raw fuel because there is numerical difficulty in how to distinguish
$^{24}$Mg from the fuel and from the ash. The replaced composition may
over-estimate the energy production. 
In previous sections we have shown that the actual results are
sensitive to multiple parameters in the configuration. Here we further 
examine how the choice of the initial composition affects the 
collapse-explode bifurcation. In particular, we study how the initial
abundance of $^{24}$Mg affects the final evolution of ECSN.

To characterize the effects of the initial composition on the final
fate of ECSN, we construct ONe cores using the uncertainties of 
the mass fraction of $^{16}$O as a model parameter at the 
same initial central density ($10^{9.95}$ g cm$^{-3}$). 
After that, we ignite ONe core with an identical initial flame
$c3$. In all previous models, the initial $^{24}$Mg, which has captured electrons
to form $^{24}$Ne, is regarded as part of the $^{16}$O.
Here we variate the initial model and treat the $^{24}$Ne 
as part of the $^{20}$Ne or both $^{16}$O and $^{20}$Ne
by half. Besides the initial model, the initial composition also 
affects the laminar flame speed (see for example the $^{12}$C-
and $^{16}$O dependence of the laminar flame speed in \cite{Timmes1992}.)
The energy production when the fuel is burnt completely to NSE
is also adjusted according to the initial composition.
For a higher $^{16}$O initial mass fraction, the laminar
flame speed is higher and the energy production is 
also higher.

\begin{figure}
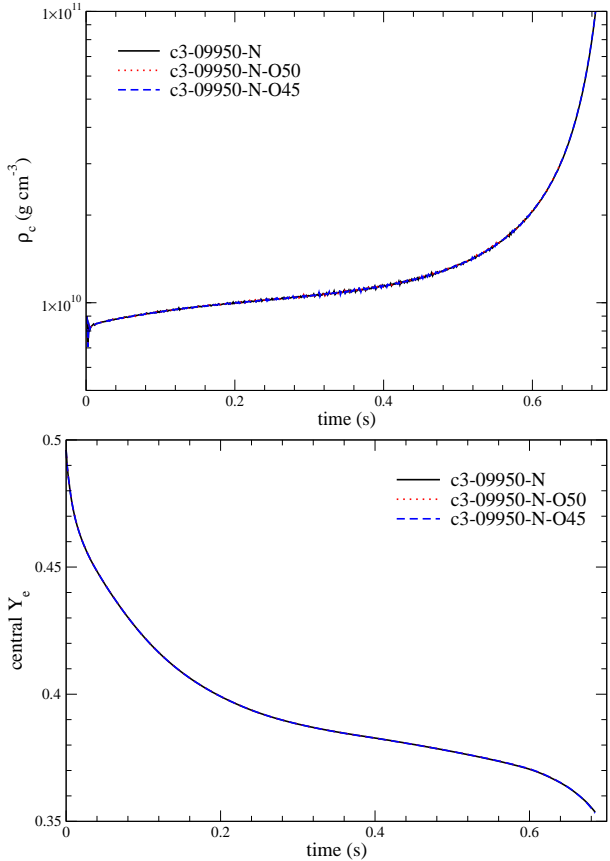

\centering
\includegraphics*[width=8cm,height=5.7cm]{rhoc_Oratio_plot.eps}
\includegraphics*[width=8cm,height=5.7cm]{yec_Oratio_plot.eps}
\caption{(upper panel) 
Evolution of the central density against time 
for Models c3-09950-N (default $^{16}$O ratio) and
c3-09950-N-O50 ($X(^{16}$O)$ = 0.50$) and
c3-09950-N-O45 ($X(^{16}$O)$ = 0.45$). 
All models share the same initial flame geometry $c3$ and 
initial central density $10^{9.95}$ g cm$^{-3}$, and they assume Newtonian gravity.
(lower panel) Same as the upper panel, 
but for the central $Y_{\rm e}$.}
\label{fig:rhoc_Oratio_plot}
\end{figure}

In Figure \ref{fig:rhoc_Oratio_plot} we plot in the upper
and lower panels the central density and $Y_{\rm e}$ 
against time respectively for the three models described 
above. In general the three curves overlap each other. 
No observable changes can be seen from the onset of 
nuclear runaway until the end of the simulations. 
This shows that the change in the initial composition
does not affect the final fate of ECSN, in the uncertainties
considered here. 
Similar results can be found for the central $Y_{\rm e}$.
The drops of $Y_{\rm e}$ of the three cases do not show 
variations from each other. From both figures, it suffices
to show that treating the neglected $^{24}$Mg as part of 
the initial $^{16}$O or as part of the initial $^{20}$Ne 
does not bring any qualitative change to our results. 

\section{Discussion}
\label{sec:discussion}

\subsection{Global Properties of ONeMg Core}
\label{sec:global}

\begin{figure}
\centering
\includegraphics*[width=8cm,height=5.7cm]{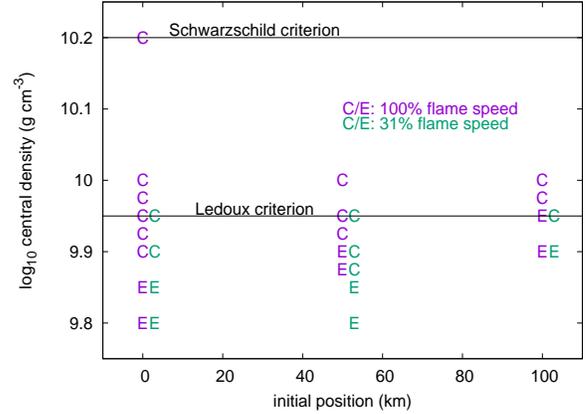}
\caption{Phase diagram of the collapse-expand bifurcation
for the models studied in this work. C and E correspond to 
the models which collapse and expand, respectively. The X- and Y-positions of the letter
correspond to the flame position (0, 50 and 100 km) and initial central density ($10^{9.8} - 10^{10.2}$).
Models for two contrasting flame speeds at 100 \% and 31 \%
are shown as the left (purple font) and the right (green font letter). 
The upper (lower) line corresponds
to the runaway density predicted by the Schwarzschild (Ledoux)
convection criteria.}
\label{fig:meanYe_plot}
\end{figure}

In previous sections, we have compared the final 
evolution of ONeMg cores with different input physics. 
We find that the initial central density, flame position,
and flame speed are important for determining the final fate of the star.
In this section, we summarize the models by building 
a phase diagram of them.

In Figure \ref{fig:meanYe_plot} we plot the phase diagram
of the collapse-expand bifurcation of our models
with the initial flame position and the initial central density
as the $x-$ and $y-$coordinates. Two contrasting flame 
speeds, the default one and a reduced one, at an asymptotic  
value of 31 \% of the default value, are shown.
We mark the figure with two 
horizontal lines that characterize the runaway densities
using the Ledoux $(10^{9.95}$ g cm$^{-3})$
and Schwarzschild criteria $(10^{10.2}$ g cm$^{-3})$. 
These are the expected runaway densities taken from the
literature (see e.g., \cite{Miyaji1987,Schwab2015}). We focus 
on models near the bifurcation point. Once the transition is located,
models with initial central densities above that collapse and 
those below that expand.
All models with an initial central density $> 10^{9.95}$ g cm$^{-3}$
collapse for the centered flame and off-center flame at 50 km from the origin.
A higher transition density at $10^{9.975}$ g cm$^{-3}$ 
is observed when the flame starts at 100 km from center. 
This suggests that ONeMg models using the Schwarzschild 
criterion, i.e., an ignition density of $10^{10.2}$ g cm$^{-3}$,
collapse for both centered and off-center flames.
ONeMg models using the Ledoux criterion, i.e., 
an ignition density of $10^{9.95}$ g cm$^{-3}$ collapse,
when the first flame starts within the innermost 50 km.
However, we remind that the convection after the onset
of core O-burning, even in the Ledoux criterion, could
delay the nuclear runaway substantially. The value $10^{9.95}$ g cm$^{-3}$
should be treated as a lower limit.

By examining the distribution of "C"s in the diagram,
we find that the majority of models still collapse into NSs. 
In the parameter range surveyed in this work, 
the initial flame position affects primarily 
the variations of the transition density, compared to 
the variations of the turbulent flame speed formula.
A change of the transition density in the $\log_{10}$ scale
by 0.075 can be observed for different initial flame positions
but only 0.025 for different turbulent flame speed formulas.

This diagram demonstrates the diversity of the 
possible outcomes of the ONeMg core, even when they
are prepared in a very similar way in terms of 
mass, flame geometry and flame position. 
It demonstrates the necessity of future stellar evolutionary
work in a better modeling of the convective process before
the runaway. This includes (1) the pre-runaway
configuration by the detailed nuclear runaway position,
(2) its initial nuclear runaway size in mass, and 
(3) the ONeMg core $Y_{\rm e}$ profile and its composition
\footnote{In this work we have not explored in details the 
role of the initial $Y_{\rm e}$. The details of the $Y_{\rm e}$ 
depend on the treatment of convective mixing
after the hydrostatic O-burning has started. The mixing 
can compensate the drop of the $Y_{\rm e}$ by the 
electron captures of $^{16}$O and $^{20}$Ne. 
Such details can be provided by the stellar evolutionary
models, but numerical difficulties for resolving the 
hydrostatic O-burning front under convective mixing make such
prediction difficult. In fact, in Section \ref{sec:time_lapse}
by using the pre-conditioned flame with initial $Y_{\rm e}$ 
differences, the initial $Y_{\rm e}$ can affect the transition
density of ECSN.}

We also remark about the divergence of results among
models with a $c3$, $mc3$ or $bc3$ flame. They demonstrate
the importance of how the collapse depends on the 
global motion of the ONeMg core. We showed that models with a $c3$
flame have a longer time for the onset of collapse
than that with an $mc3$ flame, while those with $bc3$ flame
expand. The small $c3$ flame requires a longer time for the development
of flame until global contraction is triggered.
On the other hand, the larger $mc3$ flame allows 
more electron captures to take place in the ash.
This accelerates the global contraction.

\subsection{Comparison with Literature}
\label{sec:failed}

Since there is no explicit work in the literature 
except \cite{Jones2016} on the multi-dimensional simulations of
ECSN, we compare our
hydrodynamics results with theirs.
In Table \ref{table:compare_models}
we list the input physics and configurations
used in their work and this work. 
Overlap in microphysics is attempted to make the 
comparison of results easier. 
However, some fundamental infrastructure, including the
hydrodynamics solvers, equation of states and 
nuclear reaction schemes are
different.

First, we examine the threshold density for the 
expand-collapse bifurcation.
Our models show that a central ignited flame has a transition
density at $10^{9.9}$ g cm$^{-3}$, which increases to 
$10^{9.975}$ g cm$^{-3}$ when the flame distance from the center
increases from 0 km to 100 km. In the six models presented in \cite{Jones2016}
with a $\rho_c$ at $10^{9.90}$, $10^{9.95}$ and $10^{10.2}$ g cm$^{-3}$,
the first two models expand and the last one collapses.
Given that they use a different flame structure ($\sim 100$ flame bubbles
with a total mass $\sim 10^{-3} M_{\odot}$ burnt) at the beginning,
our results agree qualitatively with theirs by considering their representative
flame distance, initial burnt mass, and central densities. 
Also, their model with $\rho_{\rm c,ini} = 10^{10.2}$ g cm$^{-3}$
has a collapse time around 0.3 s,
which also agrees with 
ours (0.26 s) (see for example Figure \ref{fig:rhoc_GR_plot} for the evolution
of central density). 

Then we compare the flame morphology.
In their work, they show the flame structure
in Figure 6 and the cross-section cut in Figure 7. 
We compare these with our results in Figure \ref{fig:flame8_plot}.
The outburst of flame in spherical shape with Rayleigh-Taylor 
instabilities and their induced small-scale sub-structures can be
seen in both work. Because our model has a 
coarser resolution compared to their work, the flame structure 
in our model shows fewer sub-structures than theirs.

\begin{table*}
\caption{Comparison of input physics and 
numerical setting between our work and those in \cite{Jones2016}.}
\begin{center}
\label{table:compare_models}
\begin{tabular}{|c|c|c|}
\hline
Physics component & Our work & \cite{Jones2016} \\ \hline
Numerical code & \cite{Leung2015a} & LEAFS \\
Dimensionality & 2D & 3D \\
Coordinates & Cylindrical & Spherical \\
Spatial discretization scheme & WENO (5th order) & PPM (3rd order) \\ \hline
EOS & Helmholtz & Individual prescription \\
Sub-grid turbulence & \cite{Niemeyer1995b} & \cite{Schmidt2006b} \\
Energy scheme (in Hydro) & 3-step burning with NSE & 1-step burning with NSE\\
Hydro Isotope network & 7 & 5 \\ \hline
Flame capturing scheme & Level-set methods & Level-set methods \\ \hline
Post-processing Isotope network & 495 & N/A \\
Electron capture rate & Extension of \cite{Seitenzahl2010} & Extension of \cite{Seitenzahl2010} \\
Nuclear reaction rate & \cite{Langanke2001} & N/A \\ \hline

\end{tabular}
\end{center}
\end{table*}

\begin{figure}
\centering
\includegraphics*[width=8cm,height=6cm]{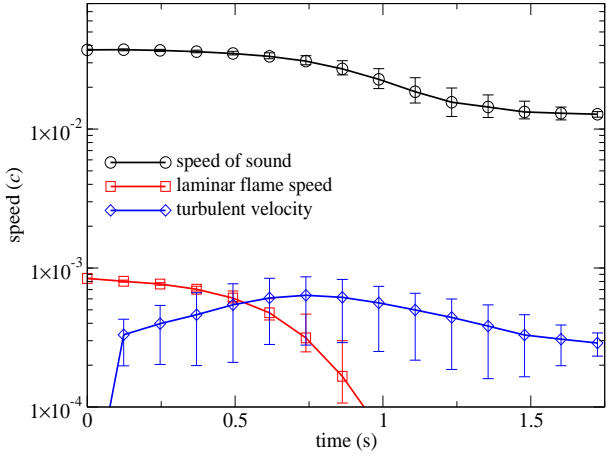}
\caption{Speed of sound, laminar flame speed and
turbulent flame speed for Model c3-09850-N shown in
Figure \ref{fig:rhoc_plot}. The lines 
stand for the mass-averaged values from the grids where the
flame surface can be found. The error bars
show the maximum and minimum flame speeds found in simulations 
at the corresponding time points.}
\label{fig:flamespeed_plot}
\end{figure}

At last, we compare the time evolution of a turbulent flame.
In Figure \ref{fig:flamespeed_plot} we plot the 
speed of sound, laminar flame speed and 
turbulent flame speed of the Model c3-09850-N. 
The data is taken from a grid point which is actively
burning by deflagration.
In the beginning, laminar flame is dominant 
because we assumed a ONeMg core in hydrostatic equilibrium.
We note that in their work the turbulent flame speed is 
slower than ours. It is because, in the formalism 
from \cite{Pocheau1994}, the minimal turbulent flame 
speed is always the laminar flame speed. We estimate that
the turbulence velocity is comparable with the flame
speed for $t < 0.2$ s.

The turbulent flame 
speed quickly exceeds the laminar flame and 
reaches an equilibrium value of $\sim O(10^{-2})$
of the speed of sound. 
This figure can be compared with Figure 4 in 
\cite{Jones2016} G13 model but with three differences. 
First, they used three-dimensional
Cartesian coordinates, and we use two-dimensional
cylindrical coordinates. Their three-dimensional simulations
may allow a more complex flame structure in the simulation.
The higher dimensional simulation allows a more flexible choice
of the initial flame with less concern of enhancement by 
a particular boundary condition.
Second, the subgrid-scale (SGS) model is 
based on the formalism in \cite{Schmidt2006b} while 
ours is based on the scheme in \cite{Niemeyer1995b}. Both models
belong to the class of one-equation model but 
with different closures. 
Third, their models start from 
a number of off-center bubbles, while, due
to symmetry, we choose a centered flame as the 
initial flame structure. Our "three-finger" structure
helps to enhance the turbulence by the initial asymmetrical flow.
This allows our model to reach the turbulent regime faster than
theirs, resulting in more vigorous nuclear burning.
On the other hand, the bubble structure, 
where bubbles are geometrically isolated at the beginning,
makes the generation of turbulence slower because of the 
initially isotropic expansion of the bubble.
Even with very different sub-grid turbulence models, 
the results are qualitatively similar, 
such as the asymptotic value and the
range of turbulent flame speed found in the simulation. 
One major difference is the time when turbulence becomes 
saturated owing to our choice of initial flame. 
We choose the $c3$ flame as was done in \cite{Reinecke1999b}. 
At last, in our simulations, the reflective inner boundaries
of both planes can create boundary flows, 
which can also enhance the SGS turbulence production.
Future extension of our work using three-dimensional simulations
and with similar flame structure and resolution, will 
provide more rigorous constraints on the collapse-expand
transition boundary.

\subsection{Electron Capture Triggered Thermonuclear Explosion}

\begin{figure}
\centering
\includegraphics*[width=8cm,height=5.7cm]{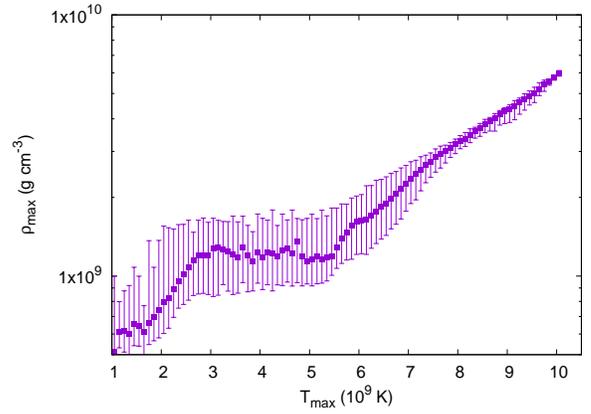}
\caption{Maximum density $\rho_{{\rm max}}$ against maximum temperature $T_{{\rm max}}$ 
for the tracer particles in Model c3-09800-N. The error bars
stand for the range of $\rho_{{\rm max}}$ of the tracers in the 
same bin of $T_{{\rm max}}$.}
\label{fig:rhoT_plot}
\end{figure}

\begin{figure}
\centering
\includegraphics*[width=8cm,height=5.7cm]{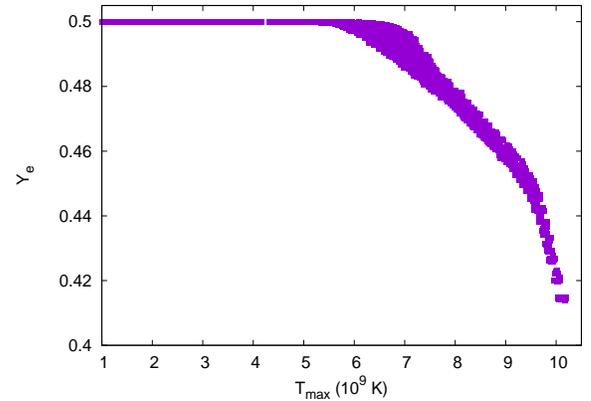}
\caption{Final $Y_{{\rm e}}$ against maximum temperature $T_{{\rm max}}$ for
the tracer particles for the Model c3-09800-N.}
\label{fig:TmaxYe_plot}
\end{figure}

Here we discuss the properties of the exploding
models, and then we analyze the possible nucleosynthesis signature
of the exploding models. We analyze the 
thermodynamics history of 
one of the expanding models c3-09800-N by studying 
the tracer particles.

First, we plot in Figure \ref{fig:rhoT_plot} $\rho_{{\rm max}}$ against 
$T_{\rm max}$ derived from the tracer particles in the simulations.
The maximum density and temperature are defined by the maximum values
experienced by the particles throughout their history from the onset 
of the flame until the expansion. The distribution is separated into three parts.
The first part is a monotonically increasing trend above high $T_{{\rm max}} \sim 6 \times 10^9$ K.
The second part is an approximately constant $\rho_{{\rm max}}$
at intermediate $T_{{\rm max}} = 3 - 6 \times 10^9$ K and the 
third part is another monotonic increasing trend at low $T_{{\rm max}} < 3 \times 10^9$ K.
The tight relation for high $T_{{\rm max}}$ is consistent with
typical Type Ia supernovae exploded by pure turbulent deflagration.
The subsonic deflagration wave does not generate any strong sound
wave which can increase the spread of $T_{{\rm max}}$ for a given
$\rho_{{\rm max}}$. Also, most inner parts of the core are burnt
at the same time by the centered flame. On the other hand,
in the intermediate $T_{{\rm max}}$ regime, the flame becomes aspherical
that the fluid elements with the same initial density
can experience different levels of time-delay when the 
flame arrives. The low $T_{{\rm max}}$ regime corresponds to
where the flame is quenching at $\rho \sim 10^9$ g cm$^{-3}$. The value is higher than 
that for CO matter because the typical energy release 
for the burning of the ONe matter is lower.

Then, we plot in Figure \ref{fig:TmaxYe_plot} the $Y_{{\rm e}}$ distribution
of the tracer particles as a function of $\rho_{{\rm max}}$.
The distribution consists of two parts. For the tracer particles
which experienced electron capture ($T > 5 \times 10^9$ K),
the final $Y_{{\rm e}}$ drops when $T_{{\rm max}}$ increases. 
The lowest $Y_{{\rm e}}$ $\sim 0.41$ are obtained by the particles 
having the highest temperature $\sim 10^{10}$ K 
in their thermodynamical history. A small spread can be seen for particles close
to the NSE transition temperature. Again, this is related
to the aspherical flame propagation.

Since the electron capture rate is much slower than
the dynamical timescale, the final $Y_{{\rm e}}$ determines the
isotopes in the ejecta. At such low $Y_{{\rm e}}$, neutron-rich
isotopes such as $^{48}$Ca ($Y_{{\rm e}} = $ 0.41), $^{54}$Cr
($Y_{{\rm e}} =  0.42$), $^{60}$Fe ($Y_{{\rm e}} = 0.43$),
and $^{64}$Zn ($Y_{{\rm e}} = 0.47$)
are the representative stable isotopes. The relative $Y_{{\rm e}}$
for Zn is high but the high entropy environment enhances the
formation of this particular isotope compared to the 
Type Ia SN counterpart. See for example \cite{Wanajo2018,Jones2018}.
As discussed in \cite{Nomoto1991,Woosley1997}, these isotopes
are not consistently produced in ordinary SNe Ia. These isotopes,
if ejected, can provide tight constraints on the relative 
rate of ECSNe to other types of supernovae.

We do not attempt to do the nucleosynthesis as in our previous 
work because a longer time after the explosion ($\sim 10$ s)
is necessary to distinguish the tracers which are
ejected and tracers which fall back to form the remnant. 
Without this information, the final yield might overestimate
the final masses of iron-peak elements, which are more likely
to fall back when the fluid elements move outwards and 
expansion, which transport their momenta from the core 
to the envelope. 

Furthermore, after the expansion takes place,
the ONeMg core ejects partially its matter. The ejecta may
contain elements from both the ONe-rich fuel and the Fe-rich ash
produced in the ONeMg core.
The remaining matter becomes a lower-mass remnant. 
In \cite{Jones2016} a typical mass of $\sim 1.2 ~M_{\odot}$
of the remnant is recorded. The lower mass remnant may coincide with
the low-mass SiFe-rich white dwarfs observed \citep{Raddi2018}.
In \cite{Jones2018} they further computed
the nucleosynthesis yield using a large nuclear reaction
network. In our future work, we will compare our nucleosynthesis yield 
with theirs and perform a detailed analysis
for different progenitor masses and flame structures.

\begin{figure}
\centering
\includegraphics*[width=8cm,height=7cm]{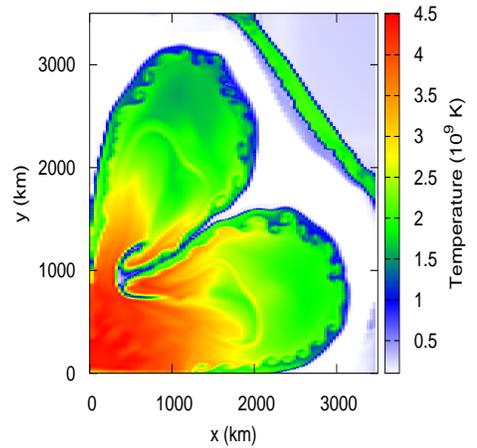}
\caption{Temperature profile of Model 
bc3-09950-N at 1 s of the simulation. Notice that the 
flame is highly irregular with the signature from
Rayleigh-Taylor instabilities and Kelvin-Helmholtz instabilities.}
\label{fig:flame8_plot}
\end{figure}

\subsection{Conclusion and Future Work}

In this work we model the final evolution of the oxygen-neon-magnesium
(ONeMg) cores using two dimensional hydrodynamical
simulations. Based on the temperature
and $Y_{\rm e}$ profile as functions of mass coordinates obtained from
stellar evolutionary models, 
we construct the ONeMg core in hydrostatic equilibrium
with a range of central densities from $10^{9.8} - 10^{10.2}$ g cm$^{-3}$.
We follow the ONe deflagration
phase to examine in which conditions the ONeMg core
can collapse into a neutron star. 

We surveyed ONeMg core models of various configurations.
They include central densities between $10^{9.80} - 10^{10.20}$
g cm$^{-3}$ and different flame structures with masses
between $10^{-4} - 10^{-2}$ $M_{\odot}$ in a centered
or off-centered ignition kernel. We also explore
the effects of input physics, which include the 
relativistic corrections in gravity, turbulent flame speed
formula and the treatment of the laminar deflagration phase.
We find that except the general relativistic effects,
the latter two can strongly affect 
the collapse condition. 
The exact transition density depends on 
the input physics but we find that 
the ONeMg core can collapse with an
initial central density with a range from $10^{9.90}$ to $10^{9.975}$ g cm$^{-3}$. 
This is consistent with
the current picture of stellar evolution
that the ECSNe evolved from stars of masses
8 -- 10 $M_{\odot}$ could be the 
origin of the lower-mass branch of the neutron star
population.

We study how the input physics affects
the bifurcation condition of the ONeMg core. Besides
the sensitivity of the models to the initial mass
as reported in the literature, for the 
models with the same initial central density, a centered flame 
favors the collapse scenario. 
Slower flame (laminar flame or less effective turbulence models)
also favors the collapse scenario. A pre-conditioned 
flame is also favorable to the collapse branch.
However, relativistic corrections in gravity and the exact abundance of 
$^{24}$Mg do not play the main role
in the evolution of the deflagration phase.

We present a phase diagram for the collapse-expansion bifurcation
for models with a range of central densities, flame positions,
and turbulent flame speeds.
We study the thermodynamics history of the ECSN 
and discussed its nucleosynthic implications.
We also carry out a detailed comparison of our models
with the representative models in the literature. 
Our results suggest that it is necessary to carefully put in treatments like
the pre-runaway convection
in the stellar evolution of ONeMg core, 
the turbulent flame modeling, and the mapping from
stellar evolutionary models to hydrodynamics
simulations to determine
the final fate of super-AGB stars after electron-capture-induced
nuclear runaway has started. 
In the stellar evolution theory, these treatments include:

\noindent (1) the exact runaway position 
of the O-Ne deflagration, whether it is centered
or off-center, and its size;

\noindent (2) the convective mixing and its velocity structure 
in the ONeMg core before nuclear runaway;

\noindent (3) the detailed $Y_{\rm e}$ profile
\footnote{In this work we do not explicitly use distinctive 
initial $Y_{\rm e}$ profiles. However, in Section \ref{sec:time_lapse}
we demonstrated how the initial laminar phase can change the bifurcation
criteria. The laminar phase contributes to a distinctive
$Y_{\rm e}$ profile when the turbulent flame is evolved 
(see Figure \ref{fig:ye_lamphase_plot}). This provides the first 
indication that even when an identical flame structure 
is used, the differences in the $Y_{\rm e}$ profile can 
alter the final fate of that stellar model.};

\noindent (4) the chemical composition, especially the 
residue $^{12}$C and $^{16}$O/$^{20}$Ne mass fraction
ratio in the ONeMg core.

The following improvements may enhance the predictability 
of our models:

\noindent (1) the empirical formula between local velocity
fluctuations and the corresponding flame propagation speed,

\noindent (2) the detailed velocity spectra of the sub-grid
scale eddy motion and its impact on flame geometry.

\section{Acknowledgment}

This work was supported by World Premier 
International Research Center Initiative 
(WPI), MEXT, Japan, and JSPS
KAKENHI Grant Numbers JP26400222, JP16H02168, JP17K05382
We acknowledge the support by the Endowed Research
Unit (Dark Side of the Universe) by Hamamatsu Photonics K.K.
We thank the developers of the stellar evolution code MESA
for making the code open-source.  

We thank the anonymous referee for the very detailed
and constructive comments for improving this article.
We also thank our
colleagues who kindly spent their time
to help us improve the writing of this article. 

We thank F. X. Timmes for
his open-source micro-phsyics algorithm
including the Helmholtz equation of state 
subroutine, the torch nuclear reaction network
designed for an arbitrary choices of isotopes,
the seven-isotope nuclear reaction 
network. We also thank Ming-Chung Chu for the 
initial inspiration of building the 
hydrodynamics code. We also thank H. Shen for 
her open source equation of state for the 
nuclear matter. We thank Christian Ott and Evan O'Connor
for the EOS driver for reading the HShen EOS
and their open source data for the parametrized 
electron capture value before bounce. We thank 
Zha Shuai for his discussion of the preliminary 
results of his pre-runaway models.

\appendix

\section{Resolution Study of the Code}

In this work we have performed a number of simulations using the same 
resolution at $\sim$ 4 km. It has been a matter of issue how the results
depend on the resolution, especially in simulations of this kind which
rely on input physics involving grid size as the input parameter (e.g. sub-grid scale
turbulence). To understand the validity of our results, we attempt 
to rerun the benchmark model (Model c3-09950-N) and its counterpart with
a slower flame speed (Model c3-09950-N-v025) in a finer resolution 
at $\sim$ 2 km. We denote this model as Model c3-09950-N-fine
and Model c3-09950-N-v025-fine respectively. We also compare the models using
slower flame because the slower flame takes longer time 
for the collapse to occur. This provides more time for the propagation
of the flame, which may amplify the resolution effects.

\begin{figure*}
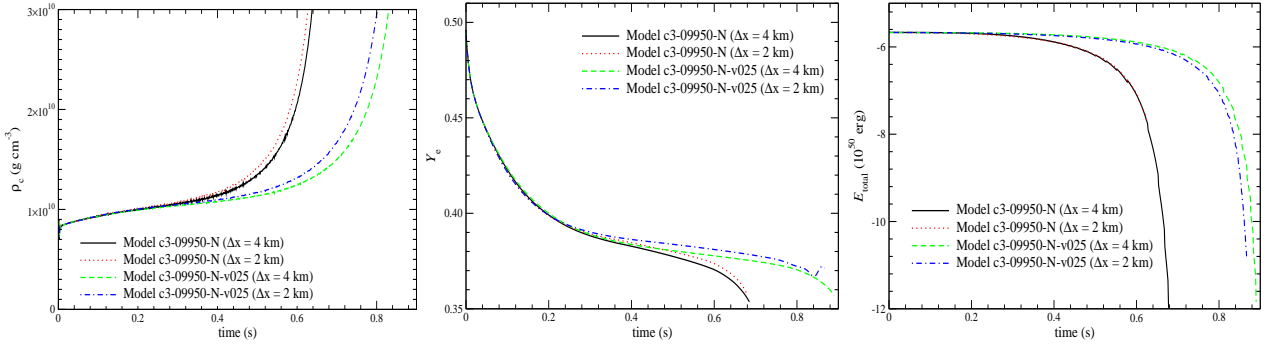

\centering
\includegraphics*[width=5.5cm,height=4.5cm]{fig22a.eps}
\includegraphics*[width=5.5cm,height=4.5cm]{fig22b.eps}
\includegraphics*[width=5.5cm,height=4.5cm]{fig22c.eps}
\caption{(left panel) The central density for the Model
c3-09950-N, c3-09950-N-fine, c3-09950-N-v025 and 
c3-09950-N-v025-fine. (middle panel) Similar to 
the left panel but for the central $Y_{{\rm e}}$. 
(right panel) Similar to the left panel but for the total energy.}
\label{fig:dx_plot}
\end{figure*}

In the left panel of Figure \ref{fig:dx_plot} we plot the central density evolution of the two models.
The evolution of the first 0.5 s of the Models c3-09950-N and c3-09950-N-fine is almost identical. 
Similar pattern can be seen for the pair of Models c3-09950-N-v025 and 
c3-09950-N-v025-fine.
However, the models deviate from each other where the central density
of the finer model grows faster. Despite that, both models stop at a time
of $\sim$ 0.62 and 0.82 s with a difference $\sim 1 \%$, 
when the central density reaches the threshold
defined in the code.

In the middle panel of Figure \ref{fig:dx_plot} we plot the central $Y_{{\rm e}}$ evolution for the four models. The central $Y_{{\rm e}}$ of the two pairs are very similar to 
each other at early time. There is a small bump for Model c3-09950-N-v025-fine, which may be originated
from resolved mixing with outer meshes, which have a higher $Y_{{\rm e}}$ in general.

In the right panel of Figure \ref{fig:dx_plot} we plot the energy evolution for the two comparison models.
The model pair based on Model c3-09950-N-v025 shows very similar evolution 
except near the end.
Both models show a sharp drop of total energy near the end 
of simulation due to the neutrino loss and energy loss by electron captures.
The model with a finer resolution shows an earlier drop in the 
energy consistent with the central density evolution.
On the contrary, the energy curves of the model pair based on Model c3-09950-N 
almost overlap each other.

\begin{figure*}
\centering
\includegraphics*[width=8cm,height=7cm]{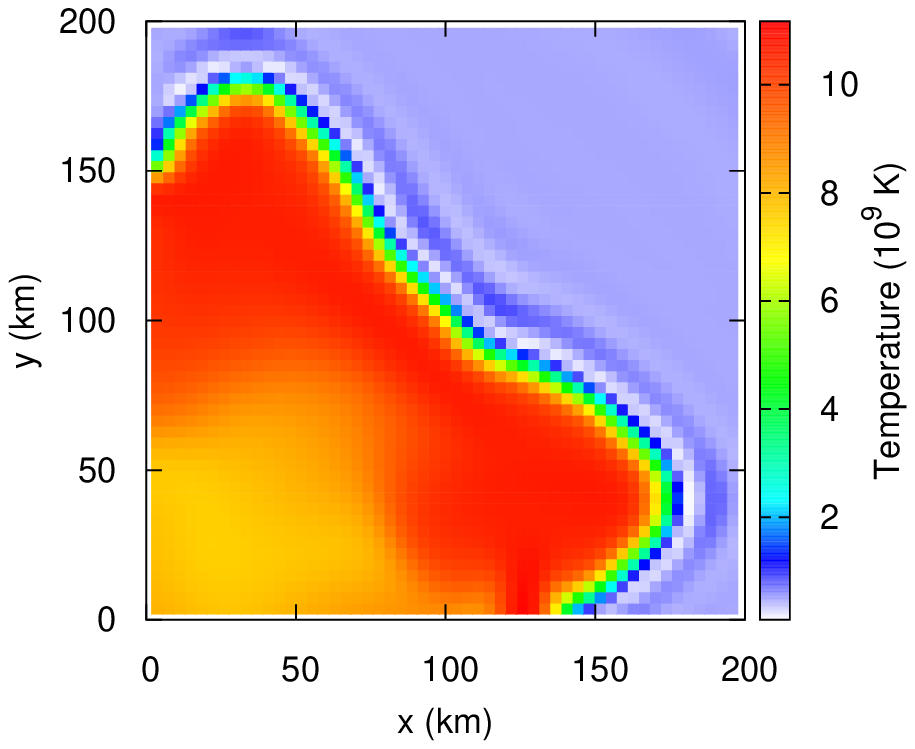}
\includegraphics*[width=8cm,height=7cm]{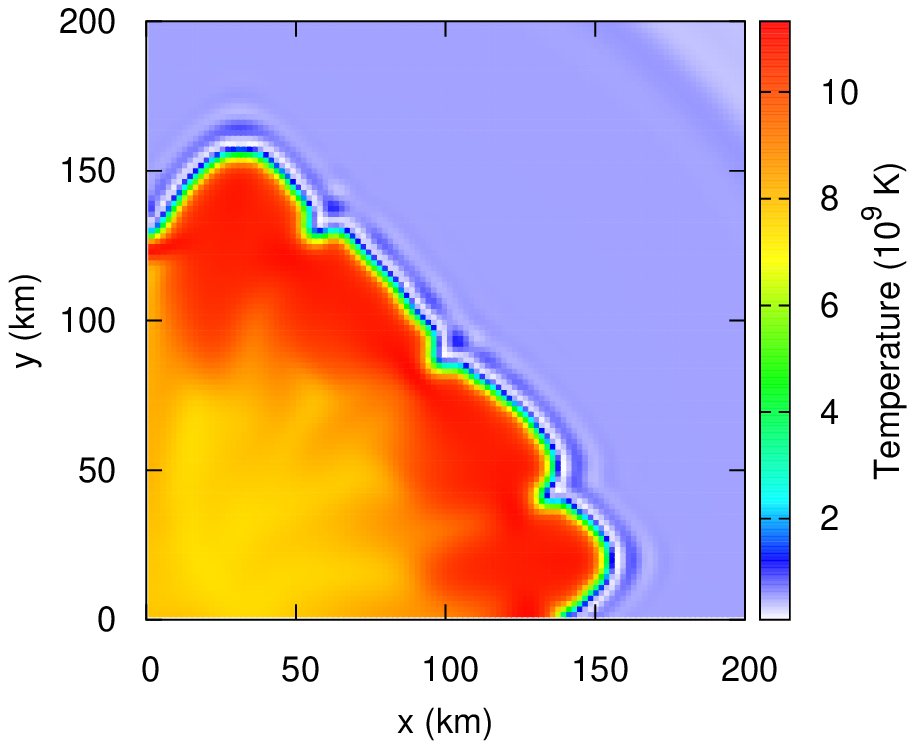}
\caption{(left panel) The temperature colour plot of the Model 
c3-09950-N-v025 at 0.75 s after the deflagration has started.
(right panel) Similar to the left panel but for the Model
at c3-09950-N-v025-fine.}
\label{fig:flame_dx_plot}
\end{figure*}

We can see from the results that a finer resolution in general allows faster
energy production. This suppresses the effects from electron capture
and hence the rate of contraction. 
To further understand the role of resolution size in our simulations, 
we plot in left and right panels of Figure \ref{fig:flame_dx_plot} 
the flame structure of the 
two models at 0.75 s after the simulations have started.
The flame structure of the two models are of similar shape. 
The initial "three-finger" structure is smoothed out by the electron captures. 
The model with a higher resolution 
shows more features on the front than the lower resolution. 
However, the flame size is slightly larger for the lower resolution model
by 20 \%. The core cooled by electron capture is on the contrary
smaller in the same model. Such difference can attribute to the different
contraction rate, where the lower resolution model, due to a more 
extended flame, needs more time for accumulating sufficient matter for 
the final collapse. Despite the difference, the flame structure 
shows that the current resolution can produce very consistent results,
despite a more rigorous proof with a resolution of even smaller $\Delta x$ 
will be needed to verify the convergence.

\section{Possible Observational Signals for the Collapsing Model}

In this section, we estimate the following evolution
for models which collapse into a neutron star.
We remap our models from the two-dimensional cylindrical grid
to the one-dimensional spherical grid by doing an angular average. 
Then we carry out one-dimensional hydrodynamics simulations
from the collapse until bounce occurs. 

In Table \ref{table:models_collapse}, we list the input physics
for doing the 1D modeling in the collapse phase. 
In the 1D simulation, we use the same
WENO $5^{{\rm th}}$ order shock-capturing scheme
and the 3-step $3^{{\rm rd}}$ order NSSP RK scheme
for spatial and temporal discretization. For the EOS,
we use the HShen EOS \citep{Shen1998},
which is based on the relativistic mean-field model to describe the 
homogeneous phase of matter. The table includes extension
with the Thomas-Fermi approximation to describe
the inhomogeneous matter composition. The parameter for the
incompressibility of nuclear matter is 281 MeV and the
symmetry energy has a value of 36.9 MeV. Before bounce
occurs, we use the parametrized neutrino transport 
scheme \citep{Liebendoerfer2005}. This scheme treats
the electron capture as the only neutrino source
and simplifies the neutrino transport by only 
including an instantaneous absorption/emission. 
The neutrino also affects the hydrodynamics
through its pressure in the neutrino-opaque region
as an ideal degenerate Fermi gas. To estimate the 
expected electron capture at high density, we use the 
fitting table in \cite{Abdikamalov2010},
which contains the $Y_{\rm e}$ as a function of density.
The electron fraction of the matter is 
instantaneously converted to the value given by
the table, where the net change of electron capture 
is treated as neutrino source. After bounce,
we switch to the Advanced Leakage Scheme 
\citep{Perego2016}. This scheme can be regarded as the extension
of the leakage scheme \citep{Rosswog2003}, but is a 
simplified scheme of the Isotropic Diffusion Source Approximation (IDSA) 
\citep{Liebendoerfer2009}. It is because this scheme treats the 
neutrino number fraction and mean energy as independent variables
as in IDSA. But in evolving to the new state, in the neutrino 
sector, it always assumes the new state inclines towards to 
the diffusion limit in the optically thin zones
or the trapped limit in the optically thick zones. This guarantees 
that the scheme can approach asymptotically to a solution
for an arbitrary timestep. This can bypass
the difficulty of finding a new state in the original version of
IDSA where occasionally no solution is found in zones where rigorous motion
or discontinuities exist. In our simulations, we use 10 energy
bands of neutrino from 3 MeV to 300 MeV in a logarithmic increasing
band size. Since we want to understand the general properties
of how the collapse takes place, we include only $\nu_e$ and $\nu_{\bar{e}}$
in our calculation with only 2 absorption/emission channels and
4 scattering channels, namely:
\begin{eqnarray}
n + \nu_e \leftrightarrow p + e^-, \\
p- + \nu_{\bar{e}} \leftrightarrow n + e^+, 
\end{eqnarray}
for the absoption/emission, and
\begin{eqnarray}
n + \nu_i \leftrightarrow n + \nu_i, \\
p + \nu_i \leftrightarrow p + \nu_i, \\
\alpha + \nu_i \leftrightarrow \alpha + \nu_i, \\
{\rm ion} + \nu_i \leftrightarrow {\rm ion} + \nu_i
\end{eqnarray}
respectively. 
We use the rate formulae given in \cite{Bruenn1985}. 
Pair neutrino and neutrino bremmstrahlung are not included
in this calculation. But these processes are less
important compare to the channels included, although we note 
that for a long term simulation such as neutron star cooling, 
these two channels gradually dominate over the first two
absorption-emission channels. 

\begin{table*}

\caption{The input physics and the choices of physics models in simulations.}
\begin{center}
\label{table:models_collapse}
\begin{tabular}{|c|c|}
\hline
Input physics & Physics model \\ \hline
Spatial discretization 			& 5$^{{\rm th}}$ order Weighted Essentially Non-Oscillatory 
								  Scheme \citep{Barth1999} \\
Time discretization 			& 5-step 3$^{{\rm rd}}$ order Non-Strong Stability Preserved  
								  Runge-Kutta Scheme \citep{Wang2007} \\ \hline
Baryonic matter EOS 			& HShen EOS \citep{Shen1998} \\ \hline
Pre-bounce electron capture 	& Fitting table from direct Boltzmann transport \citep{Dessart2006,Abdikamalov2010} \\
Pre-bounce neutrino transport 	& Parametrized neutrino transport \citep{Liebendoerfer2005} \\
Post-bounce neutrino transport	& Advanced leakage scheme (ALS) \citep{Perego2016} \\ \hline
\end{tabular}
\end{center}
\end{table*}

Since the advanced leakage scheme does not include neutrino cooling,
which is an important channel for the proto-neutron star to 
lose energy effectively after the neutrinosphere has been
settled, we only run the simulations until $\sim 200$ ms after bounce,
to extract the neutrino signals.

\begin{figure}
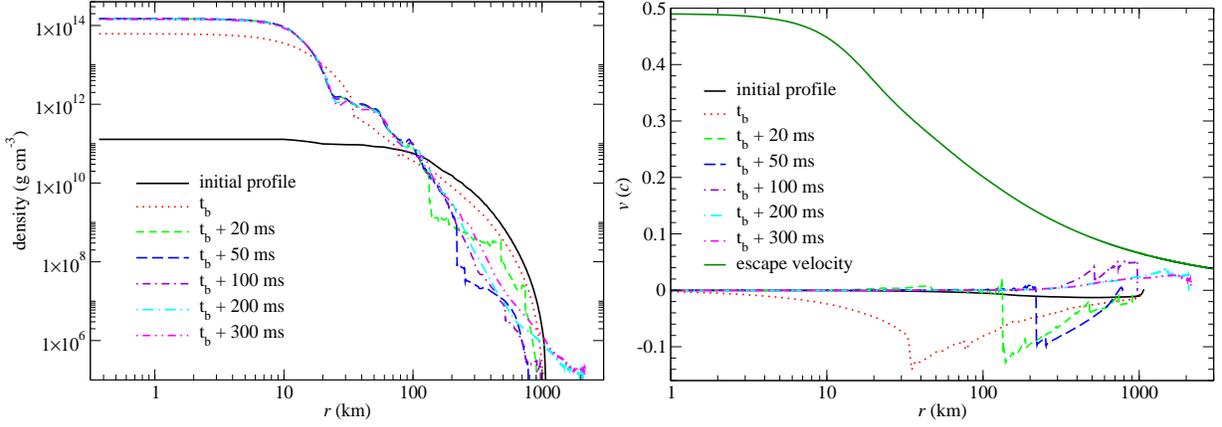

\centering
\includegraphics*[width=8cm,height=5.6cm]{fig24a.eps}
\includegraphics*[width=8cm,height=5.6cm]{fig24b.eps}
\caption{\textit{(left panel)} The density profiles of the Model 
c3-10000-N at the start of simulation (the same profile
as it ends in the 2-dimensional simulation), at the 
bounce, 20, 50, 100, 200 and 300  ms after bounce.
\textit{(right panel)} Similar to left panel, 
but for the velocity profiles.}
\label{fig:rho_rho10.0_sph_plot}
\end{figure}

In the left panel of Figure \ref{fig:rho_rho10.0_sph_plot} we plot the density 
profiles of one of the collapse models c3-10000-N at the beginning
of the one-dimensional simulation, at bounce, 25 ms and 50 ms
after bounce. At the beginning (end of the two-dimensional simulations
in the deflagration phase), the core starts with a flat density profile. 
But the inner core first contracts to reach nuclear density due to 
the loss of pressure by electron capture. At bounce, a stiff core 
made of nuclear matter at density around $3 \times 10^{14}$ g cm$^{-3}$
is formed. The inner envelope shows a steep density gradient showing
that it is still falling onto the neutron star. The outer envelope 
does not change much. At 20 ms after bounce, the neutron star 
core reaches an equilibrium state in density, while the accretion of 
matter of the inner envelope creates a layer outside the neutron star.
At around $10^{12}$ g cm$^{-3}$, strong fluctuations of density appear
due to the tension between the infalling matter from the outer envelope
and the stabilized inner envelope. At 50 ms after bounce, the neutron 
star has a static state envelope about 200 km. The remained envelope
has also contracted significantly to about 500 km, about half of its 
initial radius $\approx 1200$ km. At 100 ms after bounce onward, 
no significant change in the density profile of the neutron star
up to 200 km. But there is still observable motion of the surface showing 
expansion. The cusps in the profiles also disappear.

In the right panel of Figure \ref{fig:rho_rho10.0_sph_plot} we plot the velocity profiles
for the same model similar to the left panel.
At the beginning, the star is having a homologous contraction
with a maximum velocity about $1.3 \times 10^{-2}$ $c$  at about 500 km. 
At bounce, we can see the a neutron star core close to static is formed
with a size of about 15 km. Outside the neutron star there is 
an infalling envelope with a maximum velocity about 0.2 $c$. 
The infalling envelope preserves also the homologous velocity profile.
Through shock heating, the material fallen on the neutron star 
quickly finds a hydrostatic equilibrium state. By examining the 
velocity profile at 20 ms after bounce, the bounce
shock reaches about 100 km from the core, with a slightly lower
infalling velocity about 0.16 $c$. There is outgoing matter 
in the profile at 50 ms after bounce. This shows that the shock has reached
the region where density is low enough for the density gradient becomes
large enough, so that the shock strength increases again 
when it propagates.  
The infalling velocity has decreased to $\approx 0.12 c$. 
Once the accretion shock reaches the surface, since there
is no further matter suppress to the expansion of matter, 
it creates a high velocity flow near the surface. 
Some has a velocity exceeding 
the escape velocity.  Such ejecta is likely to make the event  
a dim and rapidly transient due to its high velocity and low ejecta mass. 
After the ejection of high velocity matter is ejected, the material becomes
bounded. 

\begin{figure}
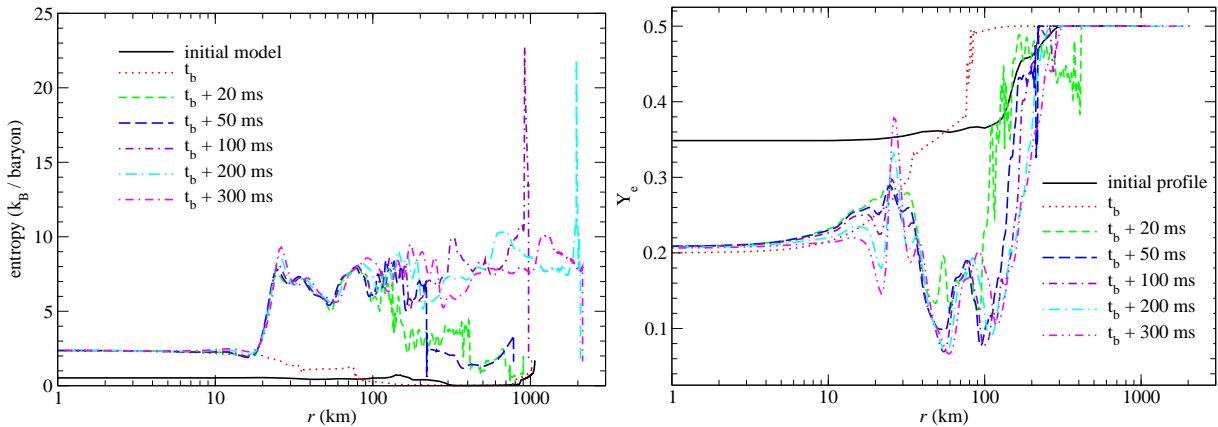

\centering
\includegraphics*[width=8cm,height=5.6cm]{fig25a.eps}
\includegraphics*[width=8cm,height=5.6cm]{fig25b.eps}
\caption{\textit{(left panel)} The entropy profiles of the benchmark
ECSN model at the beginning, at bounce, and at 20, 50, 100, 200 and 300 ms
after bounce. \textit{(right panel)} The $Y_{\rm e}$ profiles of the benchmark
ECSN model at the beginning, at bounce, and at 20, 50, 100, 200 and 300 ms
after bounce. }
\label{fig:ent_rho10.0_sph_plot}
\end{figure}

In Figure \ref{fig:ent_rho10.0_sph_plot} we plot the entropy profiles
similar to Figure \ref{fig:rho_rho10.0_sph_plot}. At the beginning,
the whole star has almost a constant entropy $\approx 0.5$ 
k$_{{\rm B}}$ per baryon, except near the surface.
This is related to the initial flame put in by hand. The initial flame
perturbs the initial hydrostatic equilibrium of the star.
At bounce, the whole star
reaches a constant entropy about 3 k$_{{\rm B}}$ per baryon. There is 
a cusp near the neutron star by the shock interaction. Similar
to the velocity profiles, the quasi-static neutron star core has a
constant entropy. At 20 ms, there is a significant rise of entropy to about 
10 k$_{{\rm B}}$ per baryon in the newly accreted layer from 10 - 80 km.
The high entropy region can be compared with the velocity profile, which is the 
region which comes to a rest after deposited on the neutron star surface. 
At 50 ms, the shock has reached 200 km and a high entropy domain
forms up to about 110 km. This is consistent with the literature
that the neutrino heating is essential in producing high entropy matter,
which is supposed to be found in the ejecta. At 100 ms onwards, there
is no significant change to the entropy profiles where a flat
constant entropy zone is created in the envelope. At 100 ms after bounce, 
the ejecta has an entropy peak as high as 
$\sim 20$ k$_{{\rm B}}$ per baryon.

In the right panel of Figure \ref{fig:ent_rho10.0_sph_plot} we plot the $Y_{\rm e}$ profiles
of the ECSN model similar to previous plot at the same time slice. 
The beginning $Y_{\rm e}$ profile is directly imported from 
the collapsing model in the main text. So, the core has reached
a minimum of $\sim 0.35$ and gradually increases at 100 km up to 0.45. 
At no electron capture takes place beyond 200 km, where 
the deflagration has not yet reached the matter. 
At bounce, the core $Y_{\rm e}$ reaches 0.2 and gradually 
increases to 0.35 at $\sim 60$ km, and up to 0.5 at 80 km.
The locally higher $Y_{\rm e}$ from 80 - 200 km is 
because of the advection of matter. The high $Y_{\rm e}$ matter
falls inwards, but has not reached the density for electron
capture, so locally it looks like the $Y_{\rm e}$ increases
by itself. After bounce, the shock and the consequent
neutrino interactions influence the $Y_{\rm e}$ distribution. 
The high temperature allows rapid neutrino emission, 
which creates a trough of $Y_{\rm e}$ from 30 - 100 km. 
Ripples of $Y_{\rm e}$ appears due to the finite 
partitioning of neutrino energy band. As the shock propagates
outwards, at 100, 200 and then 300 ms, we can see the trough 
widens. Furthermore, the neutrinos, which diffuse outwards
outside the neutrinosphere, smooth out the $Y_{\rm e}$ 
fluctuations created by acoustic waves right after bounce. 
 
\begin{figure}
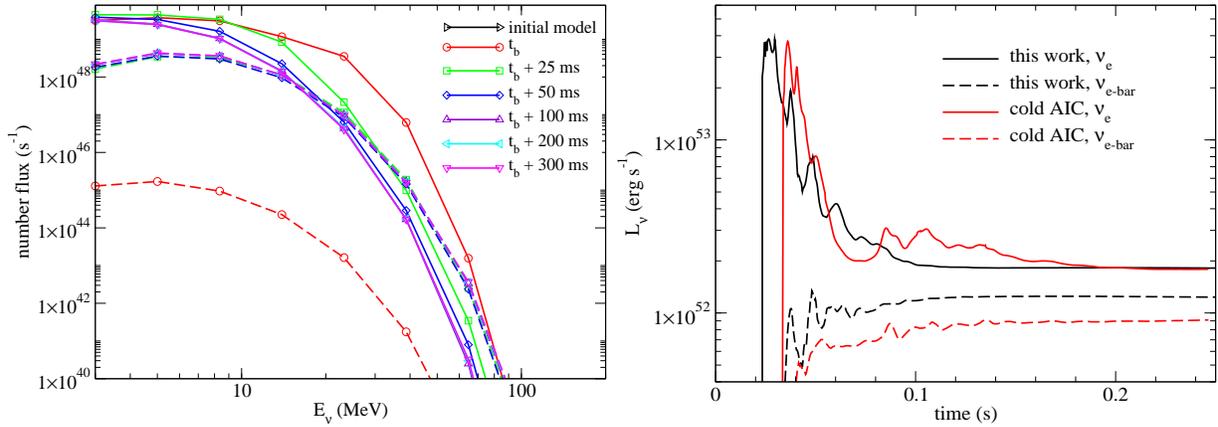

\centering
\includegraphics*[width=8cm,height=5.6cm]{fig26a.eps}
\includegraphics*[width=8cm,height=5.6cm]{fig26b.eps}
\caption{\textit{(left panel)} Similar to the left panel
but for the free streaming neutrino flux. The solid (dashed) line stands
for the $\nu_e$ ($\nu_{\bar{e}}$) flux at 300 km from the NS core.
\textit{(right panel)} The $\nu_e$ and $\nu_{\bar{e}}$ luminosity for the 
Model c3-10000-N. The sample neutrino luminosity from the 
collapse of an AIC is included for comparison.}
\label{fig:nulumin_rho10.0_sph_plot}
\end{figure}

In the left panel of Figure \ref{fig:nulumin_rho10.0_sph_plot} we plot the neutrino
energy spectra of the same model similar to 
Figure \ref{fig:rho_rho10.0_sph_plot}. The number flux is taken at
300 km from the neutron star core. The number reaching the Earth
can be scaled accordingly. There is no data for the initial model
because no matter has reached nuclear density. At bounce, one can see
the $\nu_e$ has already a spectrum comparable with the thermal spectrum.
But the $\nu_{\bar{e}}$ spectrum is still extremely low. 
At 25 ms after bounce, $\nu_e$ has relaxed with a lower 
high energy $\nu_e$ since the neutrinosphere is in general farther
from center, which has a lower temperature. The $\nu_{\bar{e}}$
has also settled down to a thermal distribution. At 50 ms,
both types of neutrinos have reached an equilibrium distribution.
There are more low energy $\nu_e$ but more high energy $\nu_{\bar{e}}$. 
 
In the right panel of Figure \ref{fig:nulumin_rho10.0_sph_plot} we plot the
$\nu_e$ and $\nu_{\bar{e}}$ luminosity against time for the same model. 
The neutrino signal from an accretion induced collapse of a
WD into a neutron star is also plotted for comparison. 
The accretion induce collapse assumes a simple collapse of a
Chandrasekhar mass isothermal WD due to an initial reduction
of $Y_{\rm e}$. It can be seen that qualitatively the two models are similar.
At the beginning, a strong pulse of $\nu_e$ is emitted.
But as the neutrinosphere of different energy bands starts to form. The
neutrino emission drops. After a few expansion of the envelope, it reaches an 
equilibrium value about $2 \times 10^{52}$ erg s$^{-1}$. One minor 
difference is that the ONeMg case shows more oscillations than the cold AIC
case. The $\nu_{\bar{e}}$ shows a similar behaviour. It has a much 
lower luminosity. Consistent to the literature, the first peak appears
later than the $\nu_e$ peak, about 20 ms after. The ONeMg model has
about 50 \% higher $\nu_{\bar{e}}$ flux than the cold AIC model. 

\begin{figure}
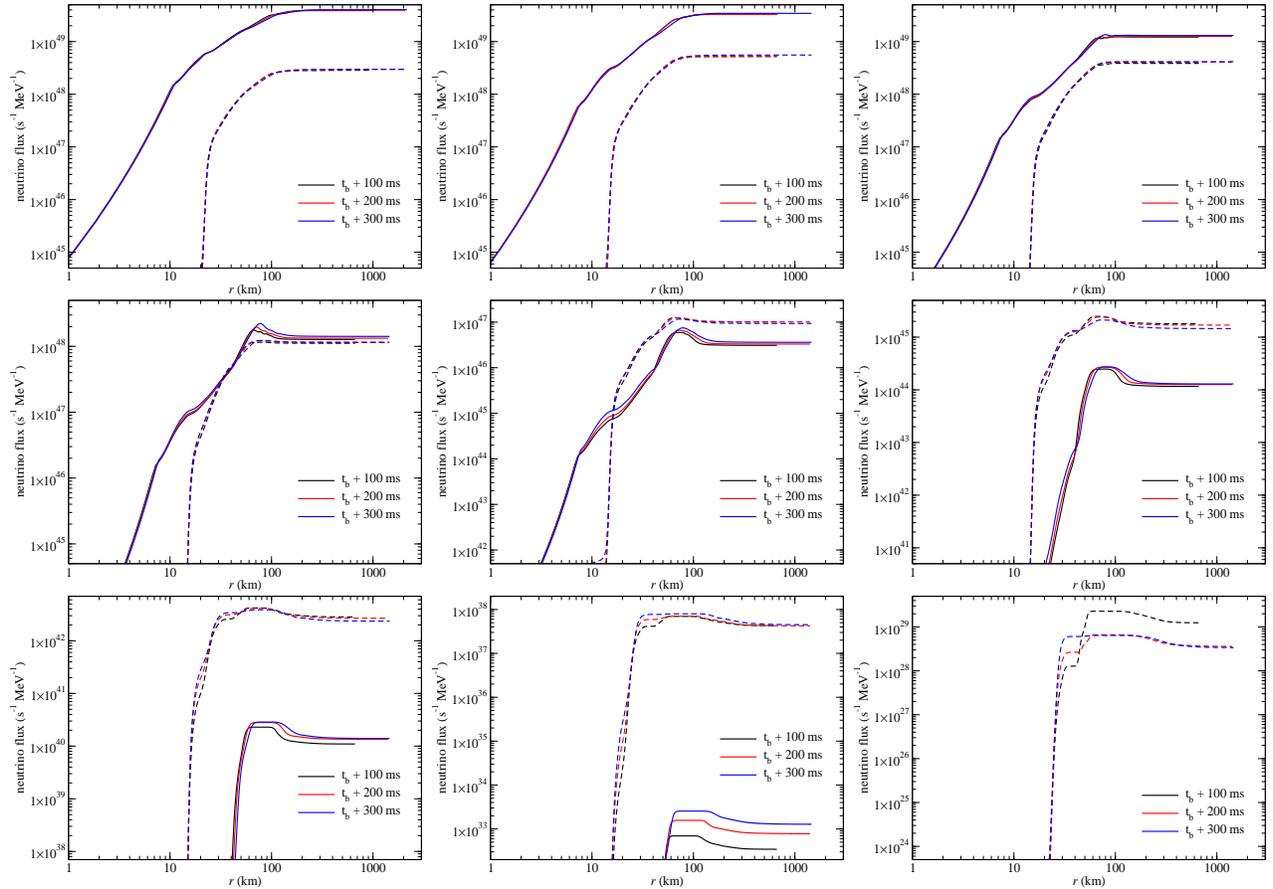

\centering
\includegraphics*[width=5.5cm,height=3.9cm]{fig27a.eps}
\includegraphics*[width=5.5cm,height=3.9cm]{fig27b.eps}
\includegraphics*[width=5.5cm,height=3.9cm]{fig27c.eps}
\includegraphics*[width=5.5cm,height=3.9cm]{fig27d.eps}
\includegraphics*[width=5.5cm,height=3.9cm]{fig27e.eps}
\includegraphics*[width=5.5cm,height=3.9cm]{fig27f.eps}
\includegraphics*[width=5.5cm,height=3.9cm]{fig27g.eps}
\includegraphics*[width=5.5cm,height=3.9cm]{fig27h.eps}
\includegraphics*[width=5.5cm,height=3.9cm]{fig27i.eps}
\caption{The $\nu_e$ (solid line) and $\nu_{\bar{e}}$ (dashed line)
number flux profile at 100 (black), 200 (red) and 300 (blue) ms 
after bounce. The neutrino energy bands include 3, 5, 8, 14, 23, 
39, 65, 108 and 180 MeV.}
\label{fig:nufluxprofile_rho10.0_sph_plot}
\end{figure}

At last we plot at Figure \ref{fig:nufluxprofile_rho10.0_sph_plot} the neutrino
number flux profile at 100, 200 and 300 ms after bounce for both 
$\nu_e$ (solid line) and $\nu_{\bar{e}}$ (dashed line). 
For low energy bands (3 MeV - 8 MeV), $\nu_e$ is the dominant 
species. They are mostly created just outside the NS,
surface. No neutrino absorption can be seen and most neutrinos
are produced within the innermost 100 km. On the contrary,
$\nu_{\bar{e}}$ is completely not produced in the NS,
and is gradually produced in the shock-heated matter
outside the NS. Its number emission is at least one order of magnitude lower. 
However, as neutrino energy increases, the drop of $\nu_e$ number flux
is faster than the drop of $\nu_{\bar{e}}$. It is because the creation
of $\nu_{\bar{e}}$ is limited to places where positron can be freely formed.
Notice that to create $\nu_e$, the electron should have a
chemical potential not only for the creation of itself,
but also the mass difference between $n$ and $p$ ($\sim$ 1.2 MeV).
At 20 - 100 km, the density has already drops below 
$10^{12}$ g cm$^{-3}$. This means the nucleons is no longer 
degenerate and thus it has a much lower chemical potential 
than those in the core. So, this leaves a strong cutoff 
in the high energy $\nu_e$. On the other hand, the 
production of $\nu_{\bar{e}}$ is aided by the 
energy difference for the same origin. 
So, its drop in number flux is less steep
than $\nu_e$. 

For a higher neutrino energy, more features can be observed.
At 14, 23 and 39 MeV, both $\nu_e$ and $\nu_{\bar{e}}$ show
a first increasing function up to 80 km and then 
slightly drop until 100 km. The change of $\nu_e$ 
is larger than that of $\nu_{\bar{e}}$, showing that 
more $\nu_e$ is absorbed. As a result, this explains 
the local bump of $Y_{\rm e}$ in the right panel 
of Figure \ref{fig:ent_rho10.0_sph_plot}. 

For even higher neutrino energy (65, 108 and 180 MeV),
the drops of $\nu_e$ becomes so rapid that it becomes
irrelevant to the neutrino transport and the 
global neutrino flux. $\nu_{\bar{e}}$ also shows 
a similar feature but with lower strength. 
But they are also unimportant to the global neutrino 
population. 


\bibliographystyle{apj}
\pagestyle{plain}
\bibliography{biblio}

\begin{thebibliography}{}
\expandafter\ifx\csname natexlab\endcsname\relax\def\natexlab#1{#1}\fi

\bibitem[{A.~Perego(2016)}]{Perego2016}
A.~Perego, R. M.~Cabezon, R.~K. 2016, ApJS, 223, 22

\bibitem[{Abdikamalov {et~al.}(2010)}]{Abdikamalov2010}
Abdikamalov, E.~B., {et~al.} 2010, Phys. Rev. D, 81, 044012

\bibitem[{{Arnett}(1996)}]{Arnett1996}
{Arnett}, D. 1996, {Supernovae and Nucleosynthesis: An Investigation of the
  History of Matter from the Big Bang to the Present}

\bibitem[{Barth \& Deconinck(1999)}]{Barth1999}
Barth, T.~J., \& Deconinck, H. 1999, Lecture Notes in Computational Science and
  Engineering 9: High-Order Methods for Computational Physics (Springer)

\bibitem[{{Bell} {et~al.}(2004{\natexlab{a}}){Bell}, {Day}, {Rendleman},
  {Woosley}, \& {Zingale}}]{Bell2004a}
{Bell}, J.~B., {Day}, M.~S., {Rendleman}, C.~A., {Woosley}, S.~E., \&
  {Zingale}, M. 2004{\natexlab{a}}, \apj, 606, 1029

\bibitem[{{Bell} {et~al.}(2004{\natexlab{b}}){Bell}, {Day}, {Rendleman},
  {Woosley}, \& {Zingale}}]{Bell2004b}
---. 2004{\natexlab{b}}, \apj, 608, 883

\bibitem[{Bruenn(1985)}]{Bruenn1985}
Bruenn, S.~W. 1985, Astrophys. J. Suppl., 58, 771

\bibitem[{Canal \& Schatzman(1976)}]{Canal1976}
Canal, R., \& Schatzman, E. 1976, Astron. Astrophys., 46, 229

\bibitem[{{Chechetkin} {et~al.}(1980){Chechetkin}, {Gershtein}, {Imshennik},
  {Ivanova}, \& {Khlopov}}]{Chechetkin1980}
{Chechetkin}, V.~M., {Gershtein}, S.~S., {Imshennik}, V.~S., {Ivanova}, L.~N.,
  \& {Khlopov}, M.~I. 1980, \apss, 67, 61

\bibitem[{Dessart {et~al.}(2006)Dessart, Burrows, Ott, {et~al.}}]{Dessart2006}
Dessart, L., Burrows, A., Ott, C., {et~al.} 2006, Astrophys. J., 644, 1063

\bibitem[{{Doherty} {et~al.}(2015){Doherty}, {Gil-Pons}, {Siess}, {Lattanzio},
  \& {Lau}}]{Doherty2015}
{Doherty}, C.~L., {Gil-Pons}, P., {Siess}, L., {Lattanzio}, J.~C., \& {Lau},
  H.~H.~B. 2015, \mnras, 446, 2599

\bibitem[{{Fink} {et~al.}(2014){Fink}, {Kromer}, {Seitenzahl},
  {Ciaraldi-Schoolmann}, {R{\"o}pke}, {Sim}, {Pakmor}, {Ruiter}, \&
  {Hillebrandt}}]{Fink2014}
{Fink}, M., {Kromer}, M., {Seitenzahl}, I.~R., {et~al.} 2014, \mnras, 438, 1762

\bibitem[{{Gershte{\v{i}}n} {et~al.}(1977){Gershte{\v{i}}n}, {Ivanova},
  {Imshennik}, {Khlopov}, \& {Chechetkin}}]{Gershtevin1977}
{Gershte{\v{i}}n}, S.~S., {Ivanova}, L.~N., {Imshennik}, V.~S., {Khlopov},
  M.~Y., \& {Chechetkin}, V.~M. 1977, Soviet Journal of Experimental and
  Theoretical Physics Letters, 26, 178

\bibitem[{Hashimoto {et~al.}(1993)Hashimoto, Iwamoto, \& Nomoto}]{hashi93}
Hashimoto, M., Iwamoto, K., \& Nomoto, K. 1993, Astrophys. J., 803, 72

\bibitem[{Hicks(2015)}]{Hicks2015}
Hicks, E.~P. 2015, Astrophys. J., 803, 72

\bibitem[{{Jones} {et~al.}(2014){Jones}, {Hirschi}, \& {Nomoto}}]{Jones2014}
{Jones}, S., {Hirschi}, R., \& {Nomoto}, K. 2014, \apj, 797, 83

\bibitem[{{Jones} {et~al.}(2016){Jones}, {R{\"o}pke}, {Pakmor}, {Seitenzahl},
  {Ohlmann}, \& {Edelmann}}]{Jones2016}
{Jones}, S., {R{\"o}pke}, F.~K., {Pakmor}, R., {et~al.} 2016, \aap, 593, A72

\bibitem[{{Jones} {et~al.}(2013){Jones}, {Hirschi}, {Nomoto}, {Fischer},
  {Timmes}, {Herwig}, {Paxton}, {Toki}, {Suzuki}, {Mart{\'{\i}}nez-Pinedo},
  {Lam}, \& {Bertolli}}]{Jones2013}
{Jones}, S., {Hirschi}, R., {Nomoto}, K., {et~al.} 2013, \apj, 772, 150

\bibitem[{{Jones} {et~al.}(2019){Jones}, {R{\"o}pke}, {Fryer}, {Ruiter},
  {Seitenzahl}, {Nittler}, {Ohlmann}, {Reifarth}, {Pignatari}, \&
  {Belczynski}}]{Jones2018}
{Jones}, S., {R{\"o}pke}, F.~K., {Fryer}, C., {et~al.} 2019, \aap, 622, A74

\bibitem[{{Karakas}(2017)}]{Karakas2017}
{Karakas}, A.~I. 2017, {in Handbook of Supernovae}, ed. A.~W. {Alsabti} \&
  P.~{Murdin} (Cham: Springer International Publishing), 461

\bibitem[{{Khokhlov} {et~al.}(1997){Khokhlov}, {Oran}, \&
  {Wheeler}}]{Khokhlov1997}
{Khokhlov}, A.~M., {Oran}, E.~S., \& {Wheeler}, J.~C. 1997, \apj, 478, 678

\bibitem[{Kim {et~al.}(2012)Kim, Kim, Choptuik, \& Lee}]{Kim2012}
Kim, J., Kim, H.~I., Choptuik, M.~W., \& Lee, H.~M. 2012, Mon. Not. R. astr.
  Soc., 424, 830

\bibitem[{Kitamura(2000)}]{Kitamura2000}
Kitamura, H. 2000, Astrophys. J., 539, 888

\bibitem[{Langanke \& Martinez-Pinedo(2001)}]{Langanke2001}
Langanke, K., \& Martinez-Pinedo, G. 2001, ADNDT, 79, 1

\bibitem[{Langer(2012)}]{langer12}
Langer, N. 2012, Ann. Rev. Astron. Astrophys., 50, 107

\bibitem[{Leung {et~al.}(2015{\natexlab{a}})Leung, Chu, \& Lin}]{Leung2015a}
Leung, S.-C., Chu, M.-C., \& Lin, L.-M. 2015{\natexlab{a}}, Mon. Not. R. astr.
  Soc., 454, 1238

\bibitem[{Leung {et~al.}(2015{\natexlab{b}})Leung, Chu, \& Lin}]{Leung2015b}
---. 2015{\natexlab{b}}, Astrophys. J., 812, 110

\bibitem[{{Leung} \& {Nomoto}(2017)}]{Leung2016}
{Leung}, S.-C., \& {Nomoto}, K. 2017, in 14th International Symposium on Nuclei
  in the Cosmos (NIC2016), ed. S.~{Kubono}, T.~{Kajino}, S.~{Nishimura},
  T.~{Isobe}, S.~{Nagataki}, T.~{Shima}, \& Y.~{Takeda}, 020506

\bibitem[{{Leung} \& {Nomoto}(2018)}]{Leung2017d}
{Leung}, S.-C., \& {Nomoto}, K. 2018, \apj, 861, 143

\bibitem[{Liebendoerfer(2005)}]{Liebendoerfer2005}
Liebendoerfer, M. 2005, Astrophys. J., 633, 1042

\bibitem[{Liebendoerfer {et~al.}(2009)Liebendoerfer, Whitehouse, \&
  Fischer}]{Liebendoerfer2009}
Liebendoerfer, M., Whitehouse, S.~C., \& Fischer, T. 2009, Astrophys. J., 698,
  1174

\bibitem[{{Livne} \& {Arnett}(1993)}]{Livne1993}
{Livne}, E., \& {Arnett}, D. 1993, \apj, 415, L107

\bibitem[{{Long} {et~al.}(2014){Long}, {Jordan}, {van Rossum}, {Diemer},
  {Graziani}, {Kessler}, {Meyer}, {Rich}, \& {Lamb}}]{Long2014}
{Long}, M., {Jordan}, George~C., I., {van Rossum}, D.~R., {et~al.} 2014, \apj,
  789, 103

\bibitem[{{Ma} {et~al.}(2013){Ma}, {Woosley}, {Malone}, {Almgren}, \&
  {Bell}}]{Ma2013}
{Ma}, H., {Woosley}, S.~E., {Malone}, C.~M., {Almgren}, A., \& {Bell}, J. 2013,
  \apj, 771, 58

\bibitem[{{Miyaji} \& {Nomoto}(1987)}]{Miyaji1987}
{Miyaji}, S., \& {Nomoto}, K. 1987, \apj, 318, 307

\bibitem[{Miyaji \& Nomoto(1987)}]{miya85}
Miyaji, S., \& Nomoto, K. 1987, Astrophys. J., 318, 307

\bibitem[{Miyaji {et~al.}(1980)Miyaji, Nomoto, Yokoi, \& Sugimoto}]{Miyaji1980}
Miyaji, S., Nomoto, K., Yokoi, K., \& Sugimoto, D. 1980, PASJ, 32, 303

\bibitem[{Mochkovitch \& Livio(1989)}]{Mochkovitch1989}
Mochkovitch, R., \& Livio, M. 1989, Astron. Astrophys., 209, 111

\bibitem[{Nabi \& Klapdor-Kleingrothaus(1999)}]{Nabi1999}
Nabi, J.-U., \& Klapdor-Kleingrothaus, H.~V. 1999, ADNDT, 71, 149

\bibitem[{Niemeyer {et~al.}(1995)Niemeyer, Hillebrandt, \&
  Woosley}]{Niemeyer1995b}
Niemeyer, J.~C., Hillebrandt, W., \& Woosley, S.~E. 1995, Astrophys. J., 452,
  979

\bibitem[{Nomoto(1982)}]{Nomoto1982}
Nomoto, K. 1982, Astrophys. J., 253, 798

\bibitem[{Nomoto(1984)}]{Nomoto1984}
---. 1984, Astrophys. J., 277, 791

\bibitem[{Nomoto(1987)}]{Nomoto1987}
---. 1987, Astrophys. J., 322, 206

\bibitem[{{Nomoto} \& {Hashimoto}(1988)}]{Nomoto1988}
{Nomoto}, K., \& {Hashimoto}, M. 1988, \physrep, 163, 13

\bibitem[{{Nomoto} {et~al.}(2013){Nomoto}, {Kobayashi}, \&
  {Tominaga}}]{Nomoto2013}
{Nomoto}, K., {Kobayashi}, C., \& {Tominaga}, N. 2013, \araa, 51, 457

\bibitem[{Nomoto \& Kondo(1991)}]{Nomoto1991}
Nomoto, K., \& Kondo, Y. 1991, Astrophys. J., 367, L19

\bibitem[{Nomoto \& Leung(2017{\natexlab{a}})}]{Nomoto2017a}
Nomoto, K., \& Leung, S.-C. 2017{\natexlab{a}}, in Handbook of Supernovae, ed.
  A.~W. {Alsabti} \& P.~{Murdin} (Cham: Springer International Publishing),
  1275

\bibitem[{Nomoto \& Leung(2017{\natexlab{b}})}]{Nomoto2017b}
---. 2017{\natexlab{b}}, in Handbook of Supernovae, ed. A.~W. {Alsabti} \&
  P.~{Murdin} (Cham: Springer International Publishing), 483

\bibitem[{Nomoto {et~al.}(1982)}]{nom82crab}
Nomoto, K., {et~al.} 1982, Nature, 299, 803

\bibitem[{{Osher} \& {Sethian}(1988)}]{Osher1988}
{Osher}, S., \& {Sethian}, J.~A. 1988, Journal of Computational Physics, 79, 12

\bibitem[{Peter(1999)}]{Peter1999}
Peter, N. 1999, JPM, 384, 107

\bibitem[{Plewa(2007)}]{Plewa2007}
Plewa, T. 2007, Astrophys. J., 657, 942

\bibitem[{Pocheau(1994)}]{Pocheau1994}
Pocheau, A. 1994, Phys. Rev. E, 49, 1109

\bibitem[{Pumo {et~al.}(2009)}]{pumo09}
Pumo, M.~L., {et~al.} 2009, Astrophys. J., 705, L138

\bibitem[{{Raddi} {et~al.}(2018){Raddi}, {Hollands}, {Koester}, {G{\"a}nsicke},
  {Gentile Fusillo}, {Hermes}, \& {Townsley}}]{Raddi2018}
{Raddi}, R., {Hollands}, M.~A., {Koester}, D., {et~al.} 2018, \apj, 858, 3

\bibitem[{{Reinecke} {et~al.}(1999){Reinecke}, {Hillebrandt}, \&
  {Niemeyer}}]{Reinecke1999b}
{Reinecke}, M., {Hillebrandt}, W., \& {Niemeyer}, J.~C. 1999, \aap, 347, 739

\bibitem[{{Reinecke} {et~al.}(2002{\natexlab{a}}){Reinecke}, {Hillebrandt}, \&
  {Niemeyer}}]{Reinecke2002a}
---. 2002{\natexlab{a}}, \aap, 386, 936

\bibitem[{{Reinecke} {et~al.}(2002{\natexlab{b}}){Reinecke}, {Hillebrandt}, \&
  {Niemeyer}}]{Reinecke2002b}
---. 2002{\natexlab{b}}, \aap, 391, 1167

\bibitem[{{R{\"o}pke}(2005)}]{Roepke2005a}
{R{\"o}pke}, F.~K. 2005, \aap, 432, 969

\bibitem[{{R{\"o}pke} \& {Hillebrandt}(2005)}]{Roepke2005b}
{R{\"o}pke}, F.~K., \& {Hillebrandt}, W. 2005, \aap, 431, 635

\bibitem[{{R{\"o}pke} {et~al.}(2004{\natexlab{a}}){R{\"o}pke}, {Hillebrandt},
  \& {Niemeyer}}]{Roepke2004a}
{R{\"o}pke}, F.~K., {Hillebrandt}, W., \& {Niemeyer}, J.~C. 2004{\natexlab{a}},
  \aap, 420, 411

\bibitem[{{R{\"o}pke} {et~al.}(2004{\natexlab{b}}){R{\"o}pke}, {Hillebrandt},
  \& {Niemeyer}}]{Roepke2004b}
---. 2004{\natexlab{b}}, \aap, 421, 783

\bibitem[{{R{\"o}pke} {et~al.}(2007){R{\"o}pke}, {Hillebrandt}, {Schmidt},
  {Niemeyer}, {Blinnikov}, \& {Mazzali}}]{Roepke2007}
{R{\"o}pke}, F.~K., {Hillebrandt}, W., {Schmidt}, W., {et~al.} 2007, \apj, 668,
  1132

\bibitem[{Rosswog \& Liebendoerfer(2003)}]{Rosswog2003}
Rosswog, S., \& Liebendoerfer, M. 2003, Mon. Not. R. astr. Soc., 342, 673

\bibitem[{Schmidt {et~al.}(2006)Schmidt, Niemeyer, Hillebrndt, \&
  Roepke}]{Schmidt2006b}
Schmidt, W., Niemeyer, J.~C., Hillebrndt, W., \& Roepke, F.~K. 2006, Astron.
  Astrophys., 450, 283

\bibitem[{Schwab {et~al.}(2017)Schwab, Martinez-Rodriguez, Piro, \&
  Badenes}]{Schwab2017}
Schwab, J., Martinez-Rodriguez, H., Piro, A.~L., \& Badenes, C. 2017,
  Astrophys. J., 851, 105

\bibitem[{Schwab {et~al.}(2015)Schwab, Quataert, \& Bildsten}]{Schwab2015}
Schwab, J., Quataert, E., \& Bildsten, L. 2015, Mon. Not. R. astr. Soc., 453,
  1910

\bibitem[{Seitenzahl {et~al.}(2010)Seitenzahl, Ropeke, Fink, \&
  Pakmor}]{Seitenzahl2010}
Seitenzahl, I.~R., Ropeke, F.~K., Fink, M., \& Pakmor, R. 2010, Mon. Not. R.
  astr. Soc., 407, 2297

\bibitem[{Shen {et~al.}(1998)Shen, Toki, Oyamatsu, \& Sumiyoshi}]{Shen1998}
Shen, H., Toki, H., Oyamatsu, K., \& Sumiyoshi, K. 1998, Nucl. Phys. A, 637,
  435

\bibitem[{Siess(2007)}]{sie07}
Siess, L. 2007, Astron. and Astropart., 476, 893

\bibitem[{Sugimoto \& Nomoto(1980)}]{Sugimoto1980}
Sugimoto, K., \& Nomoto, K. 1980, Space Sci. Rev., 25, 155

\bibitem[{Suzuki {et~al.}(2016)Suzuki, Toko, \& Nomoto}]{Suzuki2016}
Suzuki, T., Toko, H., \& Nomoto, K. 2016, Astrophys. J., 817, 163

\bibitem[{Takahashi {et~al.}(2013)}]{taka13}
Takahashi, K., {et~al.} 2013, Astrophys. J., 771, 28

\bibitem[{Timmes(1999)}]{Timmes1999a}
Timmes, F.~X. 1999, Astrophys. J., 124, 241

\bibitem[{Timmes \& Arnett(1999)}]{Timmes1999b}
Timmes, F.~X., \& Arnett, D. 1999, Astrophys. J., 125, 277

\bibitem[{Timmes \& Woosley(1992)}]{Timmes1992}
Timmes, F.~X., \& Woosley, S.~E. 1992, Astrophys. J., 396, 649

\bibitem[{Toki {et~al.}(2013)Toki, Suzuji, Nomoto, {et~al.}}]{Toki2013}
Toki, H., Suzuji, T., Nomoto, K., {et~al.} 2013, Phys. Rev. C, 88, 015806

\bibitem[{Townsley {et~al.}(2007)Townsley, Calder, Asida,
  {et~al.}}]{Townsley2007}
Townsley, D.~M., Calder, A.~M., Asida, S.~M., {et~al.} 2007, Astrophys. J.
  Suppl., 143, 201

\bibitem[{{Wanajo} {et~al.}(2018){Wanajo}, {M{\"u}ller}, {Janka}, \&
  {Heger}}]{Wanajo2018}
{Wanajo}, S., {M{\"u}ller}, B., {Janka}, H.-T., \& {Heger}, A. 2018, \apj, 852,
  40

\bibitem[{Wang \& Spiteri(2007)}]{Wang2007}
Wang, R., \& Spiteri, R.~J. 2007, SIAM J. Numer. Anal., 45, 1871

\bibitem[{{Woosley}(1997)}]{Woosley1997}
{Woosley}, S.~E. 1997, \apj, 476, 801

\bibitem[{{Woosley} \& {Heger}(2015)}]{Woosley2015}
{Woosley}, S.~E., \& {Heger}, A. 2015, \apj, 810, 34

\bibitem[{Yoon {et~al.}(2007)}]{Yoon2007}
Yoon, S.-C., {et~al.} 2007, Mon. Not. R. astr. Soc., 380, 993

\end{thebibliography}

\end{document}